\newif\ifreport
\newif\ifnotreport
\reporttrue 

\ifreport
	\notreportfalse
\else
	\notreporttrue
\fi

\ifreport
	\documentclass{article}
\else
	\documentclass[acmsmall, review, screen]{acmart}
	\setcitestyle{square}
\fi 

\usepackage{geometry}
\usepackage[T1]{fontenc}
\usepackage{ae,aecompl}

\usepackage[utf8]{inputenc}

\usepackage{graphicx}

\usepackage[english]{babel}

\usepackage{verbatim}
\usepackage{listings}

\usepackage{amsfonts}
\usepackage{amsmath}
\usepackage{amssymb}
\usepackage{upgreek}
\usepackage{stmaryrd}

\usepackage{amsthm}

\usepackage{subcaption}

\usepackage{xcolor,soul}

\usepackage{tikz}

\usepackage{tabularx}
\usepackage{multirow}
\usepackage{booktabs}
\usepackage{longtable}

\usepackage{paralist}

\ifreport
	\usepackage[sort&compress,square,comma,numbers]{natbib}
\fi

\usepackage{hyperref}

\ifreport
	\usepackage{lineno}
\fi

\ifreport
	\usepackage{cleveref}
\else
	\usepackage[nameinlink, noabbrev,capitalise]{cleveref}
\fi

\usepackage[ruled]{algorithm2e}

\graphicspath{{imgs/}}

\theoremstyle{definition}
\newtheorem{ReferenceSummary}{Ref}

\newenvironment{Reference}[2]%
{
\begin{ReferenceSummary}\label{ref:#2}
\textbf{#2}: #1~\cite{#2}.
\begin{proof}[Summary]
}{
\end{proof}
\end{ReferenceSummary}
}

\newtheorem{ToolSummary}{Tool}

{
\begin{ToolSummary}\label{tool:#2}
\textbf{\lstinline$#2$}: #1.
\begin{proof}[Summary]
}{
\end{proof}
\end{ToolSummary}
}

\newcommand{\refref}[1]{%
\hyperref[ref:#1]{\lstinline$#1$}%
}

\newcommand{\catref}[1]{%

\hyperref[#1]{\lstinline$#1$}

}

\newcommand{\toolref}[1]{%
\hyperref[tool:#1]{\lstinline$#1$}%
}

\newcommand{\claudio}[1]{\textbf{\textcolor{red}{Claudio: #1}}}
\renewcommand{\claudio}[1]{}  
\newcommand{\todocasper}[1]{\todo[inline,size=\tiny,color=blue!30]{Casper: #1}}
\renewcommand{\todocasper}[1]{}	
\newcommand{\todopeter}[1]{\todo[inline,size=\tiny,color=orange!30]{Peter: #1}}
\renewcommand{\todopeter}[1]{}	

\newcommand{\algoref}[1]{Algorithm~\ref{#1}}

\newcommand{\ra}[1]{\renewcommand{\arraystretch}{#1}}

\newcommand{\footurl}[1]{\footnote{\url{#1}}}

\newcommand{\deltaext}[0]{\ensuremath{\delta^{\mathit{ext}}}}
\newcommand{\deltaint}[0]{\ensuremath{\delta^{\mathit{int}}}}
\newcommand{\ta}[0]{\ensuremath{\mathit{ta}}}
\newcommand{\tn}[0]{\ensuremath{\mathit{tn}}}
\newcommand{\ts}[0]{\ensuremath{\mathit{ts}}}
\newcommand{\tl}[0]{\ensuremath{\mathit{tl}}}
\newcommand{\WCT}[0]{\ensuremath{\mathit{WcT}}}
\newcommand{\wct}[0]{\ensuremath{\tau}}
\newcommand{\absent}[0]{\ensuremath{\upphi}}
\newcommand{\cs}[0]{\ensuremath{\mathit{cs}}}

\newcommand{\imm}[0]{\ensuremath{\mathit{IMM}}}

\newcommand{\brackets}[1]{\ensuremath{ \left[ #1 \right] }}
\newcommand{\tuple}[1]{\ensuremath{ \left\langle #1 \right\rangle }}
\newcommand{\set}[1]{\ensuremath{ \left\{ #1 \right\}}}
\newcommand{\system}[1]{\ensuremath{ \left\{ #1 \right.}}
\newcommand{\pargroup}[1]{\ensuremath{ \left( #1 \right)}}
\newcommand{\dert}[1]{\ensuremath{ \dot{#1} }}
\newcommand{\partialder}[2]{\ensuremath{ \frac{\partial#1}{\partial#2} }}
\newcommand{\setreal}[0]{\ensuremath{\mathbb{R}}}
\newcommand{\setnat}[0]{\ensuremath{\mathbb{N}}}
\newcommand{\bigO}[1]{\ensuremath{ \mathcal{O}\left( #1 \right)}}
\newcommand{\norm}[1]{\left\lVert#1\right\rVert}
\newcommand{\vectorOne}[1]{\brackets{%
\begin{matrix}
  #1
 \end{matrix}%
}}
\newcommand{\vectorTwo}[2]{\brackets{%
\begin{matrix}
  #1 \\
  #2
 \end{matrix}%
}}
\newcommand{\vectorThree}[3]{\brackets{%
\begin{matrix}
  #1 \\
  #2 \\
  #3
 \end{matrix}%
}}
\newcommand{\vectorFour}[4]{\brackets{%
\begin{matrix}
  #1 \\
  #2 \\
  #3 \\
  #4
 \end{matrix}%
}}

\newenvironment{aligneq}%
{
\begin{equation}
\begin{aligned}
}{
\end{aligned}
\end{equation}
}

\ifreport
\else
	\acmJournal{CSUR}
	
	\setcopyright{acmcopyright} 
	
	\acmDOI{0000001.0000001}
	
\fi

\begin{document}

\ifreport
	
	\markboth{Gomes et al.}{Co-simulation: State of the art Report}
	
	\title{Co-simulation: State of the art}
	\author{Cláudio Gomes (claudio.gomes@uantwerp.be) \and Casper Thule (casper.thule@eng.au.dk) \and David Broman (dbro@kth.se) \and Peter Gorm Larsen (pgl@eng.au.dk) \and Hans Vangheluwe (hans.vangheluwe@uantwerp.be)}
	
\else
	
	\title{Co-simulation: State of the Art}
	
	\author{Cláudio Gomes}
	\orcid{0000-0003-2692-9742}
	\email{claudio.gomes@uantwerp.be}
	\affiliation{%
	  \position{PhD student}
	  \institution{University of Antwerp}
	  \department{Department of Mathematics and Computer Science}
	  \city{Antwerpen}
	  \streetaddress{Middelheimlaan 1}
	  \postcode{2000}
	  \country{Belgium}
	}
	
	\author{Casper Thule}
	\email{casper.thule@eng.au.dk}
	\affiliation{%
	  \position{PhD student}
	  \institution{Aarhus University}
	  \department{Department of Engineering}
	  \city{Aarhus}
	  \streetaddress{Nordre Ringgade 1}
	  \postcode{8200}
	  \country{Denmark}
	}
	
	\author{David Broman}
	\email{dbro@kth.se}
	\affiliation{%
	  \position{Associate professor, Docent}
	  \institution{KTH Royal Institute of Technology}
	  \department{ICT/SCS}
	  \city{Kista}
	  \streetaddress{Kistag\r{a}ngen 16}
	  \postcode{164 40}
	  \country{Sweden}
	}
	
	\author{Peter Gorm Larsen}
	\email{pgl@eng.au.dk}
	\affiliation{%
	  \position{Professor}
	  \institution{Aarhus University}
	  \department{Department of Engineering}
	  \city{Aarhus}
	  \streetaddress{Nordre Ringgade 1}
	  \postcode{8200}
	  \country{Aarhus N}
	}
	
	\author{Hans Vangheluwe}
	\email{hans.vangheluwe@uantwerp.be}
	\affiliation{%
	  \position{Professor}
	  \institution{University of Antwerp}
	  \department{Department of Mathematics and Computer Science}
	  \city{Antwerpen}
	  \streetaddress{Middelheimlaan 1}
	  \postcode{2000}
	  \country{Belgium}
	}
	\affiliation{%
	  \position{Adjunct Professor}
	  \institution{McGill University}
	  \department{School of Computer Science}
	  \city{Montréal}
	  \state{Qu\'{e}bec}
	  \streetaddress{3480 University Street}
	  \postcode{H3A 0E9}
	  \country{Canada}
	}
\fi

\newcommand{\writeAbstract}{
	It is essential to find new ways of enabling experts in different disciplines to collaborate more efficient in the development of ever more complex systems, under increasing market pressures.
	One possible solution for this challenge is to use a heterogeneous model-based approach where different teams can produce their conventional models and carry out their usual mono-disciplinary analysis, but in addition, the different models can be coupled for simulation (co-simulation), allowing the study of the global behavior of the system.
	Due to its potential, co-simulation is being studied in many different disciplines but with limited sharing of findings.
	Our aim with this work is to summarize, bridge, and enhance future research in this multidisciplinary area.
	
	We provide an overview of co-simulation approaches, research challenges, and research opportunities, together with a detailed taxonomy with different aspects of the state of the art of co-simulation and classification for the past five years.

	The main research needs identified are: finding generic approaches for modular, stable and accurate coupling of simulation units; and expressing the adaptations required to ensure that the coupling is correct.
}

\ifreport
	\maketitle
	
	\tableofcontents
	
	\begin{abstract}
	\writeAbstract
	\end{abstract}
\else	
	\begin{abstract}
	\writeAbstract
	\end{abstract}


	%
	%
	
	\terms{Co-simulation, Taxonomy}

	\thanks{
		This research was partially supported by Flanders Make vzw, the strategic research centre for the manufacturing industry, and a PhD fellowship grant from the Agency for Innovation by Science and Technology in Flanders (IWT). In addition, the work presented here is partially supported by the INTO-CPS project funded by the European Commission's Horizon 2020 programme under grant agreement number 664047. This project is financially supported by the Swedish Foundation for Strategic Research.   \\
		Author's address: 
		Cláudio Gomes, Department of Mathematics and Computer Science, University of Antwerp, Campus Middelheim, M.G.028, Middelheimlaan 1, 2020 Antwerpen, Belgium; email: claudio.gomes@uantwerp.be; 
		Casper Thule, Peter Gorm Larsen, Department of Engineering, Aarhus University, Nordre Ringgade 1, 8200 Aarhus N, Denmark; email: \{casper.thule,pgl\}@eng.au.dk;
		David Broman, ICT/SCS, KTH Royal Institute of Technology, Kistag\r{a}ngen 16, 164 40 Kista, Sweden; email: dbro@kth.se;
		Hans Vangheluwe, University of Antwerp, Department of Mathematics and Computer Science, Campus Middelheim, M.G.116, Middelheimlaan 1, 2020 Antwerpen, Belgium; email: hans.vangheluwe@uantwerp.be}

	\maketitle{}
	\renewcommand{\shortauthors}{C. Gomes et al.}
	
\fi

\section{Introduction}
\label{sec:intro}

\ifreport
	\subsection{Motivation}
\fi

Truly complex engineered systems that integrate physical, software and network aspects, are emerging \cite{Lee2008,Nielsen2015}.
Due to external pressure, the development of these systems has to be concurrent and distributed, that is, divided between different teams and/or external suppliers, each in their own domain and each with their own tools.
Each participant develops a partial solution to a constituent system, that needs to be integrated with all the other partial solutions. 
The later in the process the integration is done, the less optimal it is \cite{Tomiyama2007}.

Innovative and truly optimal multi-disciplinary solutions can only be achieved through a holistic development process \cite{VanderAuweraer2013} where the partial solutions developed independently are integrated sooner and more frequently.
Furthermore, the traditional activities carried out at the partial solution level ---such as requirements compliance check, or design space exploration--- can be repeated at the global level, and salient properties spanning multiple constituent systems can be studied.

Modeling and simulation can improve the development of the partial solutions (e.g., see \citet{Friedman2006}), but falls short in fostering this holistic development process \cite{Blochwitz2011}. 
To understand why, one has to observe that:
\begin{inparaenum}[\itshape a\upshape)]
\item models of each partial solution cannot be exchanged or integrated easily, because these are likely developed by different specialized tools; and 
\item externally supplied models have Intellectual Property (IP), hence cannot be cheaply disclosed to system integrators.
\end{inparaenum}

\ifreport
	\subsection{Co-simulation}
\fi

Co-simulation is proposed as a solution to overcome these important challenges. 
It consists of the theory and techniques to enable global simulation of a coupled system via the composition of simulators.
Each simulator is a \emph{black box} mock-up of a constituent system, developed and provided by the team that is responsible for that system.
This allows each team/supplier to work on its part of the problem with its own tools without having the coupled system in mind, and eases the relationship between system integrators and suppliers: the latter provide virtual mock-ups of the solutions they are selling without revealing their IP; the former can perform early conformance checks and evaluate different designs from multiple competing suppliers.

An alternative to co-simulation is co-modelling, were models are described in a unified language, and then simulated.
There are advantages to this approach but each domain has its own particularities when it comes to simulation (e.g., see \cite{Cellier2006,McCalla1987,Vangheluwe2002}) making it impractical to find a language and simulation algorithm that fits all.

\begin{figure}[htb]
    \centering
    	\includegraphics[width=0.8\textwidth]{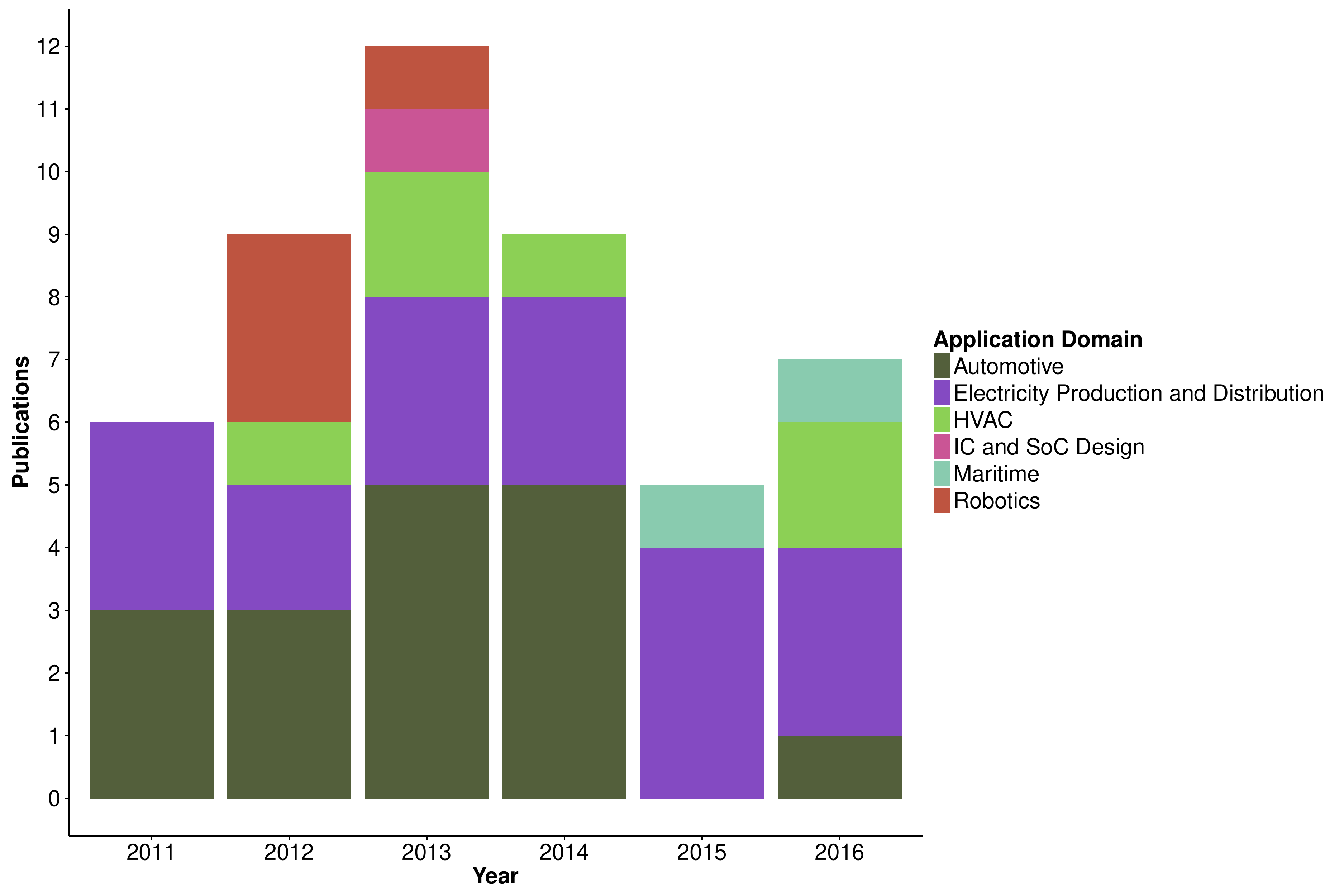}
    \caption{Research publications of co-simulation applications over the past five years. \ifnotreport See the technical report \cite{Gomes2016b} for further details. \fi}
    \label{fig:applications_time_line}
\end{figure}

As part of the systematic review that led to the current document (see \cref{sec:methodology} for details), we took note of the approaches to co-simulation and the publications in applications of co-simulation. The approaches to co-simulation shaped the taxonomy in \cref{sec:taxonomy} and the applications of co-simulation shows that in the last five years, co-simulation has been applied in many different engineering domains, as \cref{fig:applications_time_line} shows.
\ifreport
	In concrete, the publications are:
	\begin{compactdesc}
	\item[Automotive] - \cite{Schneider2014,Benedikt2013b,Karner2013,BenKhaled2014,Bombino2013,VanderAuweraer2013,Abel2012,Mews2012,Zehetner2013,Belmon2014,Drenth2014,Datar2012,Zhang2016,Zhang2014,Brezina2011,Li2011a,Faure2011}
	\item[Electricity Production and Distribution] - \cite{Li2011c,Elsheikh2013,Li2014,Lin2011,Kounev2015,Vanfretti2014,Galtier2015,Abad2015,Bian2015,Manbachi2016,Xie2016,Vaubourg2015,Gurusinghe2016,Zhao2014,Fuller2013,Sun2011,Li2013,Roche2012,Al-Hammouri2012}
	\item[HVAC] - \cite{Nouidui2014a,Pang2012,Fitzgerald2016,Dols2016,Hafner2013,Wang2013}
	\item[IC and SoC Design] - \cite{Saleh2013}
	\item[Maritime] - \cite{Pedersen2016,Pedersen2015}
	\item[Robotics] - \cite{Pierce2012,Kudelski2013,Zhang2012,Pohlmann2012}
	\end{compactdesc}
\fi
A closer look at the publications shows, however, that the average reported co-simulation scenario includes only two simulators, each a mock-up of a constituent system from a different domain.
While this gives evidence that co-simulation enhances the development multi-domain systems, it is not yet up-to-par with the scale of Cyber-Physical Systems (CPSs).
The unexplored potential is recognized in a number of completed and ongoing projects that address co-simulation (MODELISAR\footurl{https://itea3.org/project/modelisar.html}, DESTECS\footurl{http://www.destecs.org/}, INTO-CPS\footurl{http://into-cps.au.dk/}, ACOSAR\footurl{https://itea3.org/project/acosar.html}, ACoRTA\footurl{http://www.v2c2.at/research/ee-software/projects/acorta/}), and is one of the reasons why the Functional Mock-up Interface (FMI) Standard was created.

The FMI standard\footurl{https://www.fmi-standard.org} defines the interfaces that enable modelling and simulation tools to cooperate in a simulation, while keeping the IP in the models protected.
As of December 2016, 15 tools implement the standard\footnote{This number includes only the tools that have passed the crosscheck in \url{https://www.fmi-standard.org/tools}} and many others are planning to support it.

The standard was originally created for the coupling of continuous systems, but has been used (or abused) to simulate hybrid systems.
Extensions have been proposed in the literature to facilitate and standardize this simulation (e.g., zero step size transitions \cite{Bogomolov2015,Tripakis2015}, or step size prediction \cite{Broman2013}), and we complement those by adding to the understanding of the the challenges arising in hybrid system co-simulation. 
Co-simulation is not a new concept and there are many approaches in the state of the art, in particular, discrete event based approaches, that have been studied and may shed light on how hybrid system co-simulation can be performed.

\textbf{Contribution.}
We present a survey and a taxonomy, as an attempt to bridge, relate and classify the many co-simulation approaches in the state of the art.
\ifreport
	\subsection{Need for the Survey}
\fi
Despite the growing interest in the benefits and scientific challenges of co-simulation, to the best of our knowledge, no existing survey attempts to cover the heterogeneous communities in which it is being studied.
The lack of such a survey means that the same techniques are being proposed independently with limited sharing of findings.
To give an example, the use of dead-reckoning models is a well known technique in discrete event co-simulation \cite{Lee2000}, but only very recently it was used in a continuous time co-simulation approach \cite{Stettinger2014}.
\textbf{Our hypothesis} is that bridging these different co-simulation approaches means that solutions and techniques can be exchanged, and a deeper understanding of hybrid system co-simulation can be attained.

\ifreport
	Scientifically, co-simulation ---multi-disciplinary by its very nature--- mixes the following fields of research:
	\begin{enumerate}
	  \item Numerical analysis -- Accuracy and stability of the coupled system have to be studied \cite{Gunther2001,Gu2001,Gu2004,Chen2007,Pfau2006,Busch2010,Arnold2010,Kalmar-Nagy2014}.
	  \item Differential Algebraic System Simulation -- The composition of co-simulation units is, in the most general sense, made through algebraic constraints \cite{Tomulik2011,Schweizer2015,Schweizer2016}.
	  \item Hybrid Systems -- co-simulation scenarios, in the most general sense, are hybrid systems \cite{Zhang2008,Ni2012,Wang2013,Mosterman1999,Broman2015,Lee2005,Mosterman1998a,Mosterman2002}
	  \item Optimization -- the heterogeneous capabilities of co-simulation units pose interesting tradeoffs \cite{Broman2013,Acker2015}.
	  \item Hierarchy -- Systems of systems are hierarchical and the corresponding co-simulation scenarios should be hierarchical as well \cite{Galtier2015}. Compositionality properties of co-simulations becomes an interesting research challenge.
	  \item Formal Verification -- The co-simulation orchestration algorithm, also known as the master, can be certified to be correct under certain assumptions about the co-simulation scenarios \cite{Gheorghe2006,Broman2013,Gheorghe2007,Fitzgerald2013,Mustafiz2016a}.
	  \item System Testing -- Co-simulation can be used for exhaustively testing a set of black-box constituent systems, with a non-deterministic environment \cite{Lawrence2016,Mancini2013}.
	  \item Dynamic Structure Systems -- Subsystems can have different dependencies depending on whom, and which level of abstraction, they interact with \cite{Uhrmacher2001,Barros1997,Barros2008,Barros2003}.
	  \item Multi-paradigm Modeling -- Subsystems can have have different models at different levels of abstraction \cite{Vangheluwe2002}. The relationships between the multiple levels have to be known so that correct dynamic switching between levels abstraction can be made.
	\end{enumerate}
\fi

\ifreport
	\subsection{Outline}
\fi

To help structure the characteristics of the simulators and how they interact, we distinguish two main approaches for co-simulation: Discrete Event (DE), described in \cref{sec:de_cosim}, and Continuous Time (CT), described in \cref{sec:ct_cosim}.
Both of these can be, and are, used for the co-simulation of continuous, discrete, or hybrid coupled systems.
We call Hybrid co-simulation, described in \cref{sec:hybrid_cosimulation}, a co-simulation approach that mixes the DE and CT approaches \footnote{Note that in this survey, we are focusing on timed formalisms (also called models of computation) and how they interact in a hybrid co-simulation environment. 
Other formalisms, with no or only logical notion of time (such as dataflow and synchronous reactive), are not
discussed in this survey. 
For an overview of formalisms and models of computation, please see the book on Ptolemy II~\cite{Ptolemaeus2014} and the following survey~\cite{Broman2012}.
}.
\cref{sec:classification_overview} categorizes the features provided by co-simulation frameworks, and classifies the state of the art with that taxonomy.
Finally, \cref{sec:conclusion} concludes this publication.
The section below provides the terminology used in the rest of the survey.

\section{Modeling, Simulation, and Co-simulation}
\label{sec:background}

\ifreport
	\subsection{Dynamical Systems -- Models of Real Systems}
\fi

In this section we present, in an informal manner, the concepts that will be used throughout the document.

A \emph{dynamical system} is a model of a real system (for instance a physical system or a computer system) characterized by a state and a notion of evolution rules.
\ifreport
	The state is a set of point values in a state space.
	The evolution rules describe how the state evolves over an independent variable, usually time.
\fi
For example, a traffic light, the real system, can be modeled by a dynamical system that, at any time, can be in one of four possible states (\texttt{red}, \texttt{yellow}, \texttt{green}, or \texttt{off}). 
The evolution rules may dictate that it changes from \texttt{red} to \texttt{green} after some time (e.g., 60 seconds). 
Another example is a mass-spring-damper, modeled by a set of 
first order Ordinary Differential Equations (ODEs).
\ifreport
	The equations describe how the state ---position and velocity of the mass--- changes continuously over the simulated time. 
	In contrast with the traffic light system, where the state cannot take an infinite number of different values over a finite duration of simulated time, the state of the mass-spring-damper can.
\fi

The \emph{behavior trace} is the set of trajectories followed by the state (and outputs) of a dynamical system.
For example, a state trajectory $x$ can be defined as a mapping between a time base $T$ and the set of reals $\setreal$, that is, $x : T \to \setreal$.
\ifreport
	\cref{fig:behavior_trace_examples} shows a possible behavior trace for each of the example systems described before.
	In this example, the time base is $\setreal$. 
	
	\begin{figure}[htb]
		\begin{center}
		\includegraphics[width=0.6\textwidth]{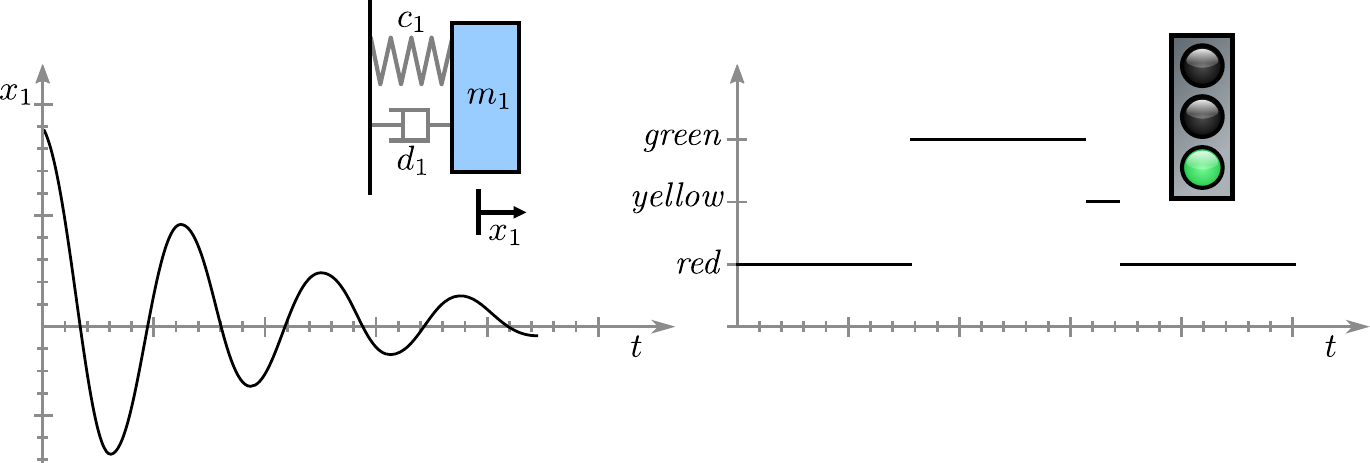}
		\caption{Examples of behavior traces.}
		\label{fig:behavior_trace_examples}
		\end{center}
	\end{figure}
\fi

\begin{figure}
\centering
\begin{subfigure}{0.3\textwidth}
  \centering
  \vspace{75pt}
  \includegraphics[width=1\textwidth]{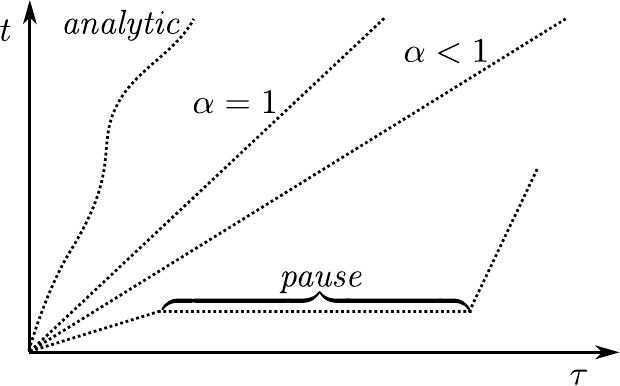}
  \caption{Classification of time constraints in simulation. Based on \cite{Mustafiz2013,VanMierlo2015}.}
  \label{fig:relation_wctime_time}
\end{subfigure}\hspace{1em}%
\begin{subfigure}{0.65\textwidth}
  \centering
  \includegraphics[width=1\textwidth]{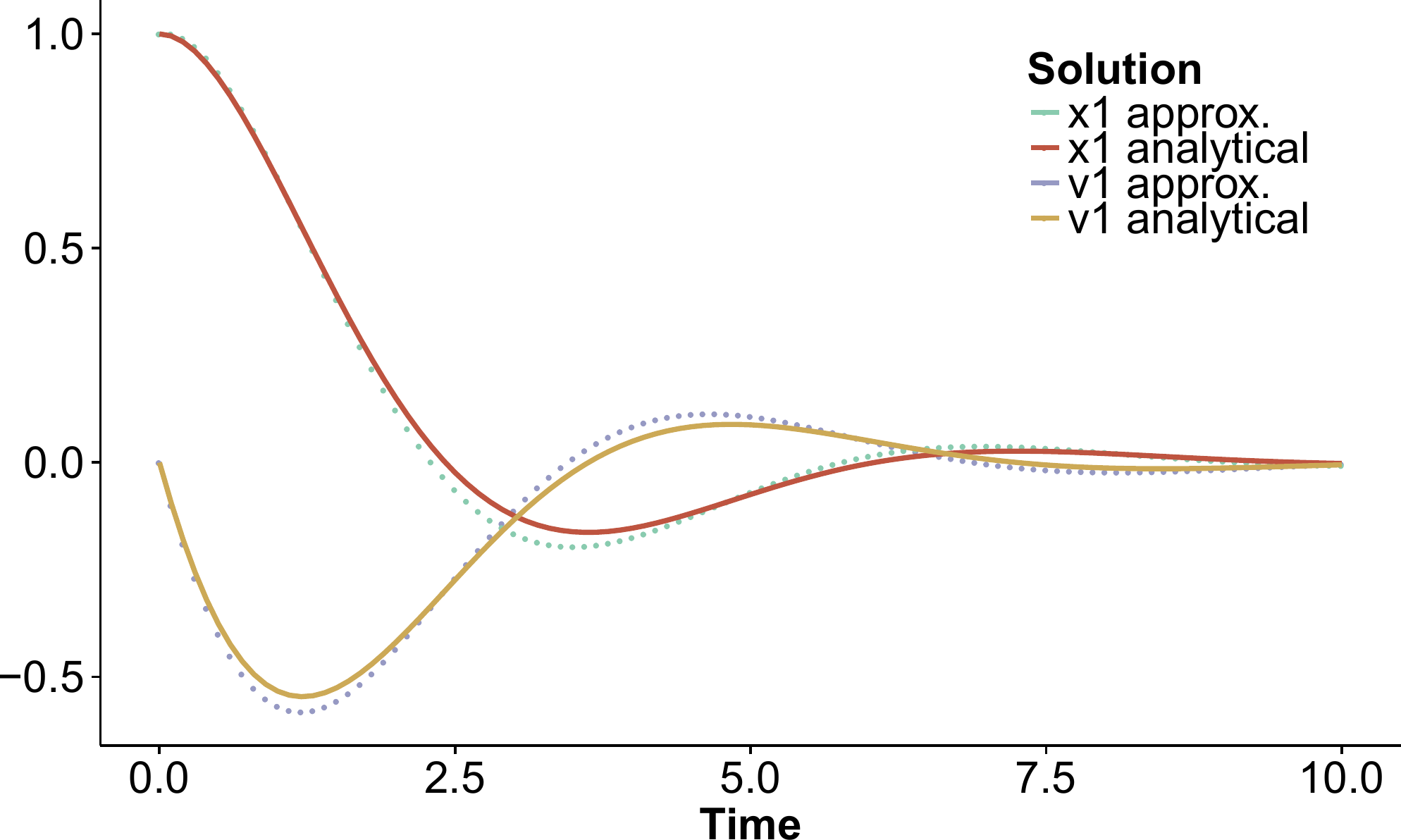}
  \caption{Approximate behavior trace of the mass-spring-damper system computed by a Forward Euler solver. Parameters are: $m=1$, $p=1$, $s=0$, $c_1=1$, $d_1=1$, and $F_e(t)=0$. $x$ is the displacement and $v$ is the velocity. The dotted lines are the approximated behaviour traces and the solid lines are the analytical behaviour traces.}
  \label{fig:approx_trace_msd_fw_euler}
\end{subfigure}
\caption{ }
\label{fig:test}
\end{figure}

We refer to the time variable $t \in T$ by \emph{simulated time} ---or simply \emph{time}, when no ambiguity exists---  defined over a time base $T$ (typical the real numbers $\setreal$), as opposed to the \emph{wall-clock time} $\wct \in \WCT$, which is the time that passes in the real world \cite{Fujimoto2000}.
When computing the behavior trace of a dynamical system over an interval $\brackets{0, t}$ of simulated time, a computer takes $\wct$ units of wall-clock time that depend on $t$. 
$\wct$ can therefore be used to measure the run-time performance of simulators.
\cref{fig:relation_wctime_time} highlights different kinds of simulation, based on the relationship between $\wct$ and $t$.
\ifreport
	In \emph{real-time simulation}, the relationship between $t$ and $\wct$ is $t  = \alpha \wct$, for a given $\alpha > 0$. 
	In most cases $\alpha = 1$ is required, but making sure this is obeyed by the simulation algorithm is one of the main challenges in real-time simulation, and by extension, of co-simulation.
\else
	In \emph{real-time simulation}, this relationship should be linear but enforcing that is one of the main challenges in real-time simulation, and by extension, of co-simulation.
\fi
\ifreport
	In as-fast-as-possible ---or \emph{analytical}--- simulation, the relationship between $\wct$ and $t$ is not restricted.
\fi
Simulation tools that offer interactive visualization allow the user to pause the simulation and/or set the relationship between $\wct$ and $t$.

Knowing when a dynamical system can be used to predict the behavior of a real system is crucial.
The \emph{experimental frame} describes, in an abstract way, a set of assumptions in which the behavior trace of the dynamical system can be compared with the one of the real system \cite{Zeigler2000,Barton1994,Vangheluwe2008,Vangheluwe2002,Traore2006}.
By real system we mean either an existing physical system, or a fictitious ---to be engineered--- one.
\emph{Validity} is then the difference between the behavior trace of the dynamical system and the behavior trace of the real system, measured under the assumptions specified by the experimental frame.
This is what conveys predictive power to dynamical systems.
\ifreport
	For example, Hooke's law, in the mass-spring-damper system, can only be used to predict the reaction force of the spring for small deformations.
	For the traffic light dynamical system, the experimental frame includes the assumption that the transition from the red light to the green light is instantaneous. It is a valid assumption, provided that the executing platform in which the controller software runs, has enough computing power \cite{Vanherpen2015,Eidson2012,Naderlinger2013}.
A model is invalid when, within the experimental frame assumptions, its behavior trace is so different than the one of the real system, that it cannot be used to predict properties of the real system.
\else
	As examples, consider the small deformation assumption for Hooke's law, or the instantaneous transitions of state in the traffic light, both valid models (to some degree) of the corresponding real systems.
\fi 

\ifreport
	In order to be practical, the behavior trace of a dynamical system has to highlight just the features of interest of the real system that are relevant for the tasks at hand \cite{Kuhne2005a}.
	In the traffic light model, the precise amount of wall-clock time a transition from red to green takes is unknown, but deemed small enough to be neglected.
	In the mass-spring-damper, Hooke's law was chosen because the maximum displacement of the mass will not be large when the context in which the system will be used is taken into account.
\fi

\ifreport
	Finally, we consider only those dynamical systems for which it is possible to obtain its meaning, i.e. the behavior trace, even if only an approximation.
\fi 

\ifreport
	\subsection{Simulators -- Computing the Behavior Trace}
	\label{sec:solver_computing_behavior}
\fi

There are two generally accepted ways of obtaining the behavior trace of a dynamical system:
\ifreport
	\begin{compactdesc}
  		\item[Translational] Translate the dynamical system into another model, which can be readily used to obtain the behavior trace. Obtaining the analytical solution of the mass-spring-damper equations is an example of this approach.
For instance, if the traffic light model is expressed in the Statechart formalism, it can be translated into a DEVS model, as done in \citet{Borland2003}, which can be used to obtain the behavior trace.
  		\item[Operational] Use of a solver -- an algorithm that takes the dynamical system as input, and outputs a behavior trace. For the mass-spring-damper example, a numerical solver can be used to obtain an approximation of the behavior trace.
	\end{compactdesc}
\else
	translational (e.g., obtaining the analytical solution of an ODE) and operational (e.g., using a simulator to approximate the solution of an ODE).
\fi
We focus on the latter.

A \emph{simulator} is an algorithm that takes a dynamical system and computes its behavior trace.
Running in a digital computer, it is often the case that a simulator will only be able to approximate that trace.
Two aspects contribute to the error in these approximations: inability to calculate a trajectory over a continuum, and the finite representation of infinitely small quantities. 
\ifreport
	Simulators of discrete dynamical systems may also tolerate some inaccuracies in the behavior traces as well (e.g., if that brings a performance benefit). 
\fi
\cref{fig:approx_trace_msd_fw_euler} shows an example approximation of the behavior trace of the mass-spring-damper system, computed by the Forward Euler. The inaccuracies are clear when compared to the analytical trace.

In order to define what an \emph{accurate simulator} is, or even be able to talk about error, we need to postulate that every dynamical system has an analytical behavior trace.
The error can then be defined as the absolute difference between the behavior trace computed by a simulator and the analytical trace.
An accurate simulator is one that produces traces that are very close\footnote{The meaning of ``very close'' depends on the numerical tolerance of the observer.} to the analytical behaviour.
Even if it is not possible to obtain the analytical behavior of every dynamical system, there are theoretical results that allow simulators to control the error they make.
These techniques are applied to co-simulation in \cref{sec:ctcosim:error_control}.
\ifreport
	For the mass-spring-damper, and linear ODEs in general, the analytical trace follows a known structure \cite{Burden2011}.
	For the traffic light, and timed statemachine models in general, the analytical behavior trace can be obtained with a sequential solver, that respects the causality of events.
\fi
In short, validity is a property of a dynamical system whereas accuracy is a property of a simulator \cite{Cellier2006}.
\ifreport
	It is perfectly possible to have an accurate behaviour trace of a model that is invalid, and vice versa \cite{Cellier1991}.
	For continuous time systems, the choice of an appropriate solver is important and should be made by domain experts \cite{Cellier2006,McCalla1987,Mosterman2012}.
\fi

\ifreport
	\subsection{Simulation Units - Mock-ups of Reality}	
\fi

In strict terms, a simulator is not readily executable: it needs a dynamical system and the input trajectories to that dynamical system, before being able to compute the behavior trace.
We use the term \emph{simulation unit} for the composition of a simulator with a dynamical system.
A simulation unit is a replacement of the real system, ready to take inputs and produce a behavior trace.
In the FMI standard, the analogous term is the Functional Mock-up Unit (FMU) for co-simulation.

A \emph{simulation} is the behavior trace obtained with a simulation unit.
The correctness of a simulation unit is dictated by the correctness of the simulation, which depends on the accuracy of the simulator and the validity of the dynamical system.

\ifreport
	\subsection{Compositional Co-simulation}
	\label{sec:compositional_cosim}
\fi

As described in \cref{sec:intro}, it useful to obtain correct simulations of complex, not yet existing, systems.
Since the constituent systems are developed independently by specialized teams/suppliers, simulation units can be produced for these.
The simulation units can be coupled via their inputs\-/outputs to produce a behavior trace of the coupled system.
A \emph{co-simulation}, a special kind of simulation, is the set of simulations computed by the coupled simulation units.

The simulation units can be independent black boxes, possibly running in different computers.
Hence, an \emph{orchestrator} is necessary to couple them.
The orchestrator controls how the simulated time progresses in each simulation unit and moves data from outputs to inputs according to a co-simulation scenario.
A \emph{co-simulation scenario} is the information necessary to ensure that a correct co-simulation can be obtained.
It includes how the inputs of each simulation unit are computed from other outputs, their experimental frames, etc.
In the FMI Standard, the orchestrator is called a master algorithm.

Analogously to the simulator and simulation unit concepts, the composition of a specific orchestrator with a co-simulation scenario, yields a \emph{co-simulation unit}, which is a special kind of simulation unit, and a substitute of the real coupled system.
It follows that a co-simulation is the simulation trace computed by a co-simulation unit.
This characterization enables hierarchical co-simulation scenarios, where co-simulation units are coupled.
 
The \emph{main purpose of the orchestrator} is to obtain a correct behavior trace of the coupled system, assuming that each simulation unit in itself is correctly defined. We make this assumption because: (1) the simulation units are often well known parts of the system, and (2) they are developed by experienced and specialized teams/suppliers.

\ifreport
	Co-simulation enables design decisions to be tried out in the model (\emph{what-if} analysis), cheaply \footnote{Another aspect to consider is the balance between insights gained and resources spent \cite{Fitzgerald&07e}.}, early in the process, and possibly automatically \cite{Gries2004,Fitzgerald2014}. 
\fi

In this survey, we focus on the coupling of black box simulation units, where limited knowledge of the models and simulators is available.
However, as will become clear in the sections below, the black box restriction has to be relaxed so that certain properties related to correctness can be ensured.
Understanding what kind of information should be revealed and how IP can still be protected is an active area of research in co-simulation.

Most challenges in co-simulation are related to compositionality: if every simulation unit $S_i$ in a co-simulation scenario satisfies property $P$, then the co-simulation unit, with a suitable orchestrator, must also satisfy $P$.
The correctness is a property that should be compositional in co-simulation.
Other properties include validity, or accuracy.
It is an open research question to ensure that a co-simulator is compositional for a given set of properties.
The following three sections provide an overview of the information and techniques being used throughout the state of the art, divided into three main approaches: discrete event (\cref{sec:de_cosim}), continuous time (\cref{sec:ct_cosim}), and hybrid (\cref{sec:hybrid_cosimulation}) co-simulation.

\section{Discrete Event Based Co-simulation}
\label{sec:de_cosim}

The Discrete Event (DE) based co-simulation approach describes a family of orchestrators and characteristics of simulation units that are borrowed from the discrete event system simulation domain.
We start with a description of DE systems, and then we extract the main concepts that characterize DE based co-simulation.

The traffic light is a good example of a DE system. 
It can be one of the possible modes: \texttt{red}, \texttt{yellow}, \texttt{green} or \texttt{off}.
\ifreport
	The off mode is often used by the police, which in some countries is characterized by a blinking yellow.
\fi
Initially, the traffic light can be red. Then, after 60 seconds, it changes to green.
Alternatively, before those 60 seconds pass, some external entity (e.g., a police officer) may trigger a change from red to off.
The output of this system can be an event signaling its change to a new color.
This example captures some of the essential characteristics of a DE dynamical system:
	\emph{reactivity} -- instant reaction to external stimuli (turning off by an external entity);
	and \emph{transiency} -- a DE system can change its state multiple times in the same simulated time point, and receive simultaneous stimuli.
In the traffic light, \emph{transiency} would happen if the light changes always after 0s (instead of 60s), or if the police officer would turn off and on the traffic light in the same instant.

These characteristics are embraced in DE based co-simulation, where the orchestrator acknowledges that simulation units can evolve the internal state and exchange values despite the fact that the simulated time is stopped.

\subsection{DE Simulation units}

A DE simulation unit is a black box that exhibits the characteristics of a DE dynamical system, but the dynamical system it stands for does not need to be a DE one.
Furthermore, it is typical to assume that DE simulation units communicate with the environment via time-stamped \emph{events}, as opposed to signals. This means that the outputs of simulation units can be absent at times where no event is produced.

We adapt the definition of DEVS\footnote{In the original DEVS definition, the initial state and the absent value in the output function are left implicit. Here we make them explicit, to be consistent with \cref{sec:ct_cosim}. Note also that there are many other variants of discrete-event formalisms. For instance, DE in hardware description languages (VHDL and Verilog) and actor based systems (for instance the DE director in Ptolemy II~\cite{Ptolemaeus2014}). } in \cite{VanTendeloo2017} (originally proposed in \cite{Zeigler1976}) to formally define a DE simulation unit $S_i$, where $i$ denotes the reference of the unit:
\begin{aligneq}\label{eq:de_causal_model}
&	S_i = \tuple{X_i, U_i, Y_i, \deltaext_i, \deltaint_i, \lambda_i, \ta_i, q_i(0)} \\
&	\deltaext_i: Q_i \times U_i \to X_i \\
&	\deltaint_i: X_i \to X_i \\
&	\lambda_i: X_i \to Y_i \cup \set{\absent} \\
&	\ta_i: X_i \to \setreal^+_0 \cup \infty \\
&	q_i(0) \in Q_i	\\
&	Q_i = \set{ (x,e) | x \in X_i \text{ and } 0 \leq e \leq \ta_i(x) } \\
\end{aligneq}
where:
\begin{compactitem}
\item $X_i$, $U_i$, and $Y_i$ are the set of possible discrete states, input events, and output events, respectively;
\item $\deltaext_i(q_i, u_i)=x'_i$ is the external transition function that computes a new total state $(x'_i, 0) \in Q_i$ based on the current total state $q_i$ and an input event $u_i$; 
\item $\deltaint_i(x_i)=x'_i$ is the internal transition function that computes a new total state $(x'_i, 0) \in Q_i$ when the current total state is $(x_i, \ta_i(x_i)) \in Q_i$;
\item $e$ denotes the elapsed units of time since the last transition (internal or external);
\item $\lambda_i(x_i) = y_i \in Y_i \cup \set{\absent}$ is the output event function, invoked right before an internal transition takes place and $\absent$ encodes an absent value;
\item $\ta_i(x_i) \in \setreal$ is the time advance function that indicates how much time passes until the next state change occurs, assuming that no external events arrive; 
\item $q_i(0)$ is the initial state; 
\end{compactitem}

The execution of a DEVS simulation unit is described informally as follows.
Suppose that the simulation unit is at time $t_i \in \setreal^+_0$ and marks the current discrete state as $x_i$ for $e \geq 0$ elapsed units of time. Since $e \leq \ta_i(x_i) $, the total state is $(x_i, e) \in Q_i$.
Let $\tn = t_i + \ta_i(x_i) - e$. 
If no input event happens until $\tn$, then at time $\tn$ an output event is computed as $y_i := \lambda_i(x_i)$ and the new discrete state $x_i$ is computed as 
$x_i :=(\deltaint_i(x_i), 0)$.
If, on the other hand, there is an event at time $\ts < \tn$, that is, $u_i$ is not absent at that time, then the solver changes to state $x_i := (\deltaext_i( (x_i, e + \ts - t_i), u_i ), 0)$ instead.

In the above description, if two events happen at the same time, both are processed before the simulated time progresses.
Due to the transiency and reactivity properties, the state and output trajectories of a DE simulation unit can only be well identified if the time base, traditionally the positive real numbers, includes a way to order simultaneous events, and simultaneous state changes.
An example of such a time base is the notion of superdense time \cite{Lee2005,Manna1993,Maler1992}, where each time point is a pair $(t,n) \in \mathcal{T} \times \mathcal{N}$, with $\mathcal{T}$ typically being the positive real numbers and $\mathcal{N}$, called the index, is the natural numbers.
In this time base, a state trajectory is a function $x_i: \mathcal{T} \times \mathcal{N} \to V_{x_i}$, where $V_{x_i}$ is the set of values for the state, and an output/input trajectory is $u_i: \mathcal{T} \times \mathcal{N} \to V_{u_i} \cup \set{\absent}$.
Simultaneous states and events can be formally represented with incrementing indexes. See \cite{Broman2015} for an introduction.
	
Equations \ref{eq:de_traffic_light_model} and \ref{eq:de_policeman_model} show examples of simulation units.

\ifreport
	A DE simulation unit is passive: it expects some external coordinator to set the inputs and call the transition functions. 
	This passivity enables an easier composition of simulation units in a co-simulation, by means of a coordination algorithm, as will be shown later in \cref{sec:de_orchestration}.
\fi
\algoref{alg:de_simulator_single} shows a trivial orchestrator which computes the behavior trace of a single DE simulation unit, as specified in \cref{eq:de_causal_model}, that has no inputs.
Remarks:
\begin{compactitem}
\item $\tl$ holds the time of the last transition;
\item the initial elapsed time satisfies $0 \leq e \leq \ta_i(x_i(0))$;
\end{compactitem}

\ifreport
	If \algoref{alg:de_simulator_single} is used to coordinate the execution of the traffic light simulation unit in \cref{eq:de_traffic_light_model}, then the resulting behavior trace is the piecewise constant traffic light state $x_1(t)$, together with the output events. The latter is represented as a trajectory $y_i(t)$ that is mostly undefined (or absent), except for the single points where an output is produced, according to $\ta_1$.
\fi

\begin{algorithm}[htb]
\KwData{A simulation unit $S_i = \tuple{X_i, \emptyset, Y_i, \deltaext_i, \deltaint_i, \lambda_i, \ta_i, (x_i(0), e_i)}$.}
$t_i := 0$ \;
$x_i := x_i(0)$ \tcp*{Initial discrete state}
$\tl := - e_i$ \tcp*{Account for initial elapsed time}
\While{true}{
	$t_i := \tl + \ta_i(x_i)$ \tcp*{Compute time of the next transition}
	$y_i := \lambda_i(x_i)$ \tcp*{Output}
	$x_i := \deltaint_i(x_i)$ \tcp*{Take internal transition}
	$\tl := t_i$ \;
}
\caption{Single autonomous DE simulation unit orchestration. Based on \cite{VanTendeloo2017} and originally proposed in \cite{Zeigler1976}.}
\label{alg:de_simulator_single}
\end{algorithm}

\subsection{DE Co-simulation Orchestration}
\label{sec:de_orchestration}

\ifreport
	DEVS simulation units communicate with their environment exclusively through inputs and outputs. 
\fi
DE co-simulation scenarios are comprised of multiple DE units $S_i$ (\cref{eq:de_causal_model}) coupled through output to input connections, which map output events of one unit to external events in other unit.

Consider the following DE simulation units of a traffic light and a police office, respectively:
\begin{minipage}[t]{0.45\textwidth}
	\ifreport
		\small
	\fi
	
	\begin{aligneq}\label{eq:de_traffic_light_model}
	&	S_1 = \tuple{X_1, U_1, Y_1, \deltaext_1, \deltaint_1, \lambda_1, \ta_1, q_1(0)} \\
	&	X_1 = \set{\mathit{red},\mathit{yellow},\mathit{green},\mathit{off}} \\
	&	U_1 = \set{\mathit{toAuto},\mathit{toOff}} \\
	&	Y_1 = X_1 \\
	&	\deltaext_1 ( (x_1, e) , u_1) = 
			\begin{cases}
				\mathit{off} & \text{ if } u_1 = \mathit{toOff}	\\
				\mathit{red} & \text{ if } u_1 = \mathit{toAuto} \text{ and } x_1 = \mathit{off}
			\end{cases}
							\\
	&	\deltaint_1 (x_1) = 
			\begin{cases}
				\mathit{green} & \text{ if } x_1 = \mathit{red}	\\
				\mathit{yellow} & \text{ if } x_1 = \mathit{green} 	\\
				\mathit{red} & \text{ if } x_1 = \mathit{yellow}	\\
			\end{cases}
							\\
	&	\lambda_1 (x_1) = 
			\begin{cases}
				\mathit{green} & \text{ if } x_1 = \mathit{red}	\\
				\mathit{yellow} & \text{ if } x_1 = \mathit{green} 	\\
				\mathit{red} & \text{ if } x_1 = \mathit{yellow}	\\
			\end{cases}
							\\
	&	\ta_1 (x_1) = 
			\begin{cases}
				60 & \text{ if } x_1 = \mathit{red}	\\
				50 & \text{ if } x_1 = \mathit{green} 	\\
				10 & \text{ if } x_1 = \mathit{yellow}	\\
				\infty & \text{ if } x_1 = \mathit{off}	\\
			\end{cases}
							\\
	&	q_1(0) = (\mathit{red}, 0) \\
	\end{aligneq}
\end{minipage}\hfill
\begin{minipage}[t]{0.45\textwidth}
	\ifreport
		\small
	\fi
	
	\begin{aligneq}\label{eq:de_policeman_model}
	&	S_2 = \tuple{X_2, U_2, Y_2, \deltaext_2, \deltaint_2, \lambda_2, \ta_2, q_2(0)} \\
	&	X_2 = \set{\mathit{working},\mathit{idle}} \\
	&	U_2 = \emptyset \\
	&	Y_2 = \set{\mathit{toWork},\mathit{toIdle}} \\
	&	\deltaint_2 (x_2) = 
			\begin{cases}
				\mathit{idle} & \text{ if } x_2 = \mathit{working}	\\
				\mathit{working} & \text{ if } x_2 = \mathit{idle}
			\end{cases}
							\\
	&	\lambda_2 (x_2) = 
			\begin{cases}
				\mathit{toIdle} & \text{ if } x_2 = \mathit{working}	\\
				\mathit{toWork} & \text{ if } x_2 = \mathit{idle}
			\end{cases}
							\\
	&	\ta_2 (x_2) = 
			\begin{cases}
				200 & \text{ if } x_2 = \mathit{working} \\
				100 & \text{ if } x_2 = \mathit{idle}
			\end{cases}
							\\
	&	q_2(0) = (\mathit{idle}, 0) \\
	\end{aligneq}
\end{minipage}

\ifreport
	With the following remarks:
	\begin{compactitem}
	\item The current state of the model in the definition of $\deltaext_1$ is $q_1 = (x_1, e)$ with $e$ being the elapsed time since the last transition.
	\item The output event function $\lambda_1$ is executed \emph{immediately before} the internal transition takes place. It must then publish the next state instead of the current.
	\end{compactitem}
\fi

To model a scenario where the police officer interacts with a traffic light, the output events $Y_2$ have to be mapped into the external events 
\ifreport
	$U_1$ of the traffic light simulation unit (\cref{eq:de_traffic_light_model}).
\else
	of the traffic light simulation unit.
\fi
In this example, if $U_1 = \set{\mathit{toAuto}, \mathit{toOff}}$ are the external input events handled by the traffic light simulation unit, the mapping 
$Z_{2,1}: Y_2 \to U_1$ is defined by:
\begin{aligneq}\label{eq:mapping_example_traffic}
&	Z_{2,1} (y_2) = 
		\begin{cases}
			\mathit{toAuto} & \text{ if } y_2 = \mathit{toIdle}	\\
			\mathit{toOff} & \text{ if } y_2 = \mathit{toWork}
		\end{cases}
						\\
\end{aligneq}
\ifreport
	This way, if the police officer changes to $\mathit{working}$ state at time $\tn$, then the output signal $y_2 := \mathit{toWork}$ will be translated by 
$Z_{2,1}$ into an input event $u_1 := \mathit{toOff}$ of the traffic light simulation unit. 
\fi

Based on the idea of abstract simulation units~\cite{Zeigler2000}, we formalize and illustrate the idea of a DE co-simulation scenario with reference $\cs$ as follows:
\begin{aligneq}\label{eq:de_cosim_definition}
&	\tuple{U_\cs, Y_\cs, D, \set{S_d : d \in D}, \set{I_d : d \in D \cup \set{\cs}}, \set{Z_{i,d} : d \in D \wedge i \in I_d} , \text{Select}}
\end{aligneq}
where: \\
\begin{compactitem}
\item $U_\cs$ is the set of possible input events, external to the scenario;
\item $Y_\cs$ is the set of possible output events from the scenario to the environment; 
\item $D$ is an ordered set of simulation unit references; 
\item For each $d \in D$, $S_d$ denotes a DE unit, as defined in \cref{eq:de_causal_model}; 
\item For each $d \in D \cup \set{\cs}$, $I_d \subseteq \pargroup{D \setminus \set{d}} \cup \set{\cs}$ is the set of simulation units that can influence simulation unit $S_d$, possibly including the environment external to the scenario ($\cs$), but excluding itself ($d$); 
\item For each $i \in I_d$, $Z_{i,d}$ specifies the mapping of events:
$
\begin{aligned}
Z_{i,d}:&  U_i \to U_d \text{, if } i = \cs \\
Z_{i,d}:&  Y_i \to Y_d \text{, if } d = \cs \\
Z_{i,d}:&  Y_i \to U_d \text{, if } i \neq \cs \text{ and } d \neq \cs \\
\end{aligned}
$
\item $\text{Select} : 2^D \to D$ is used to deterministically select one simulation unit, among multiple simulation units ready to produce output events simultaneously, i.e., when at time $t$, the set of simulation units 
\begin{equation}\label{eq:imm_definition}
\imm(t) = \set{d | d \in D \wedge q_d(t) = (x_d,\ta_d(x_d))}
\end{equation}
has more than one simulation unit reference. 
\ifreport
	This function is restricted to select one from among the set $\imm(t)$, i.e.,
\else
	In addition, 
\fi
$\text{Select}(\imm(t)) \in \imm(t)$.
\end{compactitem}

The following co-simulation scenario $\cs$ couples the traffic light simulation unit to the police officer simulation unit:
\begin{aligneq}\label{eq:de_example_scenario}
&	\tuple{\emptyset, Y_\cs, \set{1,2}, \set{S_1, S_2}, \set{I_1, I_2, I_\cs}, \set{Z_{2,1}, Z_{1,\cs}} , \text{Select}} \\
&	Y_\cs = Y_1 \\
&	I_1 = \set{2} \\
&	I_2 = \emptyset \\
&	I_\cs = \set{1} \\
&	Z_{1,\cs}(y_1) = y_1 \\
\end{aligneq}
where:
\begin{compactitem}
\item $S_1$ is the traffic light simulation unit
\ifreport
	(\cref{eq:de_traffic_light_model})
\fi
and $S_2$ the police officer simulation unit (\cref{eq:de_policeman_model});
\item $Y_1$ is the output of $S_1$;
\item $Z_{2,1}$ is defined in \cref{eq:mapping_example_traffic}; and
\item The omitted $Z_{i,d}$ functions map anything to absent ($\absent$).
\end{compactitem}

The $\text{Select}$ function is particularly important to ensure that the co-simulation trace is unique. 
For example, consider the co-simulation scenario of \cref{eq:de_example_scenario}, and suppose that at time $\tn$ both simulation units are ready to output an event and perform an internal transition. 
Should the traffic light output the event and perform the internal transition first, or should it be the police office to do it first? 
In general, the order in which these output/transition actions are performed matters.
\ifreport
	The reason is that the way one simulation unit reacts to the other simulation unit's output may be different, depending on the internal state of the former.
	In the example co-simulation scenario, the end result is always the same but this is not the general case.
\fi

\algoref{alg:de_scenario_autonomous} illustrates the orchestrator of an autonomous (without inputs) DE co-simulation scenario.
\ifreport
	It assumes that the co-simulation scenario does not expect external events, that is, all events that can affect the simulation units are produced by other simulation units in the same scenario.
	External output events are possible though.
\fi
Remarks:
\begin{compactitem}
\item $t_\cs$ holds the most recent time of the last transition in the scenario; 
\item $e_d$ is the elapsed time of the current state $q_d=\pargroup{x_d, e_d}$ of simulation unit $S_d$; 
\item $\tn$ is the time of the next transition in the scenario;
\item $i^*$ denotes the chosen imminent simulation unit;
\item $I_\cs$ is the set of simulation units that can produce output events to the environment;
\item $y_\cs$ is the output event signal of the scenario to the environment; and
\item $\set{d | d \in D \wedge i^* \in I_d}$ holds the simulation units that $S_{i^*}$ can influence.
\end{compactitem}

\begin{algorithm}[htb]
\KwData{A co-simulation scenario $\cs = \tuple{\emptyset, Y_\cs, D, \set{S_d}, \set{I_d}, \set{Z_{i,d}} , \text{Select}}$.}
$t_\cs := 0$ \;
$x_i := x_i(0) \text{ for all } i \in D$ \tcp*{Store initial discrete state for each unit}
\While{true}{
	$\ta_\cs := \min_{d \in D} \set{\ta_d(x_d) - e_d} $ \tcp*{Time until the next internal transition}
	$\tn := t_\cs + \ta_\cs$ \tcp*{Time of the next internal transition}
	$i^* := \text{Select}(\imm(\tn))$ \tcp*{Get next unit to execute}
	$y_{i^*} := \lambda_{i^*}(x_{i^*})$ \;
	$x_{i^*} := \deltaint_{i^*}(x_{i^*})$ \tcp*{Store new discrete state}
	$e_{i^*} := 0$ \tcp*{Reset elapsed time for the executed unit}
	\If{$i^* \in I_\cs$}{
		$y_\cs := Z_{i^*,\cs}(y_{i^*})$ \tcp*{Compute output of the scenario}
	}
	\For{$d \in \set{d | d \in D \wedge i^* \in I_d}$}{ 
		$u_d := Z_{i^*,d}(y_{i^*})$ \tcp*{Trigger internal units that are influenced by unit $i^*$}
		$x_d := \deltaext_d( (x_d, e_d + \ta_\cs) , u_d)$ \; 
		$e_d := 0$ \;
	}
	\For{$d \in \set{d | d \in D \wedge i^* \not \in I_d}$}{
		$e_d := e_d + \ta_\cs$ \tcp*{Update the elapsed time of the remaining units}
	}
	$t_\cs := \tn$ \tcp*{Advance time}
}
\caption{Autonomous DE co-simulation scenario orchestration. Based on \cite{VanTendeloo2017}.}
\label{alg:de_scenario_autonomous}
\end{algorithm}

\begin{figure}[htb]
\begin{center}
  \includegraphics[width=0.4\textwidth]{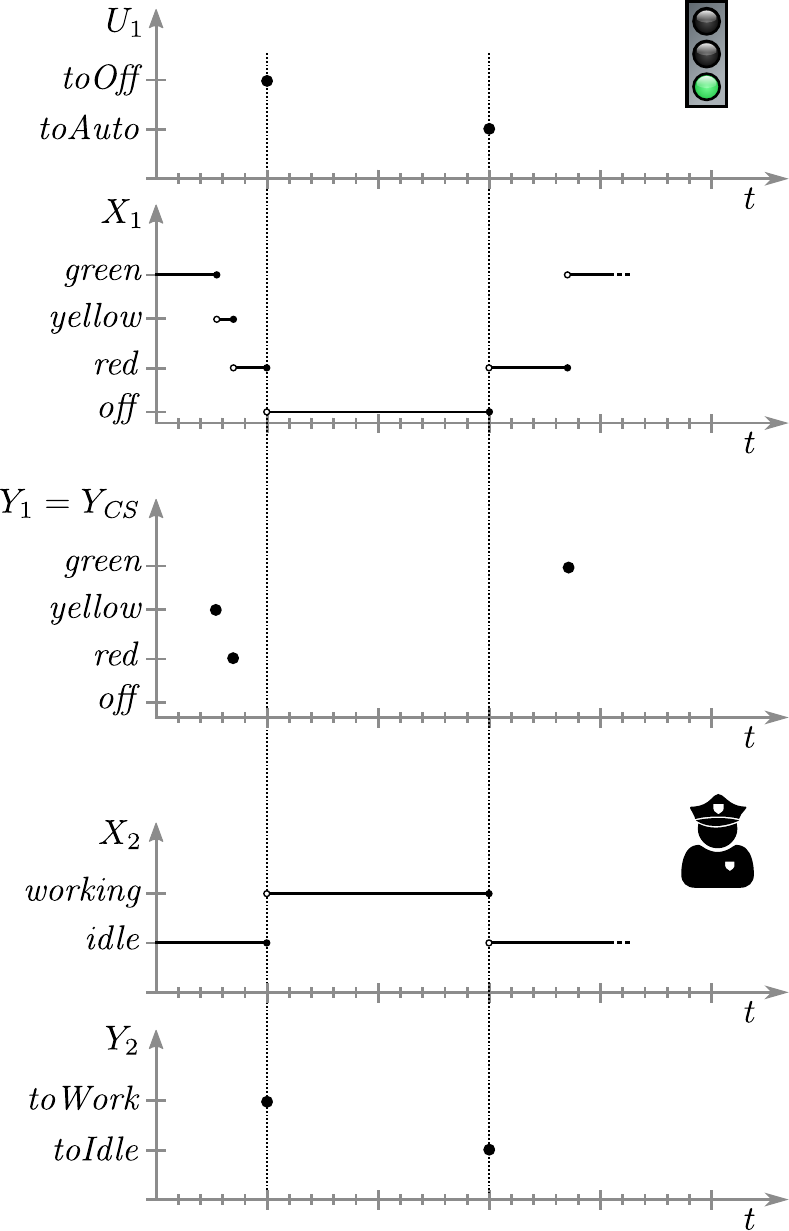}
  \caption{Example co-simulation trace of the traffic light and police officer scenario.}
  \label{fig:traffic_light_cosim_scenario_trace}
\end{center}
\end{figure}

\cref{fig:traffic_light_cosim_scenario_trace} shows the behavior trace of the co-simulation scenario in \cref{eq:de_example_scenario}.

\algoref{alg:de_scenario_autonomous} is similar to \algoref{alg:de_simulator_single}:
\begin{compactitem}
\item The time advance of the scenario $\ta_\cs$ corresponds to the time advance of a single simulation unit.
\item The output produced by the state transition is analogous to the $\lambda$ function of a single simulation unit.
\item The output and state transition of child $S_{i^*}$, together with the external transitions of the simulation units influenced by $S_{i^*}$, are analogous to the internal transition of a single simulation unit.
\end{compactitem}
It is natural then that a co-simulation scenario $\cs$ as specified in \cref{eq:de_cosim_definition}, can be made to behave as a single DE simulation unit $S_\cs$. 
Intuitively, 
the state of $S_\cs$ is the set product of the total states of each child DE unit; 
$\ta_\cs$ is the minimum time until one of the DE units executes an internal transition;
the internal transition of $S_\cs$ gets the output event of the imminent unit, executes the external transitions of all the affected units, updates the elapsed time of all unaffected units, and computes the next state of the imminent unit;
the external transition of $S_\cs$ gets an event from the environment, executes the external transition of all the affected units, and updates the elapsed time of all the unaffected units \cite{Zeigler2000}.
\ifnotreport
	In \cite{Gomes2016b}, the formal construction of $S_\cs$ is provided.
\else
	Formally:
	\begin{aligneq}\label{eq:closure_under_coupling}
	&	S_\cs = \tuple{X_\cs, U_\cs, Y_\cs, \deltaext_\cs, \deltaint_\cs, \lambda_\cs, \ta_\cs, q_\cs(0)} \\
	&	X_\cs = \times_{d \in D} Q_d \\
	&	q_\cs(0) = (\times_{d \in D} q_i(0), \min_{d \in D} e_d ) \\
	&	\ta_\cs( \pargroup{\ldots, (x_d, e_d) , \ldots} ) =  \min_{d \in D} \set{\ta_d(x_d) - e_d}  \\
	&	i^* = \text{Select}(\imm(t)) \\
	&	\lambda_\cs(x_\cs) = 	\begin{cases}
									Z_{i^*,\cs}(y_{i^*}(\tn)) & \text{ if } i^* \in I_\cs \\
									\absent & \text{ otherwise}
								\end{cases}
	\\
	& 	\deltaint_\cs(x_\cs) = (\ldots, (x'_d, e'_d) , \ldots), \text{ for all } d \in D, \text{ where:}  \\
	& \ \ \ x_\cs = \pargroup{\ldots, (x_d, e_d) , \ldots} \\
	& \ \ \ (x'_d, e'_d) =
					\begin{cases}
						(\deltaint_d(x_d), 0) \text{ if } i^* = d \\
						(\deltaext_d( (x_d, e_d + \ta_\cs(x_\cs)) , Z_{i^*,d}(\lambda_{i^*}(x_{i^*})) , 0) \text{ if }  i^* \in I_d	\\
						(x_d, e_d + \ta_\cs(x_\cs)) \text{ otherwise }
					\end{cases} \\
	& 	\deltaext_\cs(\pargroup{x_\cs, e_\cs}, u_\cs) = (\ldots, (x'_d, e'_d) , \ldots), \text{ for all } d \in D, \text{ where:}  \\
	& \ \ \ x_\cs = \pargroup{\ldots, (x_d, e_d) , \ldots} \\
	& \ \ \ (x'_d, e'_d) =
					\begin{cases}
						(\deltaext_d( \pargroup{x_d, e_d + e_\cs} , Z_{\cs,d}(u_\cs)) , 0) \text{ if }  \cs \in I_d	\\
						(x_d, e_d + e_\cs) \text{ otherwise }
					\end{cases} \\
	\end{aligneq}
	
	Remarks:\\
	It is the Cartesian product of the total state of each child simulation unit that makes the discrete state of the co-simulation unit; \\
	The elapsed times of each child simulation unit are managed solely by the co-simulation unit, whenever there is a transition (internal or external); \\
	The external transition functions of each child are executed with the mapping of the events produced by the current state of the imminent child, and not the next one computed by $(\deltaint_d(x_d), 0)$;\\
	An internal transition of a child simulation unit may cause an output event to the environment of the co-simulation unit, if the child is connected to the output of the co-simulation unit.
	
	The same internal transition causes not only a change in the child discrete state, but also, due to its output event, may cause external transitions in other child simulation units. 
	This is not a recursive nor iterative process: 
		at most one external transition will occur in all the affected child simulation units; 
		if any of the affected simulation units becomes ready for an internal transition, it waits for the next internal transition invoked from the coordinator of the co-simulation unit; \\
\fi

The resulting co-simulation unit $S_\cs$ behaves exactly as a DE unit specified in \cref{eq:de_causal_model}. 
It can thus be executed with \algoref{alg:de_simulator_single} (in case of no inputs), or composed with other units in hierarchical co-simulation scenarios.
Hierarchical co-simulation scenarios can elegantly correspond to real hierarchical systems, a natural way to deal with their complexity \cite{Kossiakoff&2011}.

In summary, DE based co-simulation exhibits the following characteristics:
\begin{compactdesc}
\item[reactivity:]  A DE simulation unit (analogously, a DE co-simulation unit) has to process an event at the moment it occurs.
\item[transiency:]  In both \algoref{alg:de_scenario_autonomous} and in a DE co-simulation unit, the time advance $\ta_\cs$ to the next imminent child internal transition can be zero for successive iterations, so an orchestrator has to be able to tolerate the fact that simulated time may not advance for several iterations.
\item[predictable step sizes:] In a DE co-simulation scenario without inputs, the orchestrator, as shown in \algoref{alg:de_scenario_autonomous}, can always predict the next simulated time step. In a scenario with inputs, if the environment provides the time of the next event, then the next simulated time step can be predicted too. For this to be possible, black box DE simulation units have to be able to inform the orchestrator what their time advance is, not a trivial task for DE units that simulate continuous systems whose future behavior trace, especially reacting to future inputs, is not easily predicted without actually computing it.
\end{compactdesc}

In the next sub-section, the main challenges in DE based co-simulation, and the requirements (or capabilities) their solutions impose in DE simulation units, are made explicit.

\subsection{Challenges}

\subsubsection{Causality}
\label{sec:causality}

For the sake of simplicity, \algoref{alg:de_scenario_autonomous} is sequential.
In a hierarchical co-simulation unit, the imminent simulation unit (closest to performing an internal transition) will be the one to execute, thus inducing that there is a global order in the events that are exchanged.
This global order avoids causality violations but is too pessimistic. 
\ifreport
	If an event $y_1(t_1)$ causes another event ---by changing the internal state of some other simulation unit, which in turn changes its next output event--- $y_2(t_2)$, then $t_1 \leq t_2$, which is ok.
	However, the converse is not true: $t_1 \leq t_2$ does not necessarily imply that $y_1(t_1)$ has caused $y_2(t_2)$, which means that simulation unit $S_2$ could execute before ---in the wall-clock time sense--- $y_1(t_1)$ without violating causality, at least within a small window of simulated time. 
	To see why, suppose that $S_1$ and $S_2$ do not influence each other in the scenario. 
	Then $y_2(t_2)$ would happen anyway, regardless of $y_1(t_1)$ occurring or not.
\else
	Not every event $y_2(t_2)$ occurring after some event $y_1(t_1)$ has been caused necessarily by $y_1(t_1)$.
\fi
Moreover, the co-simulation scenario holds information ---the dependencies $\set{I_d}$---  that can be used to determine who influences what \cite{Lamport1978,Chandy1979}.

A parallel optimistic orchestrator that takes $\set{I_d}$ into account is, in general, faster in the wall clock time sense, than a pessimistic, sequential one.
However, most of these, the Time-warp algorithm \cite{Jefferson1982,Jefferson1985} being a well known example, require rollback capabilities of simulation units.
\ifreport
	This is because simulation units proceed to advance their own time optimistically, assuming that any other simulation units will not affect them, until they are proven wrong by receiving an event which occurs before their own internal time. 
	When that happens, the simulation unit has to rollback to a state prior to the time of timestamp of the event that just arrived.
	This may in turn cause a cascade of rollbacks in other affected simulation units.
\fi
Moreover, in parallel optimistic DE co-simulation, any of the units in the scenario needs (theoretically) to support multiple rollbacks and have enough memory to do so for an arbitrary distant point in the past \cite{Fujimoto2000}.
\ifreport
	This point in the past is limited in Time-warp by the Global Virtual Time (GVT). 
	The GVT represents the minimum internal time of all simulation units. 
	By definition, no event that is yet to be produced (in wall-clock time) can have a timestamp smaller than the GVT.
\fi

We make the distinction between multiple rollback, from single rollback capabilities. 
To support single rollback, a simulation unit needs to store only the last committed state, thereby saving memory.

Causality is a compositionality property:
if each child simulation unit does not violate causality, then any orchestrator has to ensure that the causality is not violated when these units are coupled.
\ifreport
	Optimistic orchestration algorithms do so by requiring rollback capabilities from child simulation units, whereas pessimistic algorithms do so at the cost of performance.
\fi

\subsubsection{Determinism and Confluence}

Determinism is also a compositional property. 
The Select function, in the co-simulation scenario definition of \cref{eq:de_cosim_definition}, is paramount to ensure the compositionality of deterministic behavior.
\ifreport
	This function is used to ensure that a unique behavior trace can be obtained when the co-simulation scenario is executed by \algoref{alg:de_scenario_autonomous} or when it is turned into a co-simulation unit, as in \cref{eq:closure_under_coupling}.
\fi
The alternative to the Select function is to ensure that all possible interleavings of executions always lead to the same behavior trace -- this is known as ``confluence''.
Intuitively, if a co-simulation unit is compositional with respect to confluence, then it is also compositional with respect to determinism.

Proving confluence is hard in general black box DE co-simulation because it depends on how the child simulation units react to external events: potentially valuable IP.
Parallel-DEVS \cite{Chow1994} is an approach, which leaves the confluence property to be satisfied by the modeler.

\subsubsection{Dynamic Structure}
\label{sec:dynamic_structure}

Until now, the dependencies $\set{I_d}$, in \cref{eq:de_cosim_definition}, have been assumed to be fixed over time.
From a performance perspective, a static sequence of dependencies may be too conservative, especially if used to ensure causality in optimistic parallel co-simulation.
\ifreport
	To see why, consider that in a large scale simulation, there is a simulation unit $S_1$ which may influence simulation unit $S_2$ but only under a very specific set of conditions, which may not be verified until a large amount of simulated time has passed.
	A pessimistic co-simulation unit assumes that $S_1$ may always affect $S_2$ and hence, tries to ensure that the simulated time of $S_2$ is always smaller than $S_1$, to minimize possible rollbacks.
	This incurs an unnecessary performance toll in the overall co-simulation because $S_1$ does not affect $S_2$ most of the time.
	This is where making $I_2$ dynamic can improve the performance of the co-simulation since the co-simulation unit will know that most of the time, $S_1$ does not affect $S_2$.
\else
	If, throughout a parallel co-simulation, a simulation unit $S_1$ seldom outputs events that influence $S_2$, it makes sense that most of the time $t$, $S_1 \not\in I_2$.
\fi
Dynamic structure co-simulation allows for $\set{I_d}$ to change over time, depending on the behavior trace of the simulation units.
It can be used to study self-organizing systems \cite{Uhrmacher1993,Barros1997}.

\subsubsection{Distribution}

Co-simulation units whose child simulation units are geographically distributed are common \cite{Fujimoto2000}.
Interesting solutions like computation allocation \cite{Muzy2010,VanTendeloo2014}, bridging the hierarchical encapsulation \cite{VanTendeloo2015},  and the use of dead-reckoning models \cite{Lee2000}, have been proposed to mitigate the additional communication cost.
Moreover, security becomes important, and solutions such as \cite{Norling2007} address it.


\section{Continuous Time Based Co-simulation}
\label{sec:ct_cosim}

In the continuous time (CT) based co-simulation approach, the orchestrators' and simulation units' behavior and assumptions are borrowed from the continuous time system simulation domain.
We describe these below.

\subsection{CT Simulation Units}

A continuous time simulation unit is assumed to have a state that evolves continuously over time.
It is easier to get the intuitive idea of this by considering a simulation unit of a continuous time dynamical system, such as a mass-spring-damper, depicted in the left hand side of \cref{fig:mass-spring-damper_multibody_system}.
The state is given by the displacement $x_1$ and velocity $v_1$ of the mass, and the evolution by:
\begin{aligneq}\label{eq:mass-spring-damper-ode}
\dert{x_1}				&=	v_1\\
m_1 \cdot \dert{v_1} 	&= - c_1 \cdot x_1 - d_1 \cdot v_1 + F_e \\
x_1(0)					&=	p_1\\
v_1(0)					&=	s_1
\end{aligneq}
\noindent where $\dert{x}$ denotes the time derivative of $x$; 
$c_1$ is the spring stiffness constant and $d_1$ the damping coefficient; 
$m_1$ is the mass; 
\ifreport
	$p_1$ and $s_1$ the initial position and velocity;
\fi
and $F_e$ denotes an external input force acting on the mass over time.
\ifreport
	The solutions $x_1(t)$ and $v_1(t)$ that satisfy \cref{eq:mass-spring-damper-ode} constitute the behavior trace of the dynamical system. \cref{fig:approx_trace_msd_fw_euler} shows an example of such trace.
\else
	\cref{fig:approx_trace_msd_fw_euler} shows an example of a behavior trace.
\fi

\cref{eq:mass-spring-damper-ode} can be generalized to the state space form:
\begin{aligneq}\label{eq:dynamical_system_state_space_form}
\dert{x}	=	f(x, u)  \ \ ; \ \ 
y 			= 	g(x, u)  \ \ ;\ \ 
x(0)		=	x_0
\end{aligneq}
where $x$ is the state vector, $u$ the input and $y$ the output vectors, and $x_0$ is the initial state.
\ifreport

	A solution $\brackets{x(t), y(t)}^T$ that obeys \cref{eq:dynamical_system_state_space_form} is the behavior trace of the system.
	If $f$ is linear and time-invariant, an analytical form for $x(t)$ can be obtained \cite{Astrom2010}. 
	An analytical solution obtained by the application of mathematical identities is an example of a behavior trace obtained via the translational approach, described in \cref{sec:solver_computing_behavior}.
	Alternatively, the behavior trace can be computed.
	
\fi
If $f(x, u)$ is sufficiently differentiable, $x$ can be approximated with a truncated Taylor series \cite{Taylor1715,Cellier2006}:
\begin{aligneq}\label{eq:taylor_series_ode}
x(t+h)=x(t)+f(x(t),u(t))\cdot h + \bigO{h^2}
\end{aligneq}
where
\ifreport
	$$
	\bigO{h^{n+1}} = \max_i \pargroup{\lim_{h \to 0}\frac{x^{n+1}\pargroup{\zeta(t^{*})}}{(n+1)!}h^{n+1}} = \text{const } \cdot  h^{n+1}
	$$
	denotes the order of the truncated residual term; 
	$t^{*} \in \brackets{t, t+h}$; 
	and
\fi
$h \geq 0$ is the micro-step size.
\cref{eq:taylor_series_ode} is the basis of a family of numerical solvers that iteratively compute an approximated behavior trace $\tilde{x}$.
\ifreport
	For example, the Forward Euler method is given by:
	\begin{aligneq}\label{eq:fw_euler_solver}
	\tilde{x}(t+h)				&:=	\tilde{x}(t) + f(\tilde{x}(t), u(t)) \cdot h \\
	\tilde{x}(0)				&:=	x(0)
	\end{aligneq}
\fi

A CT simulation unit is assumed to have a behavior that is similar to one of a numerical solver computing a set of differential equations.
We reinforce that this does not restrict CT simulation units to being mockups of CT systems, even though it is easier to introduced them as such.
In the FMI Standard, a simulation unit is analogous to a Functional Mock-up Unit (FMU) for co-simulation.
\ifreport
	For example, a simulation unit $S_1$ of the mass-spring-damper, using the Forward Euler solver, can be written by embedding the solver (\cref{eq:fw_euler_solver}) into \cref{eq:mass-spring-damper-ode}:
	\begin{aligneq}\label{eq:fw_euler_msd_simulator}
	\tilde{x}_1(t+h_1)&:=	\tilde{x}_1(t) + v_1(t) \cdot h_1 \\
	\tilde{v}_1(t+h_1)&:=	\tilde{v}_1(t) + \frac{1}{m_1} \cdot \pargroup{- c_1 \cdot \tilde{x}_1(t) - d_1 \cdot \tilde{v}_1(t) + F_e(t)} \cdot h_1\\
	\tilde{x}_1(0)	&:= p_1 \\
	\tilde{v}_1(0)	&:= s_1 
	\end{aligneq}
	where $h_1$ is the micro-step size, $F_e(t)$ is the input, and $\brackets{x(t+h), v(t+h)}^T$ is the output.
\else
	For example, a simulation unit $S_1$ that simulates the mass-spring-damper system takes as input the external force $F_e(t)$, applies \cref{eq:taylor_series_ode} to \cref{eq:mass-spring-damper-ode}, to compute the new state $\brackets{x(t+h), v(t+h)}^T$, and outputs it.	
\fi

\subsection{CT Co-simulation Orchestration}

\begin{figure}[htb]
\begin{center}
	\includegraphics[width=0.35\textwidth]{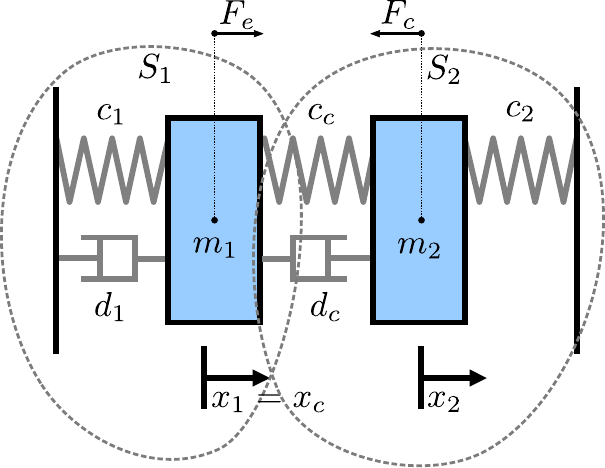}
	\caption{A multi-body system comprised of two mass-spring-damper subsystems.}
	\label{fig:mass-spring-damper_multibody_system}
\end{center}
\end{figure}

\ifreport
	Consider now a second system, depicted in the right hand side of \cref{fig:mass-spring-damper_multibody_system}.
	It is governed by the differential equations:
	\begin{aligneq}\label{eq:mass-spring-ode}
	\dert{x_2}				&=	v_2\\
	m_2 \cdot \dert{v_2} 	&= - c_2 \cdot x_2 - F_c \\
	F_c						&= c_c \cdot \pargroup{x_2 - x_c} + d_c \cdot \pargroup{v_2 - \dert{x_c}}\\
	x_2(0)					&=	p_2\\
	v_2(0)					&=	s_2
	\end{aligneq}
	where $c_c$ and $d_c$ denote the stiffness and damping coefficients of the spring and damper, respectively; $x_c$ denotes the displacement of the left end of the spring-damper.
	Combining with the Forward Euler solver, yields the following simulation unit:
	\begin{aligneq}\label{eq:fw_euler_ms_simulator}
	\tilde{x}_2(t+h_2)&:=	\tilde{x}_2(t) + \tilde{v}_2(t) \cdot h_2 \\
	\tilde{v}_2(t+h_2)&:=	\tilde{v}_2(t) + \frac{1}{m_2} \cdot \pargroup{- c_2 \cdot \tilde{x}_2(t) - F_c(t)} \cdot h_2\\
	F_c(t)						&= c_c \cdot \pargroup{\tilde{x}_2(t) - x_c(t)} + d_c \cdot \pargroup{\tilde{v}_2(t) - \dert{x_c(t)}}\\
	\tilde{x}_2(0)	&:= p_2 \\
	\tilde{v}_2(0)	&:= s_2 
	\end{aligneq}
	where $h_2$ is the micro-step size, $x_c$ and $\dert{x_c}$ are inputs, and $F_c$ the output.
\else
	Consider now a simulation unit $S_2$ for the system depicted in the right hand side of \cref{fig:mass-spring-damper_multibody_system}. 
	It takes the displacement $x_c$ of the left end of the spring/damper and its derivative $\dert{x_c}$, and outputs the reaction force $F_c$.
\fi
Suppose $S_1$
\ifreport
	(\cref{eq:fw_euler_msd_simulator})
\fi
and $S_2$ are coupled, setting $x_c = x_1$, $\dert{x_c} = v_1$ and $F_e = F_c$, so that the resulting co-simulation scenario represents the multi-body system depicted in \cref{fig:mass-spring-damper_multibody_system}.

\ifreport
	In the co-modeling approach, the models in Equations \ref{eq:mass-spring-damper-ode} and \ref{eq:mass-spring-ode} would be combined to get the following coupled model:
	\begin{aligneq}
	\dert{x_1}				&=	v_1		\\
	m_1 \cdot \dert{v_1} 	&= - c_1 \cdot x_1 - d_1 \cdot v_1 + F_c	\\
	\dert{x_2}				&=	v_2		\\
	m_2 \cdot \dert{v_2} 	&= - c_2 \cdot x_2 - F_c \\
	F_c						&= c_c \cdot \pargroup{x_2 - x_1} + d_c \cdot \pargroup{v_2 - v_1}\\
	x_1(0)					&=	p_1		\\
	v_1(0)					&=	s_1		\\
	x_2(0)					&=	p_2		\\
	v_2(0)					&=	s_2
	\end{aligneq}
	which can be written in the state space form (\cref{eq:dynamical_system_state_space_form}) as:
	\begin{aligneq}\label{eq:ct_coupled_model}
	\vectorFour{\dert{x_1}}{\dert{v_1}}{\dert{x_2}}{\dert{v_2}} &=
		\vectorFour{  0&	1&	0&	0}%
					{ -\frac{c_1 + c_c}{m_1}&	-\frac{d_1 + d_c}{m_1}&	\frac{c_c}{m_1}&	\frac{d_c}{m_1}}%
					{ 0&	0&	0&	1}%
					{ \frac{c_c}{m_2}&	\frac{d_c}{m_2}&	-\frac{c_2 + c_c}{m_2}&	-\frac{d_c}{m_2} }
		\vectorFour{x_1}{v_1}{x_2}{v_2} \\
		\vectorFour{x_1(0)}{v_1(0)}{x_2(0)}{v_2(0)} &= \vectorFour{p_1}{s_1}{p_2}{s_2}
	\end{aligneq}
	
	The behavior trace of \cref{eq:ct_coupled_model} can be obtained either analytically, or with Forward Euler solver (\cref{eq:fw_euler_solver}).
\fi

In CT based co-simulation, to overcome the fact the each simulation unit's micro-step sizes are independent, a communication step size $H$ (also known as macro-step size or communication grid size) has to be defined. 
$H$ marks the times at which the simulation units exchange values of inputs/outputs.

Suppose a simulation unit $S_i$ is at time $n \cdot H$, for some natural $n$, and it being asked by an orchestrator to execute until time $(n+1) \cdot H$.
If $S_i$ only gets its inputs valued at $n \cdot H$, then extrapolation must be used to associate a value with those inputs in any of the internal micro-steps of the simulation unit.
In other words, when time $n \cdot H + m \cdot h_i$, for $m \leq \frac{H}{h_i}$ and micro-step size $h_i$, an extrapolation function 
$\phi_{u_i}(m \cdot h_i, u_i(n \cdot H), u_i((n - 1) \cdot H), \ldots )$, 
built from input values known at previous communication time points, is used to approximate the value of $u_i(n \cdot H + m \cdot h_i)$.
\ifreport
	Notice that $m=\frac{H}{h_i}$ is allowed, even though, theoretically, the value of $u_i((n+1) \cdot H)$ can be obtained from the environment. 
	The reason for this becomes clear in \cref{sec:algebraic_loops}.
\fi
Analogously, interpolation techniques have to be used when the orchestrator makes the input value available at time $(n+1) \cdot H$ but the simulation unit is still at time $n \cdot H$.

\ifreport
	For example, the input $F_e$ of the simulation unit described in \cref{eq:fw_euler_msd_simulator} can be defined as: 
	\begin{aligneq}\label{eq:ct_io_no_exchange_time_1}
	F_e(n \cdot H + m \cdot h_1) :=	\phi_{F_e}(m \cdot h_1, F_e(n \cdot H), F_e((n - 1) \cdot H), \ldots ) \text{, for } m \leq \frac{H}{h_1}
	\end{aligneq}
	Similarly, the inputs $x_c$ and $\dert{x_c}$ of the simulation unit described in \cref{eq:fw_euler_ms_simulator} can be defined as 
	\begin{aligneq}\label{eq:ct_io_no_exchange_time_2}
	x_c(n \cdot H + m \cdot h_2)			&:=	\phi_{x_c}(m \cdot h_2, x_c(n \cdot H), x_c((n - 1) \cdot H), \ldots )		\\
	\dert{x_c}(n \cdot H + m \cdot h_2)	&:=	\phi_{\dert{x_c}}(m \cdot h_2, \dert{x_c}(n \cdot H), \dert{x_c}((n - 1) \cdot H), \ldots ) \\
	\text{for } m \leq \frac{H}{h_2}
	\end{aligneq}
\fi

In the simplest case, the extrapolations can be constant. In the coupled mass-spring-dampers example:
\begin{aligneq}
	\phi_{F_e}(t, F_e(n \cdot H)))					&= F_e(n \cdot H))			\\
	\phi_{x_c}(t, x_c(n \cdot H) ) 					&= x_c(n \cdot H)			\\
	\phi_{\dert{x_c}}(t, \dert{x_c}(n \cdot H)) 	&= \dert{x_c}(n \cdot H)
\end{aligneq}

In the state of the art, input extrapolation approaches can be classified by increasing degree of complexity: Constant; Linear; Polynomial; Extrapolated-Interpolation \cite{Dronka2006,Busch2012a,Busch2016}; Context-aware \cite{Khaled2014}; and Estimated Dead-Reckoning Model \cite{Stettinger2014,Brembeck2014};
These can, and often are, combined in practical use cases.
See \citet{Busch2016,Arnold2010,Schweizer2016,Andersson2016} for an overview of linear and higher order extrapolation techniques and how these affect the accuracy of the co-simulation trace.

\ifreport
	The orchestrator for this co-simulation scenario, at a time $t=n \cdot H$, gets the outputs of both simulation units and computes their inputs. 
	Then, each simulation unit is instructed to compute its behavior trace until the next communication step size, at $t=(n+1) \cdot H$, making use of the extrapolating functions to get the inputs at each of the micro steps (Equations \ref{eq:ct_io_no_exchange_time_1} and \ref{eq:ct_io_no_exchange_time_2}).	
\fi

\claudio{$n$ is being used in many places. And that can be confusing.}

We are ready to formally define the behavior of a CT simulation unit $S_i$:
\begin{aligneq}\label{eq:ct_causal_simulator}
&	S_i = \tuple{X_i,U_i,Y_i, \delta_i, \lambda_i, x_i(0), \phi_{U_i}} \\
&	\delta_i: \setreal \times X_i \times U_i \to X_i \\
&	\lambda_i: \setreal \times X_i \times U_i \to Y_i \text{ or } \setreal \times X_i \to Y_i \\
&	x_i(0) \in X_i \\
&	\phi_{U_i}: \setreal \times U_i \times \ldots \times U_i \to U_i
\end{aligneq}
\noindent where:
\begin{compactitem}
\item $X_i$ is the state set, typically $\setreal^n$;
\item $U_i$ is the input set, typically $\setreal^m$;
\item $Y_i$ is the output set, typically $\setreal^p$;
\item $\delta_i(t, x_i(t), u_i(t))=x_i(t+H)$ or $\delta_i(t, x_i(t), u_i(t+H))=x_i(t+H)$ is the function that instructs the simulation unit to compute a behavior trace from $t$ to $t+H$, making use of the input extrapolation (or interpolation) function $\phi_{U_i}$; 
\item $\lambda_i(t, x_i(t), u_i(t))=y_i(t)$ or $\lambda_i(t, x_i(t))=y_i(t)$ is the output function; and
\item$x_i(0)$ is the initial state.
\end{compactitem}

\ifreport
	For instance, the simulation unit in \cref{eq:fw_euler_msd_simulator} can be described as follows:
	\begin{aligneq}\label{eq:ct_causal_simulator_msd_1}
	&	S_1 =\tuple{\setreal^2,\setreal,\setreal^2,\delta_1,\lambda_1, \vectorTwo{p_1}{s_1}, \phi_{F_e}} \\
	&	\delta_1(t, \vectorTwo{\tilde{x}_1(t)}{\tilde{v}_1(t)}, F_e(t)) = 
			\vectorTwo{\tilde{x}_1(t+H)}{\tilde{v}_1(t+H)} \\
	&	\lambda_1(t,\vectorTwo{\tilde{x}_1(t)}{\tilde{v}_1(t)})=\vectorTwo{\tilde{x}_1(t)}{\tilde{v}_1(t)}
	\end{aligneq}
	\noindent where $\brackets{\tilde{x}_1(t+H) , \tilde{v}_1(t+H)}^T$ is obtained by the iterative application of the simulation unit in \cref{eq:fw_euler_msd_simulator} over a finite number of micro-steps, making use of the extrapolation of $F_e$ (defined in \cref{eq:ct_io_no_exchange_time_1}):
	$$
	\vectorTwo{\tilde{x}_1(t+H)}{\tilde{v}_1(t+H)} = 
		\vectorTwo{\tilde{x}_1(t)}{\tilde{v}_1(t)} + 
				\vectorTwo{\dert{x_1}(t)}{\dert{v_1}(t,\phi_{F_e}(t,F_e(t),\ldots))} \cdot h + 
				\vectorTwo{\dert{x_1}(t+h)}{\dert{v_1}(t+h,\phi_{F_e}(t+h,F_e(t),\ldots))} \cdot h + \ldots
	$$
\fi

A continuous time co-simulation scenario with reference $\cs$ includes at least the following information\footnote{Please note that this formalization is related to the formalization proposed by \citet{Broman2013}, with the main differences: i) it is not designed to formalize a subset of the FMI Standard, ii) it accommodates algebraic coupling conditions, and iii) it does not explicitly define port variables.}:
\begin{aligneq}\label{eq:ct_causal_cosim_scenario}
&	\tuple{U_\cs, Y_\cs, D, \set{S_i  : i \in D}, L, \phi_{U_\cs}} \\
&	L : \pargroup{\Pi_{i \in D} Y_i} \times Y_\cs \times \pargroup{\Pi_{i \in D} U_i} \times U_\cs \to \setreal^m
\end{aligneq}
\noindent where 
$D$ is an ordered set of simulation unit references,
each $S_i$ is defined as in \cref{eq:ct_causal_simulator}, 
$m \in \setnat$, 
$U_\cs$ is the space of inputs external to the scenario, 
$Y_\cs$ is the space of outputs of the scenario, 
$\phi_{U_\cs}$ a set of input approximation functions,
and $L$ induces the simulation unit coupling constraints (e.g., if $D=\set{1,\ldots,n}$, then the coupling is $L(y_1, \ldots, y_n, y_\cs, u_1, \ldots, u_n, u_\cs) = \bar{0}$).

As an example, the co-simulation scenario representing the system of \cref{fig:mass-spring-damper_multibody_system} is:
\begin{aligneq}\label{eq:ct_causal_cosim_scenario_example}
&	\cs = \tuple{\emptyset,\emptyset,\set{1,2},\set{S_1, S_2},L, \emptyset} \\
&	L =  \vectorThree{x_c - v_1}{\dert{x_c} - x_1}{F_e - F_c}
\end{aligneq}
\ifreport
	\noindent where:
	\begin{compactitem}
	\item $S_1$ is the simulation unit for the constituent system on the left (\cref{eq:ct_causal_simulator_msd_1}), 
	and $S_2$ is the simulation unit for the remaining constituent system;
	\item $x_c$, $\dert{x_c}$ are the inputs of $S_2$, and $F_e$ is the input of $S_1$; and
	\item $x_1$, $v_1$ are outputs of $S_1$ and $F_c$ is the output of $S_2$.
	\end{compactitem}
\fi

\algoref{alg:ct_cosim_scenario} summarizes in a generic way the tasks of the orchestrator for computing the co-simulation of a scenario $\cs$ with no external inputs.
It represents the Jacobi communication approach: simulation units exchange values at time $t$ and independently compute the trace until the next communication time $t+H$.
The way the system in \cref{eq:system_generic_jacobi_orchestrator} is solved depends on how the simulation units are coupled, that is, the definition of $L$.
In the most trivial case, the system reduces to an assignment of an output $y_j(t)$ to each input $u_i(t)$, and so the orchestrator just gets the output of each simulation unit and copies it onto the input of some other simulation unit, in an appropriate order.
Concrete examples of \algoref{alg:ct_cosim_scenario} are described in \cite{Bastian2011a,Friedrich2011,Krammer2015,Galtier2015,Enge-Rosenblatt2011,Gu2001,Busch2011,Wetter2010}.

An alternative to the Jacobi communication approach is the Gauss-Seidel (a.k.a. sequential or zig-zag) approach, where an order of the simulation units' $\delta$ function is forced to ensure that, at time $t$, they get inputs from a simulation unit that is already at time $t+H$. 
Gauss-Seidel approach allows for interpolations of inputs, which is more accurate, but hinders the parallelization potential.
Examples are described in \cite{Arnold2001,Carstens2003,Bastian2011a,Arnold2010,Acker2015}.

\begin{algorithm}[htb]
 \KwData{An autonomous scenario $\cs = \tuple{\emptyset, Y_\cs, D=\set{1,\ldots,n}, \set{S_i}, L, \emptyset}$ and a communication step size $H$.}
 \KwResult{A co-simulation trace.}
 $t := 0$ \;
 $x_i := x_i(0) \text{ for } i=1,\ldots,n$ \;
 \While{true}{
	 Solve the following system for the unknowns: \\
	 	\begin{aligneq}\label{eq:system_generic_jacobi_orchestrator}
	 	\system{
			\begin{matrix}
				y_1=\lambda_1(t, x_1, u_1) \\
				\ldots\\
				y_n=\lambda_n(t, x_n, u_n) \\
				L(y_1, \ldots, y_n, y_\cs, u_1, \ldots, u_n) = \bar{0}
			\end{matrix}
		 }
		\end{aligneq}
	 \ $x_i := \delta_i( t, x_i, u_i ), \text{ for } i=1,\ldots,n$ \tcp*{Instruct each simulation unit to advance to the next communication step}
	 \ $t := t + H$ \tcp*{Advance time}
 }
 \caption{Generic Jacobi based orchestrator for autonomous CT co-simulation scenarios.}
 \label{alg:ct_cosim_scenario}
\end{algorithm}

Similarly to DE based co-simulation, a CT co-simulation scenario, together with an orchestrator, should behave as a (co-)simulation unit of the form of \cref{eq:ct_causal_simulator}, and thus be coupled with other simulation units, forming hierarchical co-simulation scenarios:
the state of the co-simulation unit is the set product of the states of the internal units;
the inputs are given by $U_\cs$ and the outputs by $Y_\cs$;
the transition and output functions are implemented by the orchestrator;
the communication step size $H$ used by the orchestrator is analogous to a simulation unit's micro-step sizes,
and the input extrapolation function is $\phi_{U_i}$.

\algoref{alg:ct_cosim_scenario} makes it clear that the simulation units can be coupled with very limited information about their internal details.
\ifreport
	In concrete:
	\begin{compactitem}
	\item The output $\lambda_i$ and state transition $\delta_i$ functions need to be executable but their internal details can remain hidden;
	\item the inputs $u_i$ need to be accessible;
	\item the state variables can be hidden. These are represented merely to illustrate that the internal state of the simulation unit changes when executing $\delta_i$.
	\end{compactitem}
\fi
However, the blind coupling can lead to compositionality problems, as will be discussed in the sections below.
The common trait in addressing these is to require more from the individual simulation units: either more capabilities, or more information about the internal (hidden) dynamical system.

\subsection{Challenges}

\subsubsection{Modular Composition -- Algebraic Constraints}
\label{sec:algebraic_composition}

\begin{figure}[htb]
\begin{center}
	\includegraphics[width=0.35\textwidth]{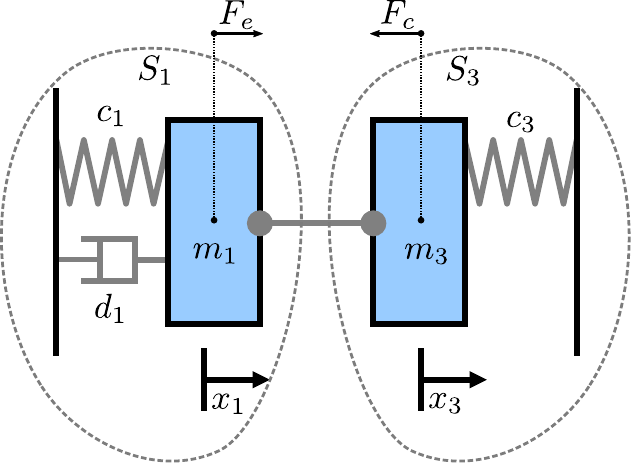}
	\caption{A multi-body system coupled by a mass-less link, based on the example provided in \citet{Schweizer2015}.}
	\label{fig:mass-spring-damper_multibody_system_algebraic}
\end{center}
\end{figure}

In the co-simulation scenario described in \cref{eq:ct_causal_cosim_scenario_example}, the coupling condition $L$ translates into a set of assignments from outputs to inputs.
\ifreport
	This is because the inputs of the simulation unit of the system in the left hand side of \cref{fig:mass-spring-damper_multibody_system} and the outputs of the simulation unit of the system represented in the right hand side of the same picture can be connected directly, and vice versa. 
\fi
In practice, the simulation units' models are not created with a specific coupling pattern in mind and $L$ can be more complex.
As an example, consider the system coupled by a massless rigid link, depicted in \cref{fig:mass-spring-damper_multibody_system_algebraic}.
\ifreport
	The first subsystem is the same as the one in the left hand side of \cref{fig:mass-spring-damper_multibody_system} and its simulation unit is in \cref{eq:fw_euler_msd_simulator}.
	The second constituent system is governed by the following differential equations:
	\begin{aligneq}\label{eq:mass-spring-ode_couplingF}
	\dert{x_3}				&=	v_3\\
	m_3 \cdot \dert{v_2} 	&= - c_3 \cdot x_3 + F_c \\
	x_3(0)					&=	p_3\\
	v_3(0)					&=	s_3
	\end{aligneq}
	And the following simulation unit:
	\begin{aligneq}\label{eq:mass-spring-ode_couplingF_simulator}
	\tilde{x}_3(t+h_3)&=	\tilde{x}_3(t) + v_3(t) \cdot h_3 \\
	\tilde{v}_3(t+h_3)&=	\tilde{v}_3(t) + \frac{1}{m_3} \cdot \pargroup{- c_3 \cdot x_3(t) + F_c(t)} \cdot h_3\\
	\tilde{x}_3(0)	&= p_3 \\
	\tilde{v}_3(0)	&= s_3 
	\end{aligneq}
\fi
The input to the simulation unit $S_3$ is the coupling force $F_c$, and the output is the state of the mass $\brackets{\tilde{x}_3, \tilde{v}_3}^T$.
The input to the simulation unit $S_1$ is the external force $F_e$ and the outputs are the state of the mass $\brackets{\tilde{x}_1, \tilde{v}_1}^T$. 
\ifreport
	Recall \cref{eq:fw_euler_msd_simulator}.
\fi
There is clearly a mismatch. 
\ifreport
	The outputs $\brackets{\tilde{x}_1, \tilde{v}_1}^T$ of the first simulation unit cannot be coupled directly to the input $F_c$ of the second simulation unit, and vice versa.
\fi
However, the massless link restricts the states and inputs of the two units to be the same. 
Whatever the input forces may be, they are equal and opposite in sign. 
Hence, any orchestration algorithm has to find inputs that ensure the coupling constraints are satisfied:
\begin{aligneq}\label{eq:ct_nontrivial_algebraic_constraints}
L = \vectorThree{\tilde{x}_1(n \cdot H)-\tilde{x}_3(n \cdot H)}{\tilde{v}_1(n \cdot H) - \tilde{v}_3(n \cdot H)}{F_e(n \cdot H) + F_c(n \cdot H)} = \bar{0}
\end{aligneq}

This problem has been addressed in \citet{Gu2001,Gu2001a,Gu2004,Arnold2001,Schweizer2014,Schweizer2015d,Arnold2010,Schweizer2016,Schweizer2015,Schweizer2015a,Sicklinger2014}. 
The approach taken in \citet{Gu2001} is worth mentioning because it defines a Boundary Condition Coordinator (BCC) which behaves as an extra simulation unit, whose inputs are the outputs of the original two simulation units, and whose outputs are $F_e$ and $F_c$.
They show that the initial co-simulation scenario with the non-trivial constraint can be translated into a co-simulation, with a trivial constraint, by adding an extra simulation unit.
\ifreport
	This is illustrated in \cref{fig:ct_cosim_scenario_translation}.
\fi

\ifreport
	\begin{figure}[htb]
	\begin{center}
	  \includegraphics[width=0.35\textwidth]{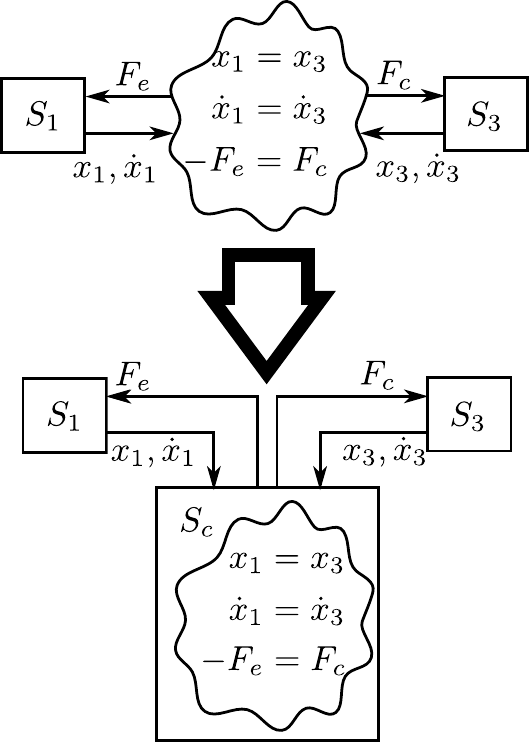}
	  \caption{Transforming a co-simulation scenario with a non-trivial constraint into a simpler scenario by adding an extra simulation unit that induces a trivial constraint. This promotes separation of concerns.}
	  \label{fig:ct_cosim_scenario_translation}
	\end{center}
	\end{figure}
\fi

\ifreport
	Transforming the co-simulation scenario to make it simpler marks an important step in separating the concerns of the orchestrator \cite{Gomes2016}.
	In fact, the newly created simulation unit can be run with a smaller internal micro-step size, required to meet stability and accuracy criteria, as shown in \citet{Gu2001}.
\fi

In many of the solutions proposed (e.g., \cite{Arnold2001,Arnold2010,Schweizer2016,Schweizer2015,Sicklinger2014}), information about the rate of change (or sensitivity) of outputs and states of each simulation unit, with respect to changes in its inputs is required to solve the non-trivial coupling condition.
This information can be either provided directly as a Jacobian matrix of the system and output functions, or estimated by finite differences, provided that the simulation units can be rolled back to previous states.
A frequent characteristic of co-simulation: the availability of certain capabilities from simulation units can mitigate the lack of other capabilities.

\ifreport
	To show why the sensitivity information is useful, one of the tasks of the BCC is to ensure that 
	$\tilde{x}_1-\tilde{x}_3$ 
	is as close to zero as possible, by finding appropriate inputs $F_e$ and $F_c$.
	This is possible since $\tilde{x}_1$ and $\tilde{x}_3$ are functions of the inputs $F_e$ and $F_c$, and $- F_e = F_c$.
	So the constraint can be written as
	\begin{aligneq}\label{eq:coupling_condition_example}
	g(F_e)=\tilde{x}_1(F_e)-\tilde{x}_3(- F_e)=0
	\end{aligneq}
	
	From one communication step to the next, $g$ can be expanded with the Taylor series:
	\begin{aligneq}\label{eq:taylor_series_coupling_constraint}
	g(F_e((n+1) \cdot H)) = g(F_e(n \cdot H) + \Delta F_e) \approx g(F_e(n \cdot H)) + \partialder{g(F_e(n \cdot H))}{F_e} \cdot \Delta F_e
	\end{aligneq}
	
	From a known input $F_e(n \cdot H)$, Equations \ref{eq:coupling_condition_example} and \ref{eq:taylor_series_coupling_constraint} can be combined to obtain the input $F_e((n+1) \cdot H)$ at the next communication step:
	\begin{aligneq}\label{eq:corrector_alg_coupling}
	g(F_e(n \cdot H) + \Delta F_e)	& \approx g(F_e(n \cdot H)) + \partialder{g(F_e(n \cdot H))}{F_e} \cdot \Delta F_e = 0 \leftrightarrow \\
	g(F_e(n \cdot H))  				& = - \partialder{g(F_e(n \cdot H))}{F_e} \cdot \Delta F_e \leftrightarrow \\
	\Delta F_e  & = - \brackets{\partialder{g(F_e(n \cdot H))}{F_e}}^{-1} \cdot g(F_e(n \cdot H))  \leftrightarrow \\
	F_e((n+1) \cdot H)  & = F_e(n \cdot H) - \brackets{\partialder{g(F_e(n \cdot H))}{F_e}}^{-1} \cdot g(F_e(n \cdot H))  \\
	\end{aligneq}
	
	with \\
	\begin{aligneq}\label{eq:constraint_derivative_alg_coupling}
	\partialder{g(F_e(n \cdot H))}{F_e} = 
		\partialder{\tilde{x}_1(F_e(n \cdot H))}{F_e} + \partialder{\tilde{x}_3(-F_e(n \cdot H))}{F_c}
	\end{aligneq}
	
	A simple orchestration algorithm will then perform the following steps, at each co-simulation step:
	\begin{compactenum}
	\item Let $\tilde{x}_1(nH),\tilde{x}_3(nH)$ be the current position outputs of the two simulation units $S_1$ and $S_3$;
	\item Perform a co-simulation step with a known $F_e$, obtaining $\tilde{x}^p_1(nH),\tilde{x}^p_3(nH)$ as new outputs.
	\item Rollback simulation units to state $\tilde{x}_1(nH),\tilde{x}_3(nH)$;
	\item Perform a co-simulation step with $F_e + \Delta F_e$, obtaining $\tilde{x}^d_1(nH),\tilde{x}^d_3(nH)$;
	\item Approximate $\partialder{g(F_e(n \cdot H))}{F_e}$ by finite differences and \cref{eq:constraint_derivative_alg_coupling};
	\item Obtain a corrected $F^c_e$ by \cref{eq:corrector_alg_coupling};
	\item Rollback simulation units to state $\tilde{x}_1(nH),\tilde{x}_3(nH)$;
	\item Perform the final co-simulation step with $F^c_e$;
	\item Commit states and advance time;
	\end{compactenum}
	
	\begin{figure}[htb]
	\begin{center}
	  \includegraphics[width=0.7\textwidth]{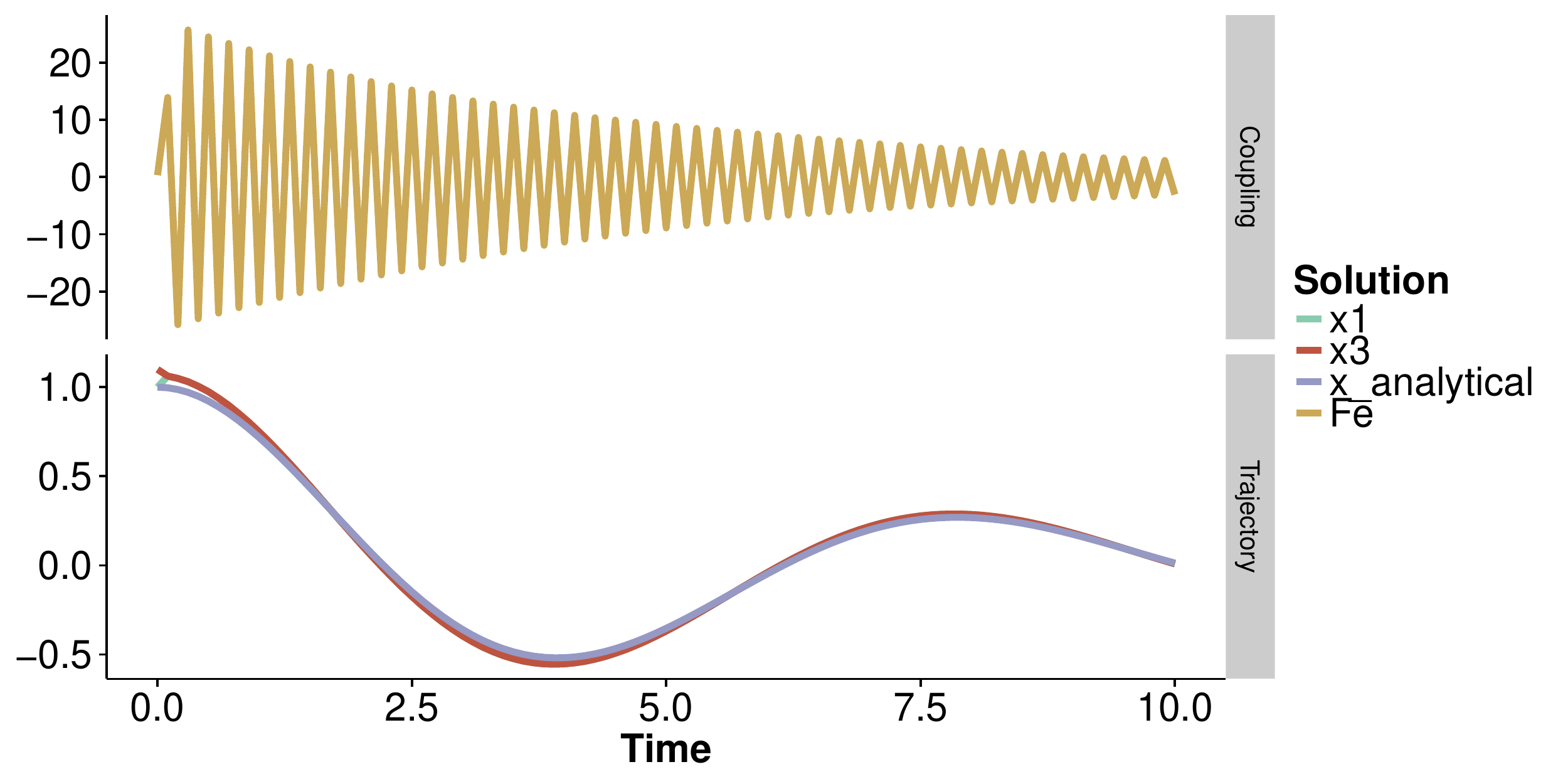}
	  \caption{Co-simulation of algebraically coupled masses. Parameters are: $m_2 = 2, m_1 = c_1 = c_3 = d_1 = c_c = 1, H=0.1, x_1(0)=1.0,x_3(0)=1.1,v_1(0)=v_3(0)=0$. Notice the small disturbance at the initial conditions.}
	  \label{fig:two_dof_algebraic}
	\end{center}
	\end{figure}
	
	As can be seen in \cref{fig:two_dof_algebraic}, this coupling cannot be carried out without errors: the constraint $g(F_e((n+1) \cdot H))$ cannot be accurately forced to zero at first try.
	Furthermore, finding initial conditions and initial inputs that satisfy Equations \ref{eq:mass-spring-damper-ode}, \ref{eq:mass-spring-ode_couplingF}, and \ref{eq:ct_nontrivial_algebraic_constraints} is very important and usually requires a fixed point iteration.
	The above algorithm could be changed to perform an arbitrary number of iterations, repeating steps 1--7 until $g(F_e((n+1) \cdot H))$ is close enough to zero. This would increase the accuracy but also increase the amount of computation.
	
	These examples show that rollback capabilities are important.
	If a simulation unit is a black box, then the rollback capability has to be provided by the simulation unit itself and there is little that the orchestrator can do to make up for the lack of the feature. See \citet{Broman2013} for an orchestrator that takes into account the existence of the rollback feature.
	If, on the other hand, the simulation unit provides access to its state, and allows the state to be set, as in \citet{Blochwitz2012}, then the orchestrator can implement the rollback by keeping track of the state of the simulation unit.
	Rollback also plays a key role when dealing with algebraic loops in the co-simulation scenario.
\fi

Finally, to explain why this sub-section refers to modular composition of simulation units, the example in \cref{fig:mass-spring-damper_multibody_system_algebraic} makes explicit one of the problems in co-simulation: the ``rigid" and protected nature of simulation units can make their coupled simulation very difficult.
To contrast, in a white box approach where the equations of both constituent systems are available, the whole system is simplified, with the two masses being lumped together, and their coupling forces canceling each other out. The simplified system is a lumped mass-spring-damper, which is easily solvable.
Such an approach is common in a-causal modeling languages, such as Modelica~\cite{Modelica2007}.
\ifreport
	To be concrete, the coupled system is obtained by combining Equations \ref{eq:mass-spring-damper-ode}, \ref{eq:mass-spring-ode_couplingF}, and \ref{eq:ct_nontrivial_algebraic_constraints}, and simplifying to:
	\begin{aligneq}
	\dert{x_1}									&=	v_1\\
	( m_1 + m_3 ) \cdot \dert{v_1} &= - (c_1 + c_3) \cdot x_1 - d_1 \cdot v_1  \\
	x_1(0)										&=	p_1\\
	v_1(0)										&=	s_1\\
	\end{aligneq}
		
	\begin{figure}[htb]
	\begin{center}
	  \includegraphics[width=0.7\textwidth]{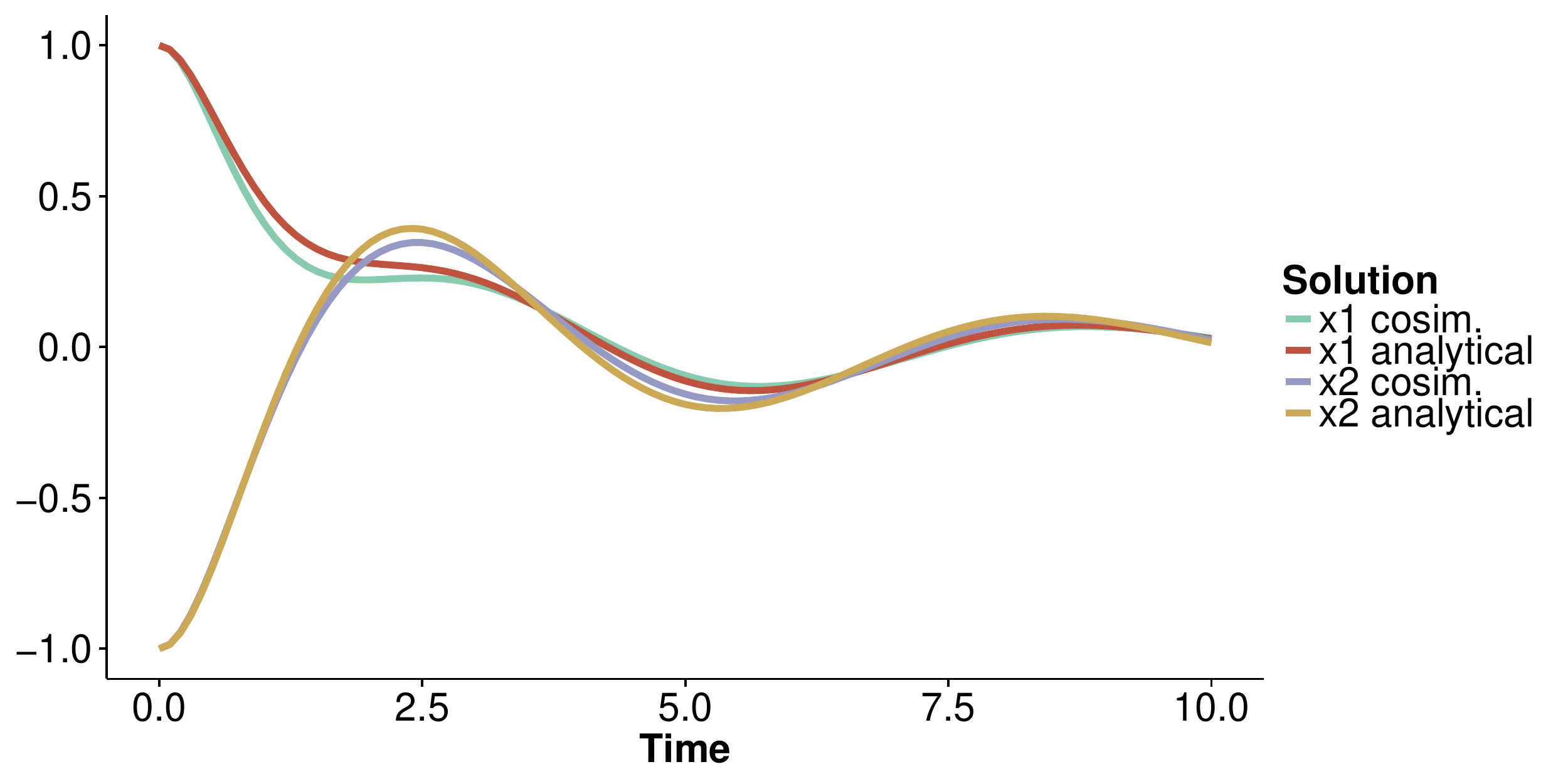}
	  \caption{Comparison of co-simulation with co-modelling for the sample coupled system. Parameters are: $m_1 = m_2 = c_1 = c_2 = d_1 = c_c = 1, H=0.1$.}
	  \label{fig:two_dof_comsim_vs_comodelling}
	\end{center}
	\end{figure}
	
	\cref{fig:two_dof_comsim_vs_comodelling} compares the behavior trace produced by \algoref{alg:ct_cosim_scenario} when applied to the co-simulation scenario described in \cref{eq:ct_causal_cosim_scenario_example}, with the analytical solution, obtained from the coupled model of \cref{eq:ct_coupled_model} (co-modelling).
	It is obvious that there is an error due to the extrapolation functions and the large communication step size $H=0.1$.
	
\fi
This is more modular because (if the equations are made available) the same constituent system can be coupled to other systems in many different contexts, without further changes.
As this sub-section shows, in co-simulation it is possible to get around the modularity aspect, but at a cost.

\subsubsection{Algebraic loops}
\label{sec:algebraic_loops}

Algebraic loops occur whenever there is a variable that indirectly depends on itself.
\ifreport
	To see how algebraic loops arise in co-simulation scenarios, recall (see \cref{eq:ct_causal_simulator}) that the state evolution and output of each simulation unit $S_i$ can be written as:
	\begin{aligneq}\label{eq:ct_causal_simulator_recall}
	x_i(t+H)	&= \delta_i(t, x_i(t), u_i(t))	\\
	y_i(t+H)	&= \lambda_i(t, x_i(t+H), u_i(t+H))	\\
	\end{aligneq}
	
	To simplify things, assume that the simulation units are coupled by a set of assignments from outputs to inputs, i.e., 
	\begin{aligneq}
	u_i(t) := y_j(t)
	\end{aligneq}
\noindent where $u_i$ is the input of simulation unit $S_i$ and $y_j$ the output of a simulation unit $S_j$, in the same co-simulation scenario.
	
	With these definitions, it is easy to see that, depending on the coupling assignments of the co-simulation scenario, the output of a simulation unit may depend on itself, that is, 
	\begin{aligneq}
	\mathbf{y_i(t+H)}	&= \lambda_i(t, x_i(t+H), u_i(t+H))		\\
	u_i(t+H) 	&= y_j(t+H) 									\\
	y_j(t+H)	&= \lambda_j(t, x_j(t+H), u_j(t+H))				\\
	u_j(t+H) 	&= y_k(t+H) 									\\
	\ldots	&													\\
	u_z(t+H)	&= \mathbf{y_i(t+H)}
	\end{aligneq}
\fi

We distinguish two kinds of algebraic loops in co-simulation \cite{Kubler2000}: the ones spanning just input variables, and the ones that include state variables as well.
The first kind can be avoided by using the input extrapolations as parameters to the output functions.
The second kind arises when implicit numerical solvers are used, or when the input approximating functions are interpolations instead of extrapolations.
\ifreport
	In the previous example, the first kind can be removed by replacing $u_i(t+H)$ in \cref{eq:ct_causal_simulator_recall} by the corresponding extrapolation $\phi_{u_i}(H, u_i(n \cdot H), u_i((n - 1) \cdot H), \ldots)$ which does not depend on $u_i((n + 1) \cdot H)$, thus breaking the algebraic loop.
\fi
These methods just ignore the algebraic loop though, and as is shown in \citet{Kubler2000,Arnold2014} (and empirically in \citet{Bastian2011a}), neglecting an algebraic loop can lead to a prohibitively high error in the co-simulation. 
A better way is to use a fixed point iteration technique.
For algebraic loops involving state variables, the same co-simulation step has to be repeated until convergence.
If the algebraic loop does not involve any state variable, then the iteration is just on the output functions.

\ifreport
	To see how algebraic loops involving state variables arise, suppose that, in the example above, $\phi_{u_i}$ is constructed from $u_i((n + 1) \cdot H)$:
	\begin{aligneq}\label{eq:interpolating_function}
	u_i(n \cdot H + m \cdot h_i) := \phi_{u_i}(m \cdot h_i, u_i((n + 1) \cdot H), u_i(n \cdot H), u_i((n - 1) \cdot H), \ldots)
	\end{aligneq}
	If an order can be imposed in the evaluation of the simulation units that ensures $u_i((n + 1) \cdot H)$ can be computed from some $\lambda_j(t, x_j((n + 1) \cdot H), u_j((n + 1) \cdot H))$ that does not indirectly depend on $u_i((n + 1) \cdot H)$, then this approach ---Gauss-Seidel--- can improve the accuracy of the co-simulation, as shown in \citet{Arnold2014,Arnold2001,Busch2016,Arnold2010,Kalmar-Nagy2014}.
	Obviously, the execution of simulation unit $S_i$ has to start after simulation unit $S_j$ has finished and its output 
	$\lambda_j(t, x_j((n + 1) \cdot H), u_j((n + 1) \cdot H))$ can be evaluated. 
	If the input $u_j((n + 1)$ depends indirectly on $u_i((n + 1) \cdot H)$, then an algebraic loop exists.
	The output function $\lambda_j(t, x_j((n + 1) \cdot H), u_j((n + 1) \cdot H))$ depends on the state of the simulation unit at $x_j((n + 1) \cdot H)$, which in turn can only be obtained by executing the simulation unit from time $n \cdot H$ to $(n + 1) \cdot H$, using the extrapolation of the input $u_j$, $\phi_{u_j}(m \cdot h_i, u_j((n + 1) \cdot H, \ldots))$;
	any improvement in the input $u_j((n + 1) \cdot H$, means that the whole co-simulation step has to be repeated, to get an improved $x_j((n + 1) \cdot H)$ and by consequence, an improved output $\lambda_j(t, x_j((n + 1) \cdot H), u_j((n + 1) \cdot H))$.
\fi

A fixed point iteration technique that makes use of rollback to repeat the co-simulation step with corrected inputs is called dynamic iteration, waveform iteration, and strong or onion coupling \cite{Hoepfer2011,Trcka2007}.
If the simulation units expose their outputs at every internal micro-step, then the waveform iteration can be improved \cite{Lelarasmee1982}.
Strong coupling approaches are typically the best in terms of accuracy, but worst in terms of performance.
\ifreport
	Approaches that do not perform any correction steps are the best in terms of performance, but worst in accuracy.
\fi
A variant that attempts to obtain the middle-ground is the so-called semi-implicit method, where a fixed limited number of correction steps is performed. 
See \citet{Schweizer2014,Schweizer2015d,Schweizer2016,Schweizer2015,Schweizer2015a} for examples of this approach.

In the current FMI Standard for co-simulation, it is not possible to perform a fixed point iteration on the output variables only, in the \emph{step mode}. 
A work-around this is to rollback the simulation units and repeat the co-simulation step, effectively treating the algebraic loop as involving the state variables too.

Until here, we have assumed full knowledge of the models being simulated in each simulation unit to explain how to identify, and deal with, algebraic loops.
In practice, with general black-box simulation units, extra information is required to identify algebraic loops.
According to \citet{Benveniste2000,Arnold2013,Broman2013}, a binary flag denoting whether an output depends directly on an input is sufficient.
A structural analysis, for example, with Tarjan's strong component algorithm \cite{Tarjan1972}, can then be performed to identify the loops.

\subsubsection{Consistent Initialization of Simulators}

The definition of a simulation unit in \cref{eq:ct_causal_simulator} assumes that an initial condition is part of the simulation unit. 
However, as seen in the example of \cref{sec:algebraic_composition}, the initial states of the simulation units can be coupled by algebraic constraints, through the output functions, which implies that the initial states of the simulation units cannot be set independently of the co-simulation in which they are used.
\ifreport
	For example, the constraint in \cref{eq:ct_nontrivial_algebraic_constraints} has to be satisfied for the initial states:
\begin{quote} 
  $\set{\tilde{x}_1(0),\tilde{v}_1(0),\tilde{x}_3(0),\tilde{v}_3(0)}$.
\end{quote}
\fi

In general, for a co-simulation scenario as defined in \cref{eq:ct_causal_cosim_scenario}, there is an extra coupling function $L_0$ that at the time $t=0$, has to be satisfied. For example:
\begin{aligneq}\label{eq:initial_constraint}
	L_0(x_1(0), \ldots, x_n(0), y_1(0), \ldots, y_n(0), y_\cs(0), u_1(0), \ldots, u_n(0), u_\cs(0)) = \bar{0}
\end{aligneq}
\ifreport
	\noindent where: 
	\begin{itemize}
	\item $x_i(0)$ denotes the initial state of simulation unit $S_i$; 
		and 
	\item $L_0: X_1 \times \ldots \times X_n \times Y_1 \times \ldots \times Y_n \times U_1 \times \ldots \times U_n \to \setreal^m$ represents the initial constraint, not necessarily equal to $L$ in \cref{eq:ct_causal_cosim_scenario}.
	\end{itemize}
\fi
\cref{eq:initial_constraint} may have an infinite number of solutions
\ifreport
	-- as in the case of the example provided in \cref{sec:algebraic_composition} --
\fi 
or have algebraic loops.
The initialization problem (or co-initialization) is identified in \citet{Blochwitz2012} and addressed in \citet{Galtier2015}.

In the FMI Standard, there is a dedicated mode for the (possible fixed point iteration based) search of a consistent initial state in all simulation units.

\subsubsection{Compositional Convergence -- Error Control}
\label{sec:ctcosim:error_control}

\ifreport
	The accuracy of a co-simulation trace is the degree to which it conforms to the real trace as described in \cref{sec:solver_computing_behavior}.
	Obtaining the real trace can be a challenge.
	Error ---the difference between the co-simulation trace and the real trace--- is then a measure of accuracy.
\fi

In the context of continuous co-simulation, the most accurate trace is the analytical solution to the coupled model that underlies the scenario.
\ifreport
	For example, the coupled model in \cref{eq:ct_coupled_model}, corresponding to the multi-body system in \cref{fig:mass-spring-damper_multibody_system}, is implicitly created from the co-simulation scenario described in \cref{eq:ct_causal_cosim_scenario_example}. 
	Fortunately, the analytical solution can be obtained for this coupled model because it forms a linear time invariant system. 
\fi
In practice, the analytical solution for a coupled model cannot be found easily.
Calculating the error precisely is impossible for most cases but getting an estimate in how the error grows is a well understood procedure in numerical analysis.

In simulation, the factors that influence the error are \cite{Cellier2006}: model, solver, micro-step size, and, naturally, the size of the time interval to be simulated.
In co-simulation, the extrapolation functions introduce error in the inputs of the simulation units, which is translated into error in the state/outputs of these, causing a feedback on the error that can increase over time.
Intuitively, the larger the co-simulation step size $H$, the larger is the error made by the extrapolation functions.

\ifreport
	For example, when the Forward Euler solver (\cref{eq:fw_euler_solver}) is used to compute the approximated behavior trace of the dynamical system in \cref{eq:dynamical_system_state_space_form}, in a single micro step, it is making an error in the order of
	$$
	\norm{\underbrace{\pargroup{x(t)+f(x(t))\cdot h + \bigO{h^2}}}_{\text{by infinite Taylor series}} - \underbrace{\pargroup{x(t) + f(x(t)) \cdot h}}_{\text{by Forward Euler}}} = \bigO{h^2}
	$$
	Obviously, the order in the error made at one step \bigO{h^2}, most commonly called the local error, depends on:
	\begin{compactitem}
	\item $f$ having no unbounded derivatives -- to see why, observe that if the derivative of $f$ is infinite, then the residual term cannot be bounded by a constant multiplied by $h^2$. Fortunately, since most continuous time dynamic systems model some real system, this assumption is satisfied.
	\item The solver used -- other solvers, such as the midpoint method, are derived by truncating higher order terms of the Taylor series. For the midpoint method, the local truncation error is \bigO{h^3};
	\item Naturally, the larger the micro step size $h$ is, the larger the local error \bigO{h^2} is.
	\end{compactitem}
	
	The local error assumes that the solver only made one step, starting from an accurate point $x(t)$.
	To compute the approximate behavior trace, the only accurate point the solver starts from is the initial value $x(0)$. The rest of the trace is approximate and the error gets compounded over the multiple steps.
	For the Forward Euler method, if there is a limit to how $f$ reacts to deviations on its parameter $\tilde{x}(t)=x(t)+e(t)$ from the true parameter $x(t)$, that is, if 
	$$ \norm{f(x(t)) - f(x(t) + e(t))} \leq \text{const } \cdot e(t) $$ 
	and $\text{const } < \infty $, then the order of the total accumulation of error can be defined in terms of the micro-step size.
	This condition is called global Lipschitz continuity \cite{Eriksson2013}.
	For the Forward Euler solver, the total (or global) error is \bigO{h}.
\fi

For a solver to be useful, it must be convergent, that is, the computed trace must coincide with the accurate trace when $h \to 0$ \cite{Wanner1991}.
It means the error can be controlled by adjusting the micro step size $h$.
The same concept of convergence applies to co-simulation but does, as the intuition suggests, decreasing the communication step size $H$ lead to a more accurate co-simulation trace?
This cannot be answered yet in general co-simulation because the behavior of the coupled model induced by the coupling of simulation units may not satisfy Lipschitz continuity.

In the context of CT co-simulation, according to \citet{Arnold2014,Arnold2001,Kubler2000,Busch2011,Hafner2013}, if the simulation units are convergent and the coupled model induced by the scenario coupling conditions can be written in the state space form of \cref{eq:dynamical_system_state_space_form}, then the co-simulation unit induced by any of the Jacobi, Gauss-Seidel, or Strong coupling methods, is convergent, with any polynomial extrapolation technique for the inputs.
Presence of algebraic loops, complex coupling constraints (such as the one shown in \cref{sec:algebraic_composition}), are factors that may make it impossible to write the coupled model in state space form. 
See \citet{Arnold2010} for more examples.

\ifreport
	The local error vector, in a co-simulation, is defined as the deviation from the true trace after one co-simulation step $H$, starting from an accurate point.
	\begin{aligneq}\label{eq:local_error_cosimulation}
	\begin{matrix}
	  x_1(t+H) - \tilde{x}_1(t+H) 	\\
	  \cdots 						\\
	  x_n(t+H) - \tilde{x}_n(t+H) 	\\
	  y_1(t+H)- \tilde{y}_1(t+H)	\\
	  \cdots 						\\
	  y_n(t+H)- \tilde{y}_n(t+H)	
	 \end{matrix}
	\end{aligneq}
\noindent where $\tilde{x}_i(t+H)=\delta_i(t, x_i(t), \phi_{u_i}(t))$, $\tilde{y}_i(t+H)=\lambda_i(t, \tilde{x}_i(t+H), \phi_{u_i}(t+H))$, and $x_i(t+H)$ and $y_i(t+H)$ are the true state vectors and outputs, respectively, for simulation unit $S_i$.
\fi

For a convergent co-simulation unit, some of the techniques used traditionally in simulation to \emph{estimate} the error, have been applied in co-simulation:
\begin{compactitem}
	\item[\textbf{Richardson extrapolation:}] 
		This well-known technique is compatible with black-box simulation units as long as these provide rollback and state saving/restore capabilities \cite{Galtier2015,Arnold2013,Arnold2014a}.
		\ifreport
			The essential idea is to get an estimate of the local error by comparing $\brackets{\tilde{x}_i(t+H), \tilde{y}_i(t+H)}^T$ with a less accurate point $\brackets{\bar{x}_i(t+H), \bar{y}_i(t+H)}^T$. 
			The less accurate point can be computed by the same orchestrator but using a larger communication step size. 
			We have seen that larger communication step sizes affect the accuracy so if the two points are not too far apart, it means the communication step $H$ does not need to be changed.
			It is importance to notice that the less accurate point $\brackets{\bar{x}_i(t+H), \bar{y}_i(t+H)}^T$ has to be computed from the accurate starting point $\brackets{\tilde{x}_i(t), \tilde{y}_i(t)}^T$.
		\fi
	\item[\textbf{Multi-Order Input Extrapolation:}] 
		The outputs of two different order input approximation methods are compared \cite{Busch2012,Busch2011}.
	\item[\textbf{Milne's Device:}] 
		Similar to the previous ones, but the extrapolation of the inputs is compared with its actual value, at the end of the co-simulation step.
		Iterative approaches such as the ones studied in \citet{Schweizer2014,Schweizer2015d,Schweizer2016,Schweizer2015,Schweizer2015a,Schweizer2015c,Arnold2001,Arnold2010} can readily benefit from this technique.
	\item[\textbf{Parallel Embedded Method:}] 
		This technique runs a traditional adaptive step size numerical method in parallel with the co-simulation \cite{Hoepfer2011}. 
		The purpose is to piggy back in the auxiliary method, the decisions on the step size. 
		The derivatives being integrated in each simulation unit have to be either provided, or estimated.
	\item[\textbf{Conservation Laws:}] 
		The local error is estimated based on the deviation from a known conservation law \cite{Sadjina2016}. Extra domain knowledge about the coupling between simulation units is required. 
		\ifreport
			For example, if the couplings form power bonds \cite{Paynter1961}, then energy should be conserved across a co-simulation step. In practice there is always an error due to the usual factors. The magnitude of the energy residual at a start and at end of a co-simulation step serves as an estimate of the local error. This technique has been implemented and studied in \citet{Sadjina2016}. It has the advantage that it may not require rollback functionalities.
		\fi
	\item[\textbf{Embedded Solver Method:}] 
		If the individual simulation units support adaptive step size, then the decisions made internally can be made public to help the orchestrator decide on the communication step size. 
		To the best of our knowledge, there is no orchestrator proposed that performs this, but the FMI Standard allows simulation units to reject too large communication step sizes \cite{Blochwitz2012,Broman2013}.
\end{compactitem}

After the error is deemed too large by one of the above methods, the correction can be applied pessimistically (rollback and repeating the same step) or optimistically (adapt the next step).
To mitigate the overhead of a pessimistic approach, the corrections may be applied only to sensitive simulation units, as done in \citet{Verhoeven2008}. 

Finally, the traditional simulation techniques can be applied to chose the right communication step size $H$: See \citet{Busch2011} for the  PI-controller approach, and \citet{Gustafsson1992,Gustafsson1988} for other techniques that can potentially be applied to co-simulation.

\subsubsection{Compositional Stability}

\ifreport
	In the previous section we have presented conditions in which an orchestration engine can reduce the communication step size to an arbitrarily small value in order to meet arbitrary accuracy. 
	Theoretically, this is useful as it tells the orchestrator that by reducing the local error, it also reduces the global error.
	In practice, the communication step size cannot be reduced to an arbitrarily small value without facing performance and roundoff error problems.
	Performance because, for smaller communication step sizes, it takes more steps to compute a behavior trace over a given interval of time.
	Round-off accuracy because in a digital computer, real numbers can only be represented approximately. 
	Computations involving very small real numbers incur a non-negligible round-off error.
	So that means that in practice convergence does not imply that arbitrary accuracy can be achieved.
	A better question is to analyze what happens to the global error, as the co-simulation trace is computed with a non-null communication step size $H$.
	
	Suppose that the analytical solution to the coupled model induced by the co-simulation scenario eventually goes to zero.
	This is the case for the coupled multi-body system of \cref{fig:mass-spring-damper_multibody_system}, described in \cref{eq:ct_coupled_model}, provided that at least one of the constants $d_1$ or $d_2$ is positive non-zero.
	Intuitively, this means that the system will lose energy over time, until it eventually comes to rest.
	
	Let $x_1(t)$ denote the analytical solution of the position the mass $m_1$ in the system, and let $\tilde{x_1}(t)$ be the solution computed by a co-simulation unit. Then 
	$e_{x_i}(t)=\norm{x_1(t) - \tilde{x_1}(t)}$ 
	denotes the global error at time $t$ made by the co-simulation unit.
	If 
	$\lim_{t \to \infty} x_1(t) = 0$, then 
	$\lim_{t \to \infty} e_{x_i}(t)= \tilde{x_1}(t)$.
	
	If the co-simulation unit is convergent, then for an arbitrarily small $H \to 0$, $\lim_{t \to \infty} e_{x_i}(t) \to 0$ will be arbitrarily small too.
	Since in practice we cannot take arbitrarily small $H$, we want to know whether there is some non-zero $H$ such that $\lim_{t \to \infty} \tilde{x_1}(t) = 0$, thus driving $e_{x_i}(t)$ to zero as well.
	If that is the case, then it means that, assuming the system will eventually come to rest, the co-simulation unit will too.
	This property is called numerical stability.
\fi

Contrarily to convergence, numerical stability is a property that depends on the characteristics of the system being co-simulated.
\ifreport
	Numerical stability is always studied assuming that the system being co-simulated is stable. 
	It makes no sense to show that the co-simulation trace will grow unbounded provided that the system does too. 
	It is a comparison of two infinities.
\fi
One of the ways numerical stability in co-simulation can be studied is by calculating the spectral radius of the error in the co-simulation unit, written as an autonomous linear discrete system \cite{Busch2010,Busch2012a}.

\ifreport
	To give an example, recall that the coupled model induced by the co-simulation scenario described in \cref{eq:ct_causal_cosim_scenario_example} can be written as:
	\begin{aligneq}\label{eq:analytical_solution_simulators}
	\vectorTwo{\dert{x_1}}%
			  {\dert{v_1}} &= \underbrace{\vectorTwo{ 0				   & 1 						}%
			  						   				{ -\frac{c_1}{m_1} & -\frac{d_1}{m_1} 	}}_{A_1}
			  				 \vectorTwo{x_1}%
			  				 		   {v_1} +
			  				 	\underbrace{\vectorTwo{	0 }%
			  						   	  			  {	\frac{1}{m_1} }}_{B_1}
			  					u_1																\\
	y_1 &= \underbrace{\vectorTwo{ 1 & 0 }%
			  					 { 0 & 1 }}_{C_1}
			 \vectorTwo{x_1}%
			 		   {v_1} 																	\\		  					
	\vectorTwo{\dert{x_2}}%
			  {\dert{v_2}} &= \underbrace{\vectorTwo{ 0 & 1 }%
			  						   				{ -\frac{c_2+c_c}{m_2} & -\frac{d_c}{m_2} }}_{A_2}
			  				 \vectorTwo{x_2}%
			  				 		   {v_2} +
			  				 	\underbrace{\vectorTwo{	0 &	0	}%
			  						   	  			  {	\frac{c_c}{m_2} & \frac{d_c}{m_2} }}_{B_2}
			  					u_2																\\
	y_2 &= \underbrace{\vectorOne{c_c 	&	d_c}}_{C_2}
			 \vectorTwo{x_2}%
			 		   {v_2} +
			 	\underbrace{\vectorOne{-c_c 	&	-d_c}}_{D_2} u_2												\\
	\end{aligneq}
	with the coupling conditions $u_1 = y_2$ and $u_2 = y_1$.
	
	In order to write the co-simulation model as an autonomous linear discrete system, we have to write what happens at a single co-simulation step $t \in \brackets{nH, (n+1)H}$ when executed by the orchestrator presented in \algoref{alg:ct_cosim_scenario}.
	Since the purpose is to analyze the stability of a co-simulation unit, and not the stability of each of the simulation units in the co-simulation, it is common to assume that the simulation units compute the analytical trace of the system. This enables the study of the stability properties of the co-simulation unit, starting from stable simulation units.
	
	From time $t \in \brackets{nH, (n+1)H}$, simulation unit $S_1$ is computing the behavior trace of the following Initial Value Problem Ordinary Differential Equation (IVP-ODE):
	\begin{aligneq}\label{eq:ode_ivp_simulator_1}
	\vectorTwo{\dert{x_1}(t)}%
			  {\dert{v_1}(t)} &= A_1	\vectorTwo{x_1(t)}%
			  				 		   			  {v_1(t)} + B_1 u_1(nH)
	\end{aligneq}
	with initial conditions $\vectorOne{x_1(nH) & v_1(nH)}^T$ given from the previous co-simulation step.
	The term $u_1(nH)$ denotes the fact that we are assuming a constant extrapolation of the input in the interval $t \in \brackets{nH, (n+1)H}$.
	
	\cref{eq:ode_ivp_simulator_1} is linear and time invariant, so the value of \vectorTwo{x_1((n+1)H)}{v_1((n+1)H)} can be given analytically \cite{Busch2012a} as:
	\begin{aligneq}
	\vectorTwo{x_1((n+1)H)}%
			  {v_1((n+1)H)} &= e^{A_1 H}\vectorTwo{x_1(nH)}{v_1(nH)} 
			  							+ 
			  					\pargroup{\int_{nH}^{(n+1)H} e^{A_1((n+1)H - \tau)} d\tau} B_1 u_1(nH) 
	\end{aligneq}
	or, replacing the integration variable with $s = \tau - nH$,
	\begin{aligneq}\label{eq:ode_ivp_simulator_1_solution}
	\vectorTwo{x_1((n+1)H)}%
			  {v_1((n+1)H)} &= e^{A_1 H}\vectorTwo{x_1(nH)}{v_1(nH)} 
			  							+ 
			  					\underbrace{\pargroup{\int_{0}^{H} e^{A_1(H - s)} ds}}_{K_1} B_1 u_1(nH) 
	\end{aligneq}
	
	\noindent where $e^{X} = \sum_{k=0}^{\infty} \frac{1}{k!}X^k$ is the matrix exponential.
	
	Rewriting \cref{eq:ode_ivp_simulator_1_solution} as a discrete time system gives us the computation performed by simulation unit $S_1$ in a single co-simulation step, that is, the state transition function $\delta_1$:
	\begin{aligneq}\label{eq:ode_ivp_simulator_1_dts}
	\vectorTwo{x_1^{(n+1)}}%
			  {v_1^{(n+1)}} &= e^{A_1 H}\vectorTwo{x_1^{(n)}}{v_1^{(n)}} + K_1 B_1 u_1^{(n)}
	\end{aligneq}
\noindent where $z^{(n)}=z(nH)$.
	
	At the end of the co-simulation step ($t = (n+1)H$) the output of the first simulation unit, that is, its output function $\lambda_1$, is given by plugging in \cref{eq:ode_ivp_simulator_1_dts} to the output $y_1$ in \cref{eq:analytical_solution_simulators}:
	\begin{aligneq}\label{eq:ode_ivp_simulator_1_output}
	y_1^{(n+1)} &= C_1 e^{A_1 H}\vectorTwo{x_1^{(n)}}{v_1^{(n)}} + C_1 K_1 B_1 u_1^{(n)}
	\end{aligneq}
	
	Repeating the same procedure for the second simulation unit, yields the state transition $\delta_2$ and output functions $\lambda_2$:
	\begin{aligneq}\label{eq:state_output_simulator_2}
	\vectorTwo{x_2^{(n+1)}}%
			  {v_2^{(n+1)}} &= e^{A_2 H}\vectorTwo{x_2^{(n)}}{v_2^{(n)}} +  K_2 B_2 u_2^{(n)}		\\
	y_2^{(n+1)} &= C_2 e^{A_2 H}\vectorTwo{x_2^{(n)}}{v_2^{(n)}} + \pargroup{C_2 K_2 B_2 + D_2} u_2^{(n)}
	\end{aligneq}
	with $K_2 = \int_{0}^{H} e^{A_2(H - u)} du$.
	
	Since the coupling conditions are $u_1 = y_2$ and $u_2 = y_1$, we can combine Equations \ref{eq:state_output_simulator_2}, \ref{eq:ode_ivp_simulator_1_output}, and \ref{eq:ode_ivp_simulator_1} into a single discrete time system:
	\begin{aligneq}\label{eq:co_simulation_dts}
	\vectorFour{\vectorTwo{x_1^{(n+1)}}%
			  			  {v_1^{(n+1)}}}%
			   {y_1^{(n+1)}}%
			   {\vectorTwo{x_2^{(n+1)}}%
			  			  {v_2^{(n+1)}}}%
			   {y_2^{(n+1)}} = 
	   \underbrace{\vectorFour{e^{A_1 H}		&\bar{0}						&\bar{0}		& K_1 B_1}%
							   {C_1 e^{A_1 H}	&\bar{0}						& \bar{0}		& C_1 K_1 B_1}%
							   {\bar{0}			&K_2 B_2				&e^{A_2 H}		& \bar{0}}%
							   {\bar{0}			&C_2 K_2 B_2 + D_2	&C_2 e^{A_2 H}	&\bar{0}}}_{A}
					\vectorFour{\vectorTwo{x_1^{(n)}}%
							  			  {v_1^{(n)}}}%
							   {y_1^{(n)}}%
							   {\vectorTwo{x_2^{(n)}}%
							  			  {v_2^{(n)}}}%
							   {y_2^{(n)}}
	\end{aligneq}
	
	The above system is stable if the behavior traces remain bounded (e.g., by going to zero) as $n \to \infty$.
	This can be checked by observing whether the spectral radius $\rho(A) < 1$.
	For parameters $m_1 = m_2 = c_1 = c_2 = d_1 = c_c = d_c = 1, d_2 = 2$, a communication step size of $H=0.001$, $\rho(A)=0.9992$, which means that the co-simulation unit is stable.
	If the damping constant were $d_k=6.0E6$, then the co-simulation unit would be unstable ($\rho(A) \approx 76.43$).
	A stable co-simulation is shown in \cref{fig:stable_unstable_cosimulator_results}.
	
	\begin{figure}[htb]
	\begin{center}
	  \includegraphics[width=0.7\textwidth]{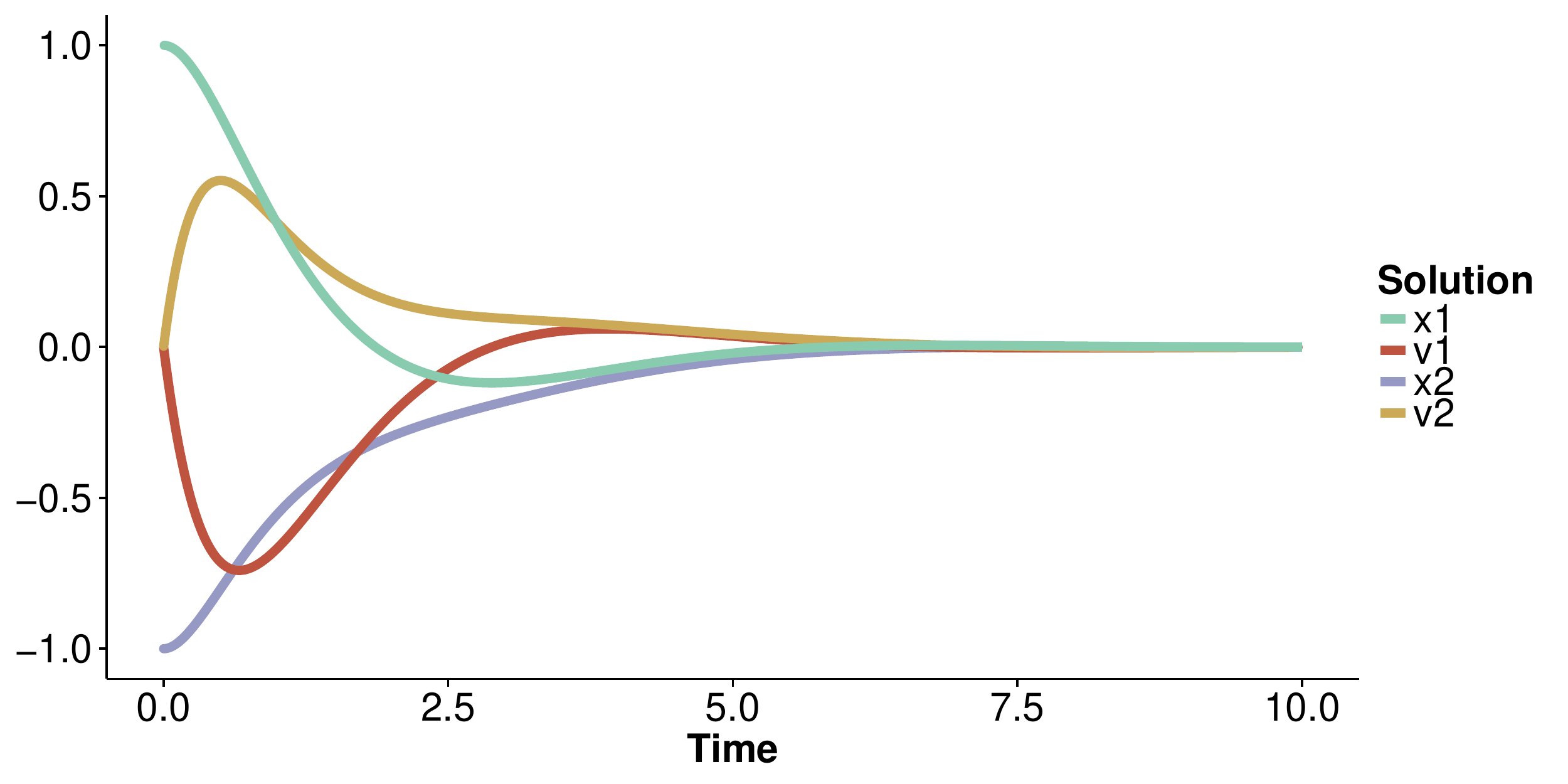}
	  \caption{Behavior trace of co-simulator described in \cref{eq:co_simulation_dts}. Parameters are: $m_1 = m_2 = c_1 = c_2 = d_1 = c_c = d_c = 1, d_2 = 2, H=0.001$.}
	  \label{fig:stable_unstable_cosimulator_results}
	\end{center}
	\end{figure}

\fi

Different coupling methods, and different approximation functions yield different stability properties.
See \citet{Busch2010,Busch2011a,Busch2012a,Busch2016} for the stability analysis of multiple coupling approaches and approximating functions.
Stability of various co-simulation units has been also studied in \citet{Arnold2010,Gu2001,Kalmar-Nagy2014,Schweizer2015c,Kubler2000a}.
The \emph{rules of thumb} drawn from these papers can be summarized as:
\begin{compactitem}
	\item Co-simulators that employ fixed point iteration techniques typically have better stability properties;
	\item Gauss-Seidel coupling approach has slightly better stability properties when the order in which the simulation units compute is appropriate. For example, the simulation unit with the highest mass should be computed first \cite{Arnold2010};
\end{compactitem}

\ifreport
	The main problem is that in co-simulation applied to industrial problems, the solvers and models may be coupled in a black box to protect IP, so there is little knowledge about the kind of solver and model being used and its stability properties.
	The best is then to always use iterative techniques that have been shown to have better stability properties.
	However, these techniques require rollback functionalities which can be difficult to support for certain simulation units. 
	Even if those functionalities are available, the cost of computing a co-simulation trace can be prohibitively  high when compared with non-iterative approaches.
	This creates a paradox where industrial co-simulation units should make use of iterative techniques but the performance toll may be too high. 
\fi

\subsubsection{Compositional Continuity}
\label{sec:continuity_behavior_traces}

Let us for now assume that each CT simulation unit is a mock-up of a continuous system.
A prerequisite is that the physical laws of continuity are obeyed.
When using extrapolation in the inputs (e.g., constant extrapolation), these laws may not be obeyed, as discussed in \citet{Busch2016,Sadjina2016}. 
\ifreport
	Consider the point of view of a simulation unit $S_i$ in co-simulation. 
	Throughout a co-simulation step $t \in \brackets{nH, (n+1)H}$ the input $\phi_{u_i}(t, u_i(nH))=u_i(nH)$ is kept constant. 
	At the next co-simulation step $t \in \brackets{(n+1)H, (n+2)H}$, the input $\phi_{u_i}(t, u_i((n+1)H))=u_i((n+1)H)$ may change radically if $u_i((n+1)H)$ is too far away from $u_i(nH)$.
\fi

The discontinuities in the inputs may wreak havoc in the performance of the simulation unit $S_i$, causing it to reduce inappropriately the micro step size, to reinitialize the solver \cite{Cellier2006}, to discard useful information about the past (in multi-step solvers \cite{Andersson2016b,Andersson2016}), and/or produce inaccurate values in its input extrapolation \cite{Oh2016}.
Furthermore, a discontinuity may be propagated to other simulation units, aggravating the problem.

\ifreport
	Most numerical methods assume that the input is a discretized version of a continuous trace. That means that, when a discontinuity occurs, simulation unit $S_i$ cannot distinguish it from a very steep change in the continuous trace.
	The way traditional solvers deal with this behavior is to reduce the micro step size $h_i$ until the change is not so steep. This works with a continuous signal with a steep change, but does not work with a discontinuity: even if the micro-step size $h_i$ is reduced, the difference between $\lim_{t \to ((n+1)H)^-} \phi_{u_i}(t, u_i(nH)) = u_i(nH)$ and $\lim_{t \to ((n+1)H)^+} \phi_{u_i}(t, u_i((n+1)H)) = u_i((n+1)H)$ is still the same, as it depends on the communication step size $H$ and not on the micro step size $h_i$.
	The solver will reduce the micro step size until a minimum is reached, at which point it gives up and finally advantages the micro step \cite{Cellier2006}.
	
	Most of the times this gives acceptable results but has a huge performance toll: when the solver is repeatedly retrying a small micro-step size, it does not advance the simulated time. This means that a huge computational effort goes to waste until the solver finally gives up \cite{Cellier1979}.
	
	We defer the discussion of the correct ways to deal with discontinuities to co-simulation scenario where discontinuities are welcome, \cref{sec:hybrid_cosimulation}.
	In continuous co-simulation scenarios, discontinuities should not occur.
\fi

A solution to avoid discontinuities in the input approximations is to use the extrapolated interpolation methods instead of constant extrapolation \cite{Dronka2006,Busch2012a,Busch2016}.
\ifreport
	These methods ensure at least that $\lim_{t \to ((n+1)H)^-} \phi_{u_i}(t, u_i(nH)) = \lim_{t \to ((n+1)H)^+} \phi_{u_i}(t, u_i((n+1)H))$.
	
	To give an example, we derive one possible linear extrapolated interpolation method for $\phi_{u_i}$ over the interval $t \in \brackets{nH, (n+1)H}$.
	Since $\phi_{u_i}$ is linear, then $\phi_{u_i}(t, u_i(nH), u_i((n-1)H))= b + a(t-nH)$, for some constants $a,b$.
	Let $\bar{u}_i(nH)=\phi_{u_i}(nH, u_i((n-1)H), u_i((n-2)H))$.
	To avoid discontinuities, we require that $\phi_{u_i}(nH, u_i(nH), u_i((n-1)H)) = \bar{u}_i(nH)$.
	And we want that $\phi_{u_i}((n+1)H, u_i(nH), u_i((n-1)H)) = u_i(nH)$.
	
	So putting these constraints together gives
	\begin{aligneq}
	\phi_{u_i}(t, u_i(nH), u_i((n-1)H)) 	&= b + a(t-nH) \\
	\bar{u}_i(nH)							&= \phi_{u_i}(nH, u_i((n-1)H), u_i((n-2)H)) \\
	\phi_{u_i}(nH, u_i(nH), u_i((n-1)H)) 	&= \bar{u}_i(nH) \\
	\phi_{u_i}((n+1)H, u_i(nH), u_i((n-1)H))&= u_i(nH)
	\end{aligneq}
	
	Solving this system for $\phi_{u_i}(t, u_i(nH), u_i((n-1)H))$ gives:
	\begin{aligneq}\label{eq:extrapolated_iterpolation_method}
	\phi_{u_i}(t, u_i(nH), u_i((n-1)H)) &= u_i((n-1)H) + \frac{u_i(nH) - u_i((n-1)H)}{H}(t-nH)
	\end{aligneq}
\fi

\subsubsection{Real-time Constraints}

As introduced in \cref{sec:background}, the major challenge in real-time simulation is to ensure that a simulation unit is fast-enough to satisfy the timing constraint $t  = \alpha \wct$.
In real-time co-simulation, this challenge gets aggravated due to the presence of multiple simulation units, with different capabilities \cite{Stettinger2013}.
In order to enable real-time co-simulation, every simulation unit has to be fast enough. 
Furthermore, real-time co-simulation is often needed because one of the simulation unit is actually the original system, wrapped as a simulation unit. 
This means that measurements are performed to the state of the system, and this means noise in the signals.
Therefore, the extrapolation functions used in the other simulation units have to be properly protected from the noise in the signal, using statistical techniques such as Kalman filtering \cite{Kalman1960}.
Finally, the quality of the network is important, as the real-time simulation units needs to receive their inputs in a timely manner. To mitigate this, the orchestration algorithm has to compensate for any delays in the receiving of data, and provide inputs to the real-time simulation unit \cite{Stettinger2014}.

\section{Hybrid Co-simulation Approach}
\label{sec:hybrid_cosimulation}

\newcommand{\hybridCT}{\emph{Hybrid CT}}
\newcommand{\hybridDE}{\emph{Hybrid DE}}

Sections \ref{sec:de_cosim} and \ref{sec:ct_cosim} described the essential characteristics and assumptions of simulation units for each kind of co-simulation approach.
When compared to a CT unit, whose state evolves continuously in time and whose output may have to obey to physical laws of continuity, a DE unit state can assume multiple values at the same time (transiency) and its output can dramatically change over a short period of time.
For an orchestrator, a CT unit has some flexibility (safe for algebraic loops and ugly coupling conditions) in computing the co-simulation.
In contrast, a DE simulation unit has to get inputs and produce outputs at the precise time some event occurs.
And due to the potentially drastic change in the outputs, there is no Lipschitz continuous condition that allows predicting how a delay in the output of the DE unit can affect the overall co-simulation trace.

\ifreport
	For example, in the simulation unit of the mass-spring-damper system, \cref{eq:ct_causal_simulator_msd_1}, with a constant extrapolation function, and running under the orchestrator in \algoref{alg:ct_cosim_scenario}, the change in the input can only affect the output after at least $H$ units of time.
	For continuous time solvers in general, as can be seen for the explicit solver in \cref{eq:fw_euler_solver}, a delayed response to the inputs is normal.
\fi

These differences between CT and DE units are at the heart of many challenges in hybrid co-simulation scenarios, mixing the two.

\subsection{Hybrid Co-simulation Scenarios}

We do not give a formal definition of a hybrid co-simulation scenarios because that is related to finding an appropriate standard for hybrid co-simulation, which is a non trivial challenge (see \cref{sec:standard_hybrid_cosim}) \cite{Broman2015}.

Instead, we define it broadly as mixing the characteristics and assumptions of both kinds of simulation units.
These scenarios, together with an adequate orchestrator, can be used as mock-ups of hybrid systems \cite{Cellier1977,Maler1992,Alur1995,Carloni2006}.
A thermostat regulating the temperature in a room is a classical example \cite{Lygeros2004}.
The continuous constituent system represents the temperature dynamics of the room, accounting for a source of heat (radiator).
The discrete event part is a controller that turns on/off the radiator depending on the temperature.

\begin{figure}[htb]
\begin{center}
  \includegraphics[width=0.2\textwidth]{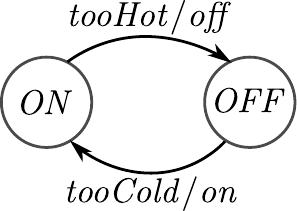}
  \caption{Statemachine model of the controller constituent system.}
  \label{fig:statechart_controller_hybridcosim}
\end{center}
\end{figure}

The continuous time simulation unit $S_1$ simulates the following dynamics:
\begin{aligneq}\label{eq:room_temperature_system}
\dot{x} &= -\alpha \pargroup{x - 30q} \\
x(0) &= x_0 \\
\end{aligneq}
\noindent where $x$ is the output temperature in the room, $\alpha>0$ denotes how fast the room can be heated (or cooled) down, and $q \in \set{0,1}$ is the control input that turns on/off the radiator.
The discrete event simulation unit $S_2$ simulates the statemachine shown in \cref{fig:statechart_controller_hybridcosim}, where one can think of the input event $\mathit{tooHot}$ as happening when $x(t) \geq 21$ and $\mathit{tooCold}$ when $x(t) \leq 19$.
The output events $\mathit{off}$ and $\mathit{on}$ will assign the appropriate value to the input $q$ of $S_1$.
Therefore, the temperature $x(t)$ is kept within a comfort region.

Clearly, the two units cannot just be coupled together via input to output assignments.
Any orchestrator for this co-simulation scenario has to reconcile the different assumptions about the inputs and output of each simulation unit.
\ifreport
	\begin{compactitem}
	\item The CT simulation unit expects a continuous input, whereas the output of the DE simulation unit is an event signal.
	\item The output of the CT simulation unit is a continuous signal, whereas the DE simulation units expects an event signal as input.
	\end{compactitem}
\fi
The coupling of continuous time and discrete event black box simulation units has been studied in the state of the art.
In essence, two approaches are known, both based on creating a wrapper component around a simulation unit to adapt its behavior:
\begin{compactenum}
\item[\hybridDE] -- wrap every CT unit as a DE simulation unit, and use a DE based orchestration;
\item[\hybridCT] -- wrap every DE unit to become a CT unit and use a CT based orchestrator.
\end{compactenum}

According to the formalization that we have proposed for CT and DE simulation units, the \hybridDE\ approach, applied to the thermostat example may involve: 
wrapping $S_1$ as a DE simulation unit, $S'_1$, with a time advance that matches the size of the co-simulation step; 
and keeping track of the output of $S_1$ in order to produce an output event whenever it crosses the thresholds.
Conversely, any output event from $S_2$ has to be converted into a continuous signal for the input $q(t)$ of $S_1$.

Other examples of \hybridDE\ are described in \citet{Vangheluwe2000,Quesnel2005,Nutaro2011,Kofman2001,Widl2013,Bolduc2002,Fey1997,Awais2013,Ylmaz2014,Zeigler2006,Bolduc2003,Camus2016,Camus2015,Kounev2015,Neema2014,Kuhr2013}.

The \hybridCT, in our example, can be followed by wrapping the DE unit $S_2$ as a CT unit that takes as input the temperature continuous signal, and internally reacts to an event caused by the crossing of the threshold.
Conversely, the output event of $S_2$ can be converted into a continuous signal $q(t)$.

Examples of the \hybridCT\ include \citet{Garro2015,Quaglia2012,Lawrence2016,Denil2015,Tavella2016,Feldman2014,Tripakis2015}.

Regardless of the approach taken, the properties of the constituent systems have to be retained: the fact that an otherwise discontinuous signal becomes continuous as a result of a linear or higher order extrapolation may not respect the properties of the coupled system.
Knowledge of the domain and the simulation units is paramount.

A third alternative, compared to only using Hybrid CT or Hybrid DE, is to have different mechanisms of orchestrating the simulation units depending on the semantic domain. For instance, in the actor modeling language Ptolemy II~\cite{Ptolemaeus2014}, an actor has many similarities to a simulation unit. Instead of using either Hybrid CT or Hybrid DE, a so called \emph{Director} block is used for a particular set of connected actors. In this context, the notion of superdense time is fundamental, as also discussed in~\cite{Broman2015}.

In the subsection below, different issues that arise in hybrid co-simulation will be described.
These should be read in the light of hierarchical hybrid co-simulation scenarios, where compositionality is important.

\subsection{Challenges}

\subsubsection{Semantic Adaptation}
\label{sec:semantic_adaptation}

While a generic wrapper based on the underlying model of computation of the simulation unit can be used, as done in \cite{Ptolemaeus2014,Cremona2016b}, the realization of any of the approaches \hybridDE\ or \hybridCT\ depends on the concrete co-simulation scenario and the features of the simulation units \cite{Boulanger2011,Mustafiz2016}, as shown with the thermostat example.
There is simply no best choice of wrappers for all scenarios.
Even at the technical level, the manner in which the events or signals are sent to (or obtained from) the unit may need to be adapted \cite{Tripakis2015}.
\ifreport
	To be concrete, the simulation unit $S_2$ can assume that all events are communicated by encoding them in a single string signal, as opposed to having a different signal signal to denote different events.
\fi
To account for this variability, the most common adaptations can be captured in a configuration language, as was done in \citet{Meyers2013,Denil2015}, or in a specialization of a model of computation, as done in \citet{Kuhr2013,Muller2015,Pedersen2016}.
These approaches require that a person with the domain knowledge describes how the simulation units can be adapted.

Our choice of wrapper for the \hybridDE\ approach is meant to highlight another problem with the adaptations of simulation units: the wrapper incorporates information that will ultimately have to be encoded in the software controller.
As such, we argue that the need for sophisticated semantic adaptations should be smaller in later stages of the development of the components so that, for more refined models of the thermostat, the decision about when to turn off the radiator is not made by a wrapper of $S_1$.

\subsubsection{Predictive Step Sizes}

In the \hybridDE\ approach, the time advance has to be defined (recall \cref{eq:de_causal_model}).
Setting it to whatever co-simulation step size $H$ the orchestrator decides will work, but the adapted simulation unit may produce many absent output events. Better adaptations have been proposed.
In the thermostat example, $S'_1$ can propose a time advance that coincides with the moment that $x(t)$ will leave the comfort region, thereby always being simulated at the relevant times.

Naturally, these approaches rely in information that may expose the IP of simulation units.
Others try to adaptively guess the right time advance by monitoring other conditions of interest, set over the own dynamics of the adapted simulation unit, the most common approach being the quantization of the output space \cite{Zeigler1998,Bolduc2003,Kofman2002,Kofman2001,Nutaro2016}.

The capability to predict the time advance is also useful to enhance the performance/accuracy of CT based co-simulation, as shown in \citet{Broman2013}.

\subsubsection{Event Location}

Locating the exact time at which a continuous signal crosses a threshold 
\ifnotreport
	(e.g., finding $t$ such that the temperature $x(t) \leq 19$)
\fi
is a well known problem \cite{Bouchhima2006,Zhang2008,Bombino2013} and intimately related to guessing the right time advance for predicting the step size \cite{Camus2016,Galtier2015}. 
To address this, solutions typically require derivative information of the signal that causes the event, and/or the capability to perform rollbacks.
\ifreport
	In the thermostat example, a co-simulation that shows the output $q$ of the controller changing from $0$ to $1$ at time $t_e$ while the temperature of the room $x$ actually crossed the confort zone at $t_e - k$, for $k > 0$, may not be accurate if $k$ is too large.
	Note that $k$ is a consequence of the decisions made in the orchestrator.
\fi
			
\subsubsection{Discontinuity Identification}

\ifreport
	Until here, we have based our discussion in the knowledge of what kind of simulation units comprise a co-simulation. 
\fi
In a general hierarchical co-simulation, a simulation unit's output may be an event signal coming from a wrapper of a CT unit, or vice-versa.
In any case, at runtime, a signal is often represented as a set of time-stamped points.
Observing this sequence of points alone does not make it possible to discern a steep change in a continuous signal, from a true discontinuity, that occurs in an event signal \cite{Lee2005,Broman2015,Zhang2008,Mosterman1999}.
Extra information is currently used:
\begin{inparaenum}[\itshape a\upshape)]
\item a formalization of time which include the notion of absent signal, as proposed in \citet{Tavella2016,Lee2005,Broman2015}; or
\item an extra signal can be used to discern when a discontinuity occurs, as done in the FMI for Model Exchange \cite{Blochwitz2012}, even facilitating the location of the exact time of the discontinuity; or
\item symbolic information (e.g., Dirac impulses \cite{Dirac1981}) that characterize a discontinuity can be included, as done in \citet{Nilsson2003}
\end{inparaenum}

\subsubsection{Discontinuity Handling}

Once a discontinuity is located, how it is handled depends on the nature of the simulation units and their capabilities.
If the simulation unit is a mock-up of a continuous system then, traditionally, discontinuities in the inputs should be handled by reinitializing the simulation unit \cite{Cellier2006}.
This step can incur a too high performance cost, especially with multi-step numerical methods, and \citet{Andersson2016b,Andersson2016} proposes an improvement for these solvers.
Furthermore, a discontinuity can cause other discontinuities, producing a cascade of re-initializations.
During this process, which may not finish, care must be taken to ensure that physically meaningful properties such as energy distribution, are respected \cite{Mosterman1998a}.

\subsubsection{Algebraic Loops, Legitimacy, and Zeno Behavior}

Algebraic loops are non-causal dependencies between simulation units that can be detected using feedthrough information, as explained in \cref{sec:algebraic_loops}.
In CT based co-simulation, the solution to algebraic loops can be attained by a fixed point iteration technique, as covered in \cref{sec:algebraic_loops}.
There is the possibility that the solution to an algebraic loop will fail to converge.
The result is that, if left unchecked, the orchestrator would move an infinite number of input and output values between simulation units, at the same point in time.

In DE based co-simulation a related property is legitimacy \cite{Zeigler2000}, which is the undesirable version of the \emph{transiency} property, explained in \cref{sec:de_cosim}.
A illegitimate co-simulation scenario will cause the co-simulation orchestrator to move an infinite number of events with the same timestamp between units, never advancing time.
Distance matrices, used to optimize parallel optimistic approaches, as explained in \citet{Fujimoto2000} and used in \citet{Ghosh1995}, can be leveraged to detect statically the presence of \emph{some} classes of illegitimacy.

A similar behavior, but more difficult to detect is Zeno behavior.
It occurs when there is an increasingly small interval of time between two consecutive events, up to the point that the the sum of all these intervals is finite \cite{VanDerSchaft2000}.
However, while illegitimate behaviors are not desired in pure DE co-simulation, at least in the theoretical sense, Zenoness is a desired feature in some hybrid co-simulation scenarios.
We say in the theoretical sense because, for the purposes of co-simulation, scenarios with Zenoness still have to be recognized and appropriate measures, such as regularization \cite{Johansson1999}, have to be taken.

\subsubsection{Stability}
\label{sec:hybrid_cosim:stability}

If a hybrid co-simulation represents a hybrid or switched system \cite{VanDerSchaft2000,Jungers2009}, then it is possible that a particular sequence of events cases the the system to become unstable, even if all the individual continuous modes of operation are stable.
New analyses are required to identify whether the CT units can yield unstable trajectories as a result of the events of wrapped DE simulation units, while keeping the IP hidden.

\subsubsection{Theory of DE Approximated States}

In a pure DE based co-simulation, if round-off errors are neglected, the computed trajectories are essentially exact.
To the best of our knowledge, only \citet{Zeigler2000} addresses theoretically how the error in a discrete event system can be propagated.
In CT based co-simulation however, error is an accepted and well studied and techniques exist to control it.

In Hybrid co-simulation, there is a need for analysis techniques that provide bounds on the error propagation in the DE simulation units, when these are coupled to sources of error.

\ifreport
	In addition, based on these analyzes, it should be possible for a DE simulation unit to recognize that its error has exceeded a given tolerance, and measures should be taken to reduce that error.
	Having these techniques in place allows a hybrid co-simulation orchestrator to take appropriate measures (e.g., adapt the communication step size, etc\ldots) the keep the error bounded in every simulation unit.
\fi

\subsubsection{Standards for Hybrid Co-simulation}
\label{sec:standard_hybrid_cosim}

While for CT co-simulation there is the Functional Mock-up Interface (FMI) standard \cite{Blochwitz2012}, and for DE co-simulation there is the High Level Architecture (HLA) \cite{HLA2010} standard, as of the time of writing, both standards have limitations for hybrid co-simulation.
\citet{Tavella2016,Bogomolov2015,Garro2015} use/propose extensions to the FMI standard and \citet{Awais2013a} proposes techniques to perform CT simulation conforming to HLA.
Recognizing that hybrid co-simulation is far from well studied, \citet{Broman2015} proposes a set of idealized test cases that any hybrid co-simulation unit, and underlying standard, should pass. In particular, it is important to have correct handling and representation of time, to achieve a sound approach for simultaneity.  

Finally, even with a standardized interface, simulation units are not all equal: a fact that makes coding an orchestration algorithm a real challenge.
A possible approach to deal with this heterogeneity, proposed in \citet{Gomes2016}, is to assume that all units implement the same set of features, code the orchestration algorithm for those features, and delegate to wrappers the responsibility of leveraging extra features (or mitigating the lack of).
In the section below, these features are classified.

\section{Classification}
\label{sec:classification_overview}

Having described the multiple facets of co-simulation, this section summarizes our classification and methodology.

\subsection{Methodology}
\label{sec:methodology}

To find an initial set of papers related to co-simulation, we used Google Scholar with the keywords ``co-simulation'', ``cosimulation'', ``coupled simulation'', and collected the first 10 pages of papers.
Every paper was then filtered by the abstract, read in detail, and its references collected.
To guide our reading to the most influential papers, we gave higher priority to most cited (from the papers that we have collected).

We read approximately 30 papers to create the initial version of the taxonomy.
Then, as we read new papers, we constantly revised the taxonomy and classified them.

After a while, new references did not cause revisions to the taxonomy, which prompted us to classify the collected papers in a more systematic fashion: all the papers that we collected from 2011 (inclusive) up to, and including, 2016 were classified.
Two main reasons justify the last 5 years interval: limited time; and most of the papers refer to, and are based on, prior work. 
\ifreport
	As a consequence, the classification would be very similar for many of the related references prior to 2011.
\fi

From the papers classified, those that report case studies where noted to create \cref{fig:applications_time_line}.

\subsection{Taxonomy}
\label{sec:taxonomy}

\begin{figure}
    \centering
    \begin{minipage}{0.25\textwidth}
        \vspace{1.45cm}
        \includegraphics[scale=0.5]{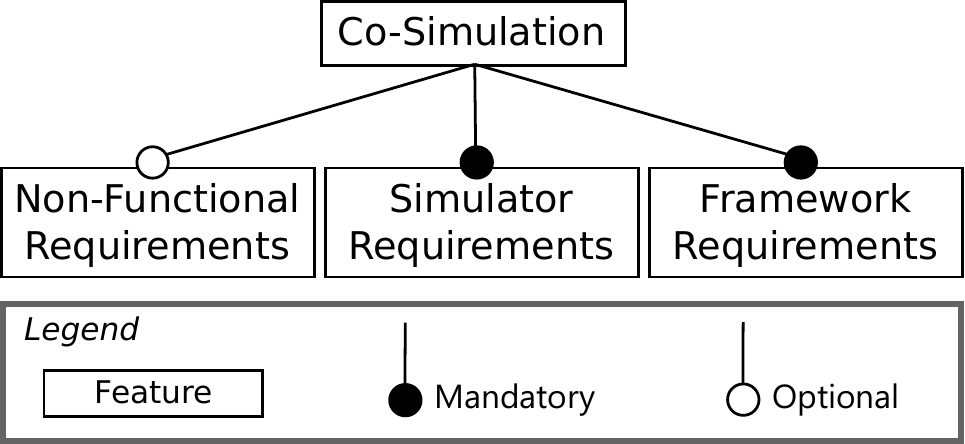}
		\caption{Top-level.}
		\label{fig:featuremodel_top}
    \end{minipage}\hfill
    \begin{minipage}{0.68\textwidth}
        \includegraphics[scale=0.5]{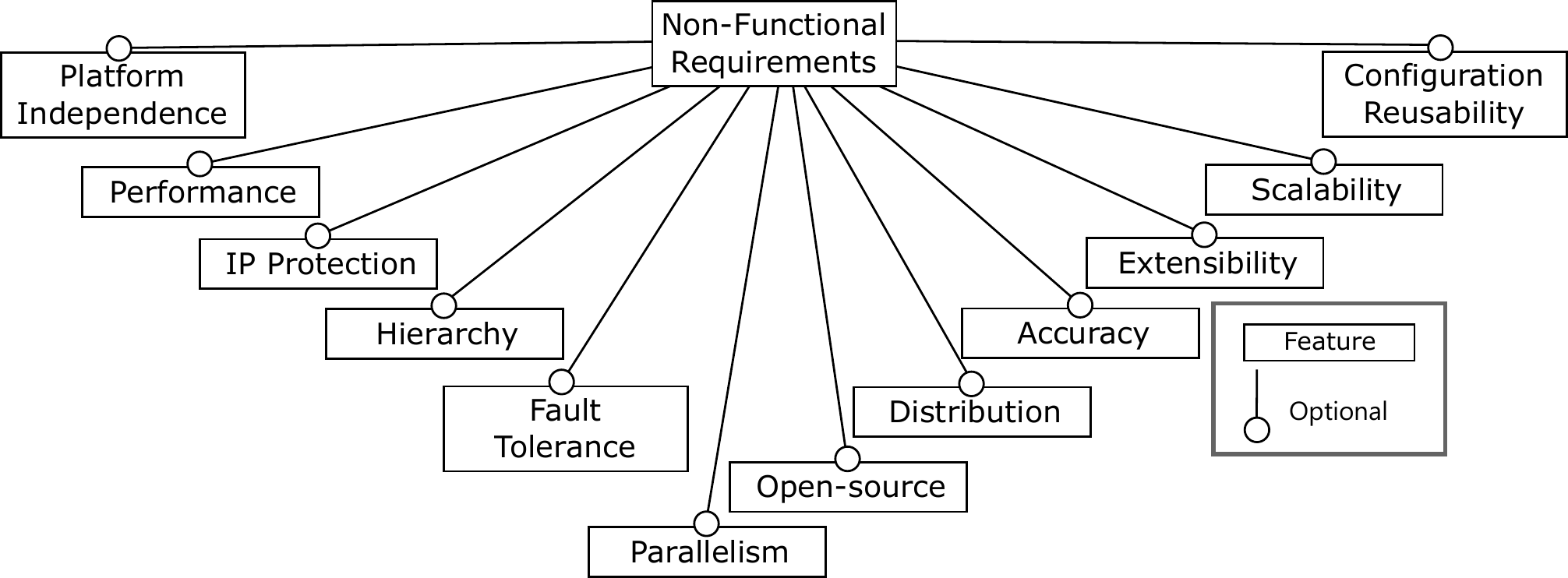}
		\caption{Non-Functional Requirements.}
		\label{fig:featuremodel_nfrs}
    \end{minipage}
\end{figure}

\begin{figure}[htb]
\begin{center}
	\includegraphics[scale=0.5]{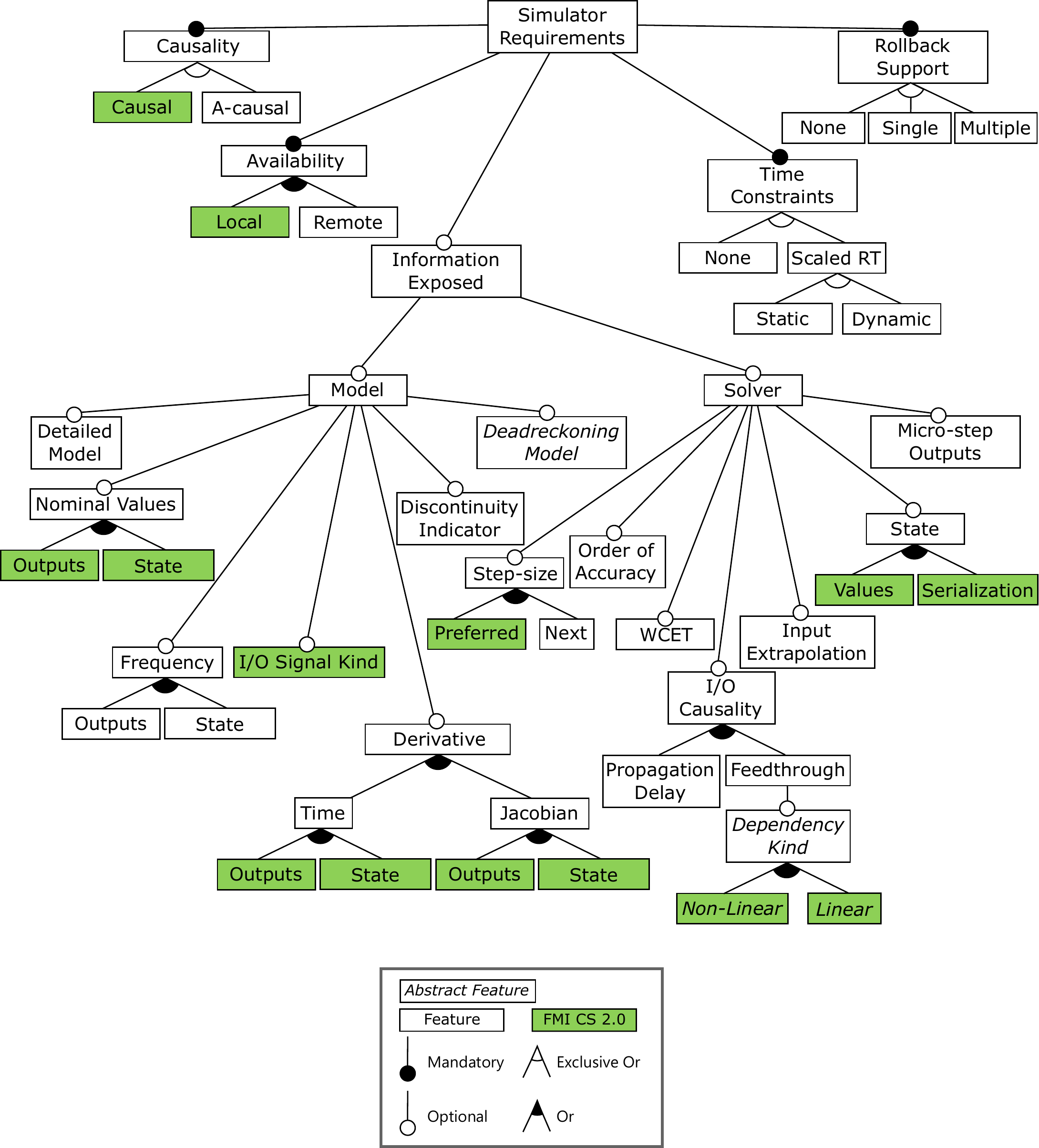}
	\caption{Simulation Unit Requirements and features provided in the FMI Standard for co-simulation, version 2.0.}
	\label{fig:featuremodel_sr}
\end{center}
\end{figure}

\begin{figure}[htb]
\begin{center}
	\includegraphics[scale=0.6]{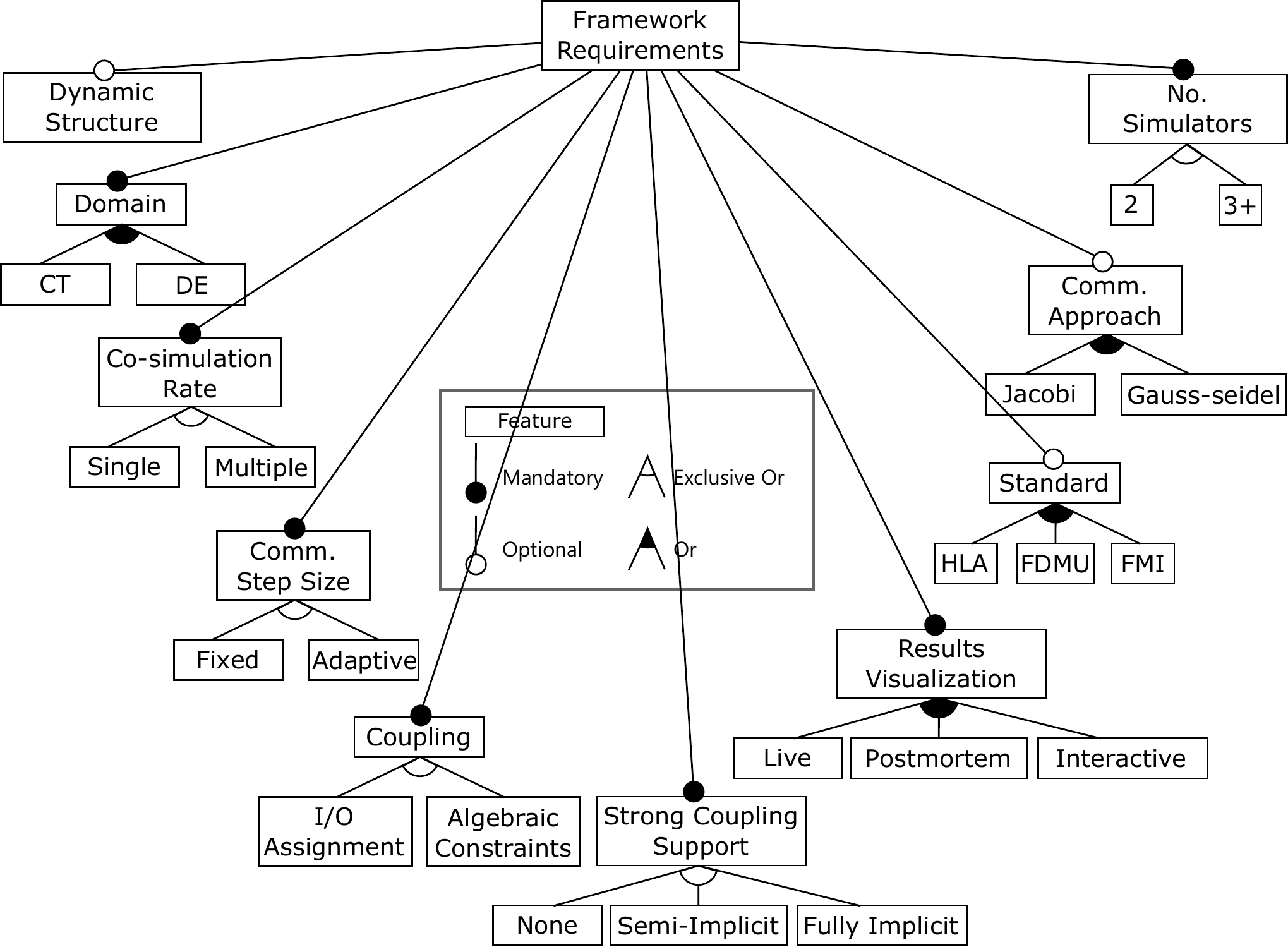}
	\caption{Framework Requirements.}
	\label{fig:featuremodel_fr}
\end{center}
\end{figure}

The taxonomy is represented as a feature model \cite{Kang1990} structured in three main categories, shown in \cref{fig:featuremodel_top}:
\begin{compactenum}
\item[\textbf{Non-Functional Requirements (NFRs):}] Groups concerns (e.g., performance, accuracy, and IP Protection) that the reference addresses.
\item[\textbf{Simulation unit Requirements (SRs):}] Features required/assumed from the simulation units by the orchestrator described in the paper. Examples: Information exposed, causality, local/remote availability, or rollback support.
\item[\textbf{Framework Requirements (FRs):}] Features provided by the orchestrator. Examples: dynamic structure, adaptive communication step size, or strong coupling support.
\end{compactenum}
Each main group is detailed in Figures \ref{fig:featuremodel_nfrs}, \ref{fig:featuremodel_sr}, and \ref{fig:featuremodel_fr}.
Abstract features denote concepts that can be easily detailed down but we chose not to, for the sake of brevity.
Mandatory features are required for the activity of co-simulation while optional are not.

\subsection{State of the Art}

To give an example on how the taxonomy is used, consider the work in \citet{Acker2015}, where an FMI based multi-rate orchestration algorithm is generated from a description of the co-simulation scenario.
In the paper, the description language introduced can be reused in a tool-agnostic manner.
The orchestration code generator analyzes the co-simulation scenario, and:
\begin{inparaenum}[a)]
	\item identifies algebraic loops using I/O feedthrough information;
	\item separates the fast moving simulation units from the slow moving ones, using the preferred step size information, and provides interpolation to the fast ones (multi-rate); and
	\item finds the largest communication step size that divides all step sizes suggested by simulation units and uses it throughout the whole co-simulation.
\end{inparaenum}
We argue that the generated orchestrator is fast because all the decisions are made at code generation stage.
The algebraic loops are solved via successive substitution of inputs, storing and restoring the state of the simulation units.

Based on these facts, \citet{Acker2015} can be classified as follows:
\begin{compactenum}
	\item[\textbf{Non-Functional requirements:}] 
		Performance, 
		IP protection, and 
		Configuration reusability;
	\item[\textbf{Simulation unit requirements:}]
		Causal simulation units, 
		Locally available simulation units, 
		Preferred step size information about the solver, 
		Feedthrough I/O causality information, 
		State values, 
		No time constraints, and 
		No rollback support;
	\item[\textbf{Framework requirements:}] 
		Continuous time domain, 
		Multi-rate simulation, 
		Fixed communication step size, 
		Input/Output coupling, 
		Fully implicit strong coupling support, 
		Postmortem visualization of results, 
		Postmortem visualization of results, 
		FMI as the underlying standard, 
		Gauss-seidel communication approach, and 
		Support for three or more simulation units.
\end{compactenum}

With similar reasoning, the FMI standard for co-simulation, version 2.0, can be classified according to the assumptions it makes about the participating simulation units. These are highlighted in \cref{fig:featuremodel_sr}.

\begin{figure}[htb]
\begin{center}
	\includegraphics[scale=0.5]{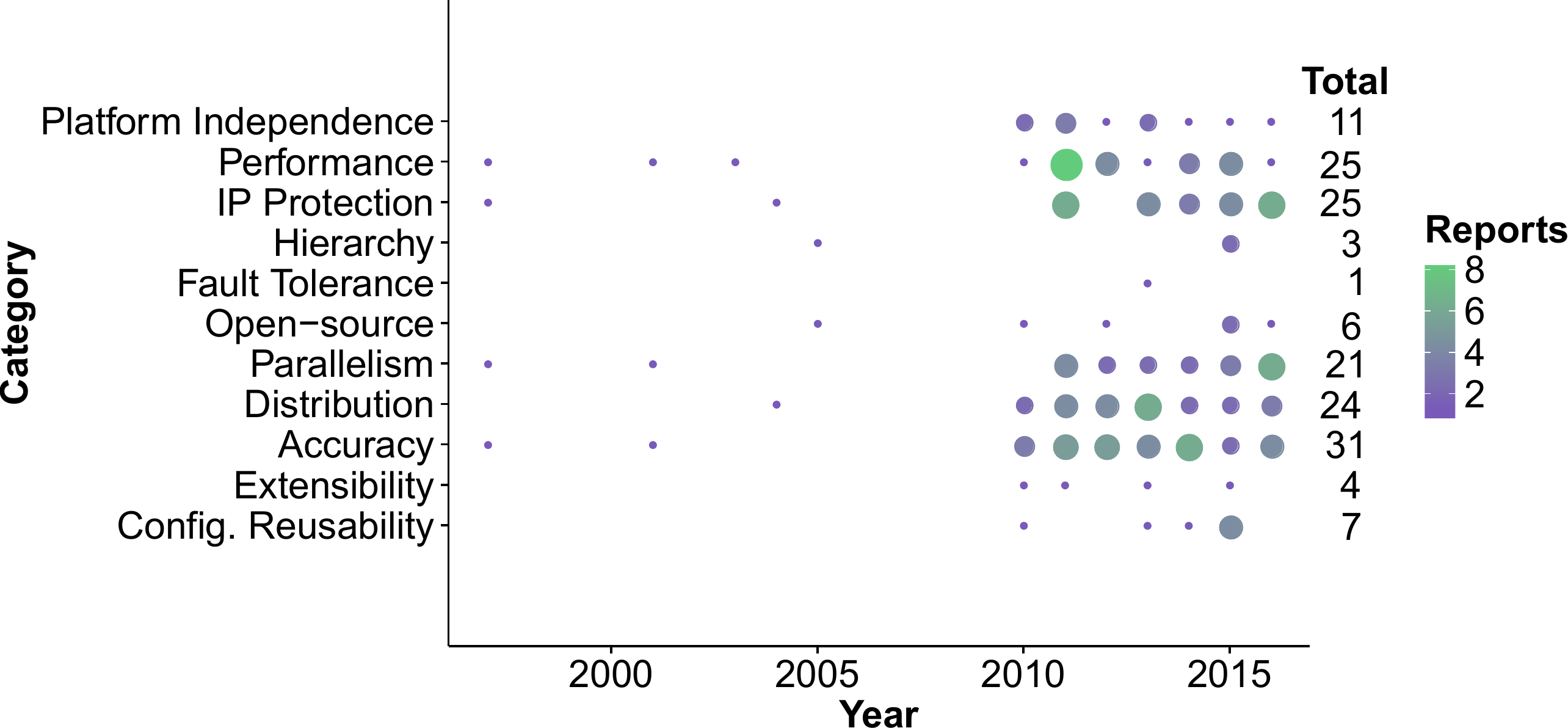}
	\caption{Classification with respect to non-functional requirements.}
	\label{fig:classification_nrf}
\end{center}
\end{figure}

\begin{figure}[htb]
\begin{center}
	\includegraphics[scale=0.5]{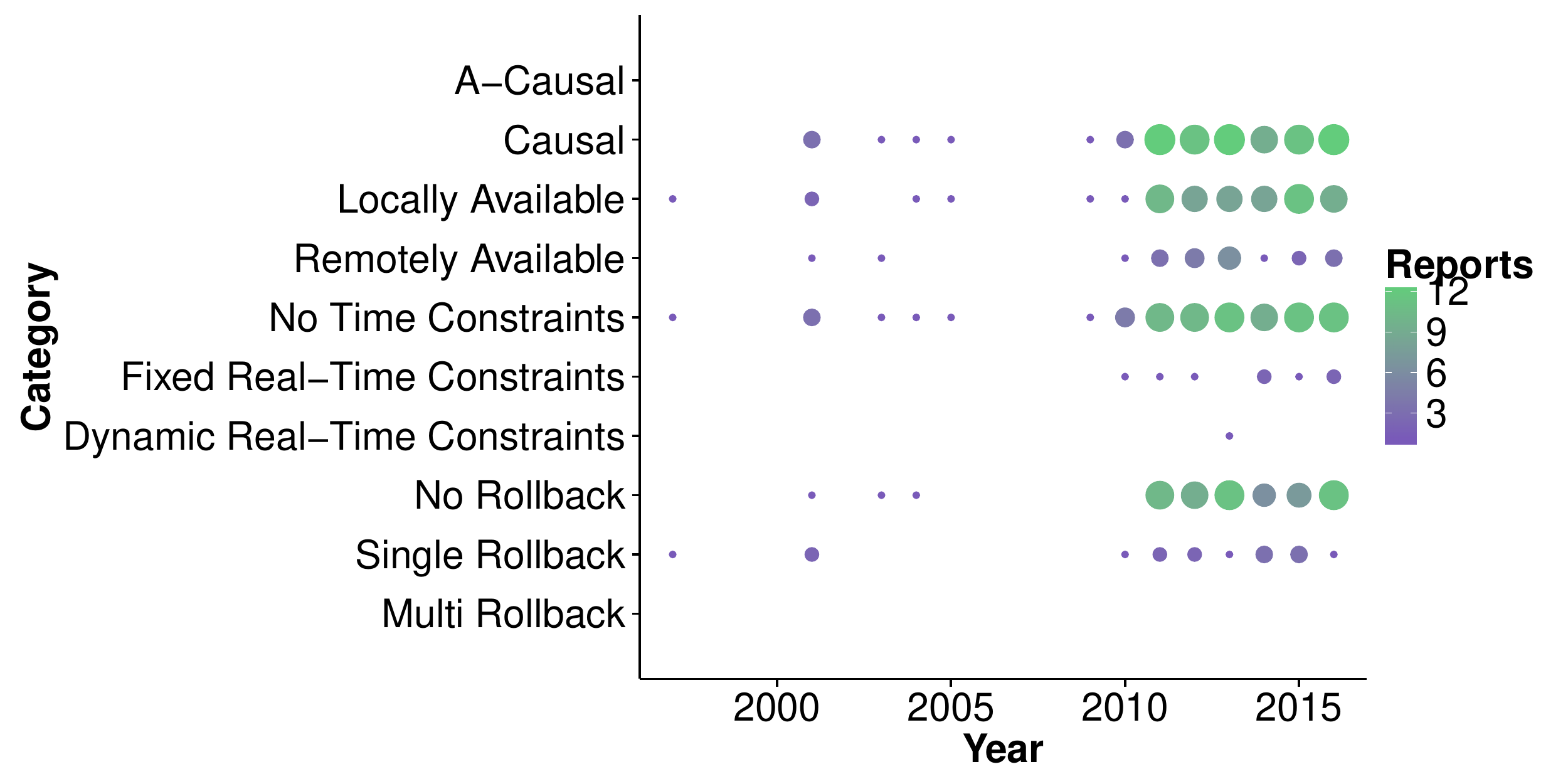}
	\caption{Classification with respect to simulation unit requirements: execution capabilities.}
	\label{fig:classification_sr_capabilities}
\end{center}
\end{figure}

\begin{figure}[htb]
\begin{center}
	\includegraphics[scale=0.5]{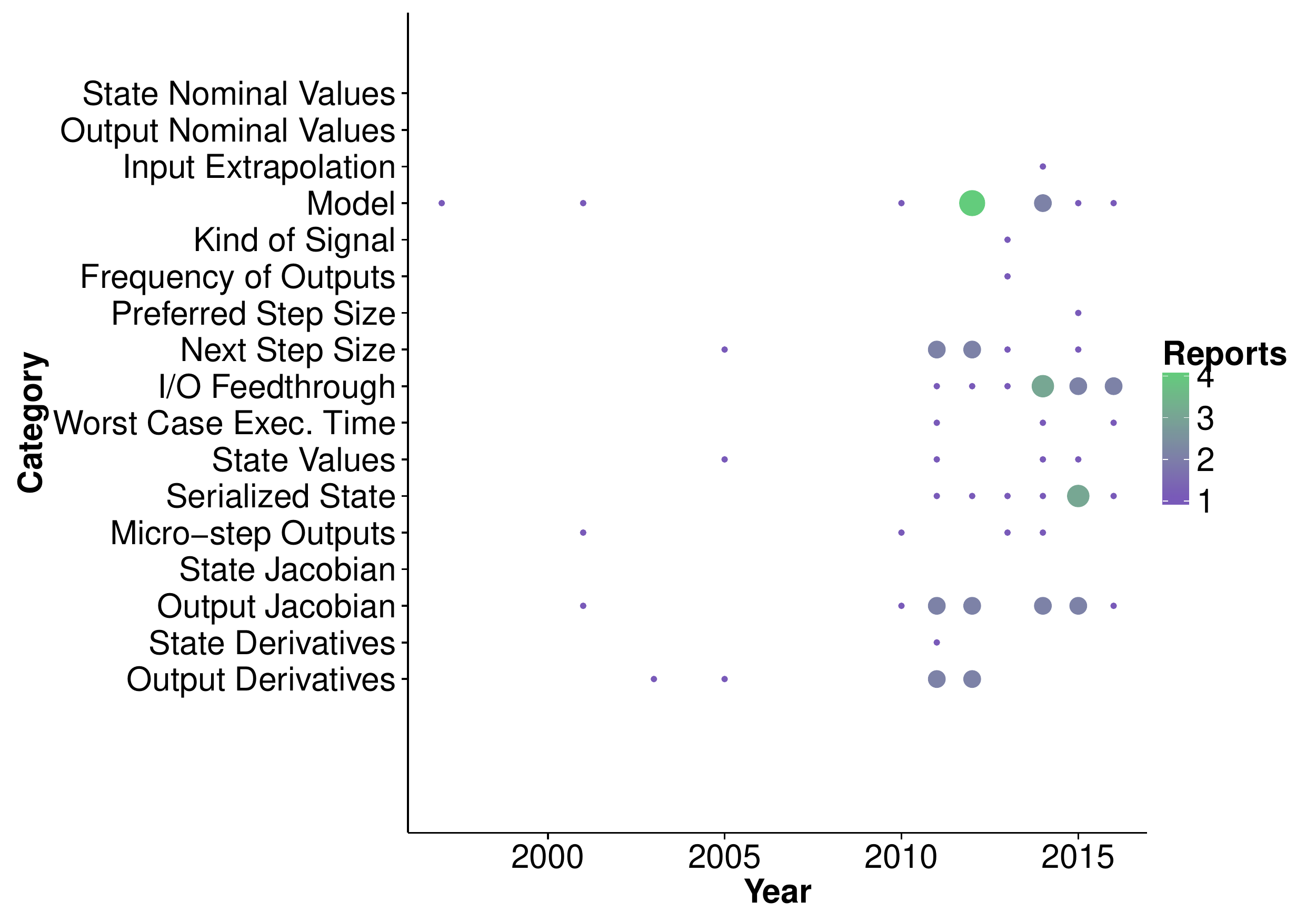}
	\caption{Classification with respect to simulation unit requirements: information exposed.}
	\label{fig:classification_sr_info}
\end{center}
\end{figure}

\begin{figure}[htb]
\begin{center}
	\includegraphics[scale=0.5]{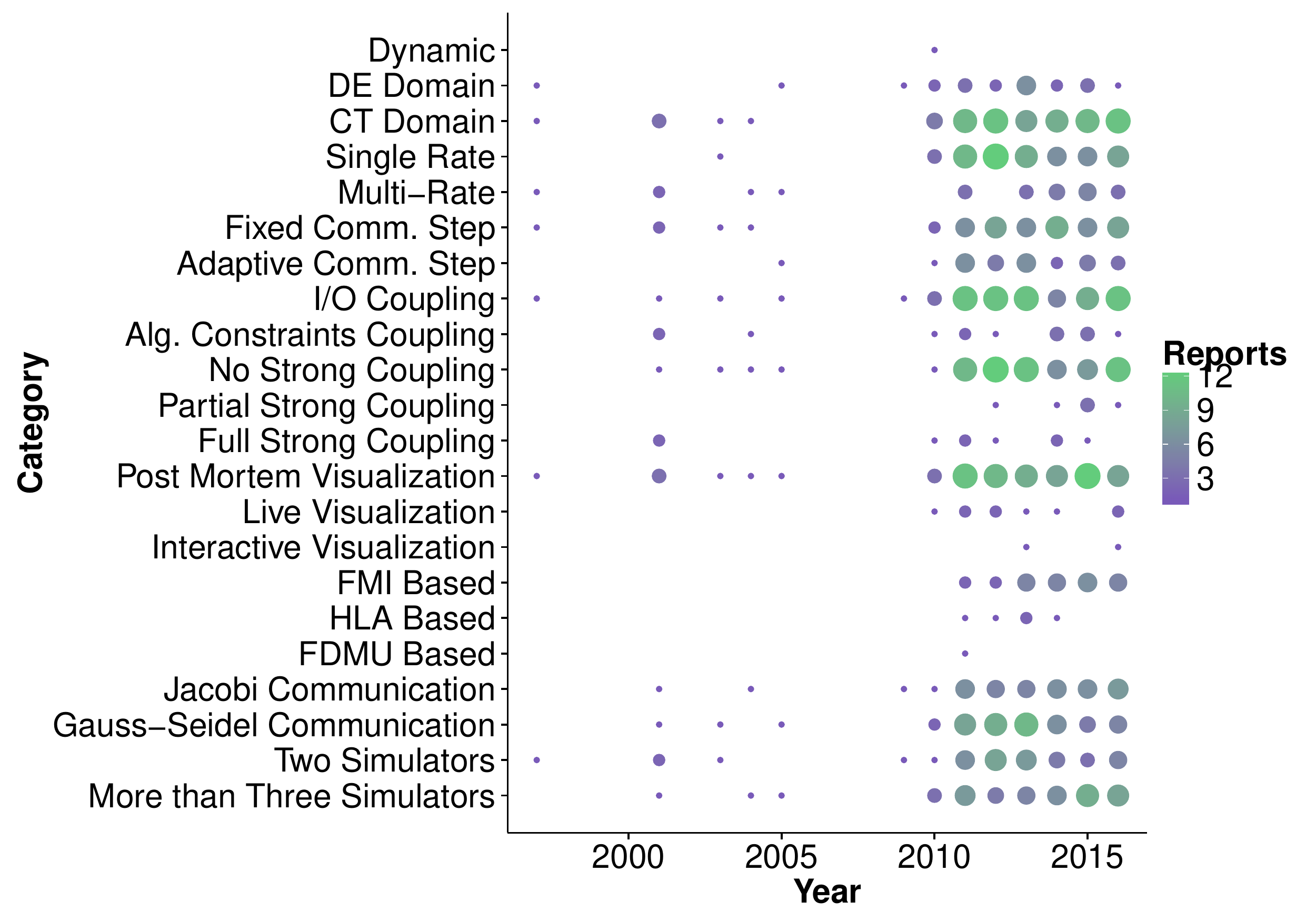}
	\caption{Classification with respect to framework requirements.}
	\label{fig:classification_fr}
\end{center}
\end{figure}

The remaining state of the art is classified in Figures \ref{fig:classification_nrf} -- \ref{fig:classification_fr}.
\ifreport
	The raw data is available online\footurl{http://msdl.cs.mcgill.ca/people/claudio/pub/Gomes2016bClassificationRawData/raw_data.zip}.
\else
	The raw data is available online\footurl{http://msdl.cs.mcgill.ca/people/claudio/pub/Gomes2016bClassificationRawData/raw_data.zip} and a more detailed description of each concept is given in the technical report \cite{Gomes2016b}.
\fi

\subsection{Discussion}

Observing \cref{fig:classification_nrf}, Accuracy is the most observed NFR, with 31 reports, followed by IP protection and Performance.
The least observed NFRs are Fault tolerance, Hierarchy and Extensibility.

Fault tolerance is especially important for long running co-simulations.
One of the industrial partners of the INTO-CPS project has running co-simulations that takes a minimum of two weeks to complete.

We argue that Extensibility (the ability to easily accomodate new features) should be given more importance:
if an heterogeneous set of simulation units participate in the same co-simulation scenario, the combination of capabilities provided (see \cref{fig:featuremodel_sr}) can be huge. 
Thus, the orchestrator can either assume a common homogeneous set of capabilities, which is the most common approach, or can leverage the capabilities provided by each one. 
The later approach can lead to an extremely complex orchestration algorithm.
In any case, extensibility is key to address new semantic adaptations (recall \cref{sec:semantic_adaptation}).

As \cref{fig:classification_sr_info} suggests, we could not find approaches that make use of the nominal values of state and output variables, even though these are capabilities supported in the FMI Standard (see \cref{fig:featuremodel_sr}), and are useful to detect co-simulations that are not valid.
A-causal approaches are important for modularity, as explained in \cref{sec:algebraic_composition}, but these are scarce too.

As for the framework requirements, in \cref{fig:classification_fr}, the least observed features are dynamic structure co-simulation, interactive visualization, multi-rate, algebraic coupling, and partial/full strong coupling support.
This can be explained by the fact that these features depend upon the capabilities of the simulation units, which may not be mature.

\begin{figure}[htb]
    \centering
    \begin{minipage}{0.32\textwidth}
        \centering
        \includegraphics[width=1\textwidth]{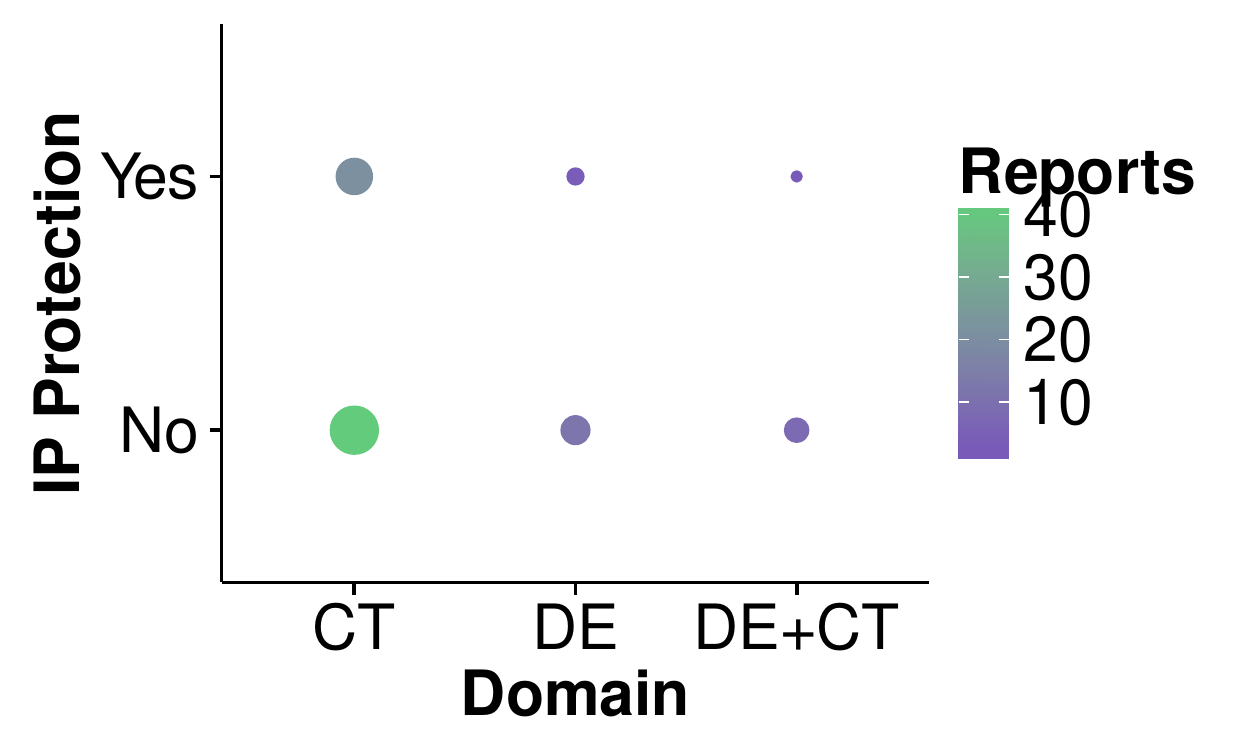}
        \caption{Formalisms vs IP Protection.}
		\label{fig:numForms_vs_ip}
    \end{minipage}\hfill
    \begin{minipage}{0.32\textwidth}
        \centering
        \includegraphics[width=1\textwidth]{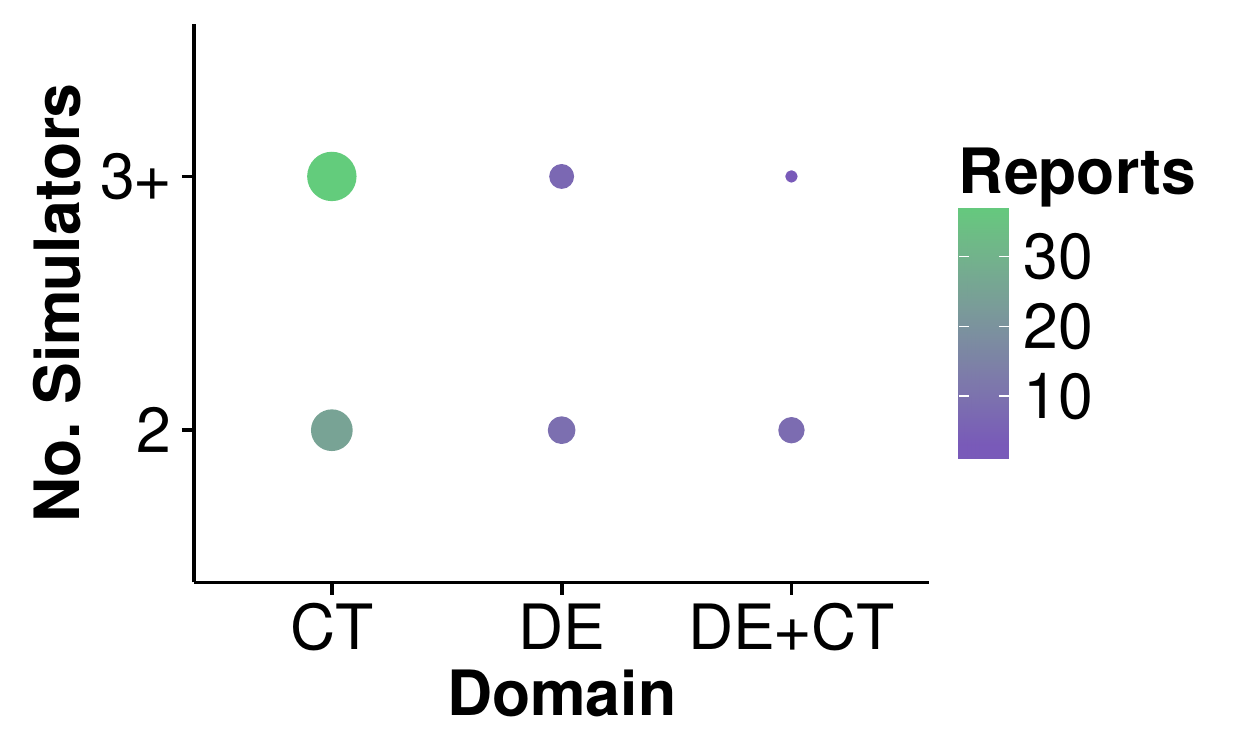}
        \caption{Formalisms vs Simulation units.}
		\label{fig:numForms_vs_numSims}
    \end{minipage}\hfill
    \begin{minipage}{0.32\textwidth}
        \centering
        \includegraphics[width=1\textwidth]{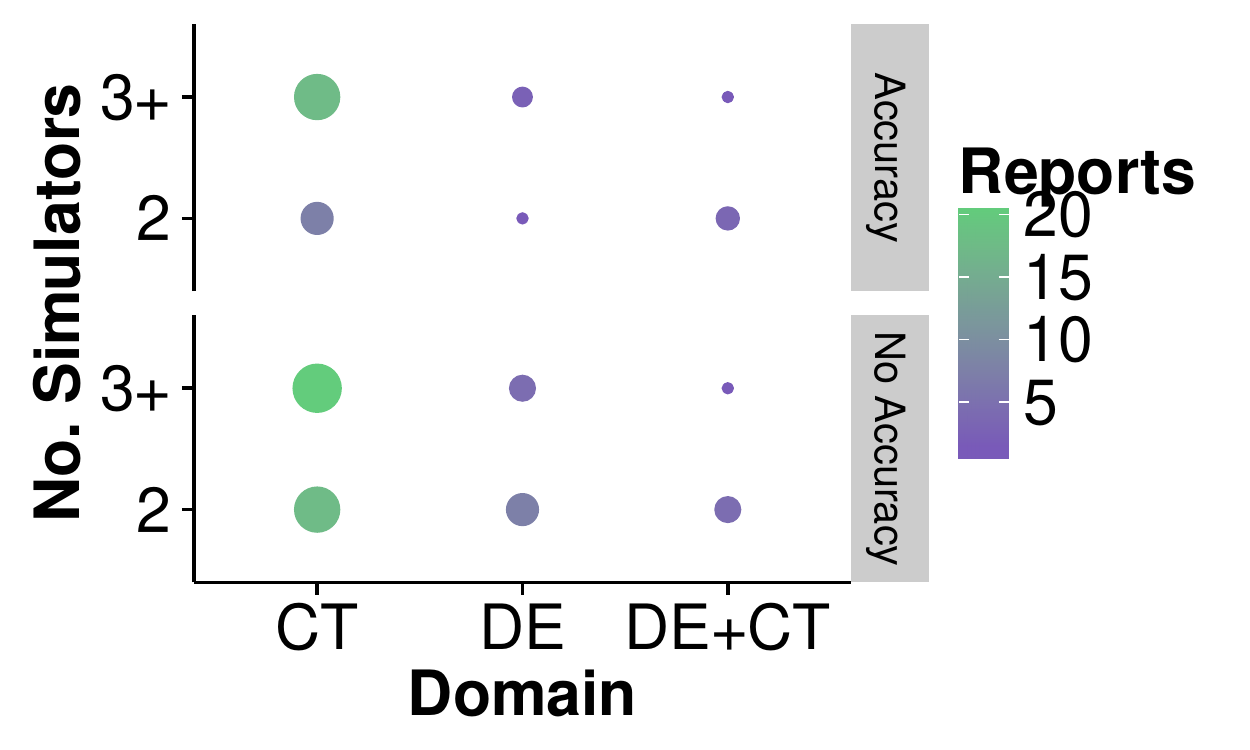}
        \caption{Accuracy vs Formalisms vs Simulation units.}
		\label{fig:accuracy_vs_others}
    \end{minipage}
\end{figure}

Figures \ref{fig:classification_nrf} -- \ref{fig:classification_fr} do not tell the full story because they isolate each feature.
Feature interaction is a common phenomenon, and among many possible interactions, we highlight the accuracy concern, domain of the co-simulation, number of simulation units supported, and IP protection.
As can be seen from \cref{fig:numForms_vs_numSims}, there is only one approach \cite{Kuhr2013} that is both CT and DE based, up to any number of simulation units. 
Note that this does not mean that the work addresses all challenges that were identified in \cref{sec:hybrid_cosimulation}. 
Accommodating the different CT and DE domains means that the approach assumes that the simulation units can behave both as a CT and as a DE unit.

The concern with IP protection is evident in \cref{fig:classification_nrf} but the number of DE and CT based approaches that provide some support for it is small, as shown in \cref{fig:numForms_vs_ip}.
Similarly, as \cref{fig:accuracy_vs_others} suggests, accuracy does not show up a lot in the DE and CT approaches, for more than two simulation units. Accuracy is particularly important in interactions between DE and CT simulation units.

In general, from the observed classification, there is a lack of research into approaches that are both DE and CT based, and that leverage the extra the features from the simulation units.

\section{Concluding Remarks}
\label{sec:conclusion}

As this work shows, there are many interesting challenges to be explored in co-simulation, which will play a key role in enabling the virtual development of complex heterogeneous systems in the decades to come.
The early success can be attributed to a large number of reported applications. However, from the application references covered (see \cref{fig:applications_time_line}), the large majority represent \emph{ad-hoc} couplings between two simulators of two different domains (e.g., a network simulator with a power grid one, or a HVAC simulator with a building envelop one).
This, however, excludes the (potentially many) unreported applications of co-simulation.
As systems become complex, the demand for co-simulation scenarios that are large, hierarchical, heterogeneous, accurate, IP protected, and so on, will increase.

This survey presents an attempt at covering the main challenges in co-simulation.
To tackle such a broad topic, we have covered the two main domains ---continuous time and discrete event based co-simulation--- separately and then surveyed the challenges that arise when the two domains are combined.
A taxonomy is proposed and a classification of the works related to co-simulation in the last five years is carried out using that taxonomy.

From the challenges, we highlight semantic adaptation, modular coupling, stability and accuracy, and finding a standard for hybrid co-simulation as being particular important.

For early system analysis, the adaptations required to combine simulators from different formalisms, even conforming to the same standard, are very difficult to generalize to any co-simulation scenario.
A possible work around this is to allow the system integrator to describe these, as proposed in \cite{Denil2015}.

One of the main conclusions of the classification is that there is lack of research into modular, stable and accurate coupling of simulators in dynamic structure scenarios.
This is where a-causal approaches for co-simulation can play a key role.
The use of bi-directional effort/flow ports can be a solution inspired by Bond-graphs~\cite{Paynter1961}, and there is some work in \cite{Sadjina2016} already in this direction.

Finally, this document is an attempt to summarize, bridge, and enhance the future research in co-simulation, wherever it may lead us to.

\ifreport
	\newcommand{\writeAcksSponsors}{
	This research was partially supported by Flanders Make vzw, the strategic research centre for the manufacturing industry, and a PhD fellowship grant from the Agency for Innovation by Science and Technology in Flanders (IWT). In addition, the work presented here is partially supported by the INTO-CPS project funded by the European Commission's Horizon 2020 programme under grant agreement number 664047. This project is financially supported by the Swedish Foundation for Strategic Research.}
\else
	\newcommand{\writeAcksSponsors}{
	This research was partially supported by \grantsponsor{}{Flanders Make vzw, the strategic research centre for the manufacturing industry}{http://www.flandersmake.be/en}, and a PhD fellowship grant from the \grantsponsor{}{Agency for Innovation by Science and Technology in Flanders (IWT)}{}. In addition, the work presented here is partially supported by the INTO-CPS project funded by the \grantsponsor{H2020}{European Commission's Horizon 2020 programme}{https://ec.europa.eu/programmes/horizon2020/} under grant agreement number \grantnum[http://cordis.europa.eu/project/rcn/194142_en.html]{H2020}{664047}. This project is financially supported by the Swedish Foundation for Strategic Research.
	}
\fi

\newcommand{\writeAcks}{%
The authors wish to thank Yentl Van Tendeloo, for the in depth review of DE based co-simulation, Kenneth Guldbrandt Lausdahl for providing valuable input and discussions throughout the making of this survey, and TWT GmbH for the valuable input on everyday challenges faced by a co-simulation master.
\writeAcksSponsors}

\ifreport
	\section*{Acknowledgment}

		\writeAcks

\else
	
	\begin{acks}
		\writeAcks
	\end{acks}
	
\fi

\clearpage

\ifreport
	\bibliographystyle{plainnat}
	\bibliography{bibliography_exceptions,bibliography_generated}
\else
	\bibliographystyle{ACM-Reference-Format}
	\bibliography{bibliography_exceptions,bibliography_generated}
\fi

\ifreport
	
	\newpage
	
	\appendix
	
	\section{Historical Perspective of Co-simulation}

This section provides an historical perspective that relates the major concepts in co-simulation to the time at which they are recognized in the studied state of the art, summarized in \cref{tab:cosim_history}.

\begin{table}
	\caption{Historical Perspective of Co-simulation.}
	\label{tab:cosim_history}
	\begin{center}
	\ra{1.5}
	\begin{tabular}{@{}p{2cm} p{3cm} p{9cm}@{}}\toprule
	Time & Concept & Description \\ 
	\midrule
	<80s & Single Formalism & The equations describing dynamic behavior are integrated together. \\
	80s & Dynamic Iteration & Large circuits are decomposed into coupled constituent systems and dynamic iteration techniques are used \cite{Gear1984a,Lelarasmee1982,Newton1983,Miekkala1987,Mitra1987,McCalla1987}. \\
	90s & Multi-Formalism & Software and Hardware are developed and simulated concurrently \cite{Gupta1992,Rowson1994,Zivojnovic1996,Chang1996} at multiple levels of abstraction \cite{Hines1997,Hines1997a,ElTahawy1993}. Orchestration methods are explored in \citet{Fey1997,Frey1998,Carothers1999a,Tseng1999}.\\ 
	Late 90s and Early 2000s & Standard Interfaces & Recognized as key for co-simulation \cite{Zwolinski1995,Kuhl1999,Petrellis1998,Hickey2006,Su2005} \\
	2010s & IP Protection, X-in-the-loop, and Scale & Important to enhance industrial applicability of co-simulation \cite{VanderAuweraer2013,AlvarezCabrera2011,Drenth2014,Galtier2015,Awais2013a,Awais2013,Awais2013b}. \\
	\bottomrule
	\end{tabular}
	\end{center}
\end{table}

\subsection{One Formalism and Dynamic Iteration}

Traditionally, the equations describing the dynamical behavior of large circuits were integrated together.
These systems are sparsely coupled, reflecting the connections of the corresponding circuits, and many techniques were developed that take advantage of this structure \cite{McCalla1987}. 

The crucial idea that improved the simulation speed in up to two orders of magnitude is to decompose the large system into a set of coupled constituent systems and integrate them independently.

The decomposition of the circuit implies the definition of inputs and outputs for each of the resulting constituent systems. 
The coupling is then the assignment of outputs to inputs.

For a subsystem $S_i$, we call the subsystems, whose outputs are assigned to any of the inputs of $S_i$, for neighbor subsystems.

The essence of the dynamic iteration approach is to integrate each subsystem independently, for a period of time $T_n \to T_{n+1}$ , using the extrapolated outputs of the neighbor subsystems as inputs \cite{Lelarasmee1982,Newton1983,Miekkala1987,Mitra1987}.

Naturally, the fact that outputs are extrapolated introduces inaccuracy in the solution of the subsystem, so the integration can be repeated for the same period of time, with corrected outputs, until some form of convergence criteria is met \cite{Jackiewicz1996}.
The extrapolated outputs of a subsystem $S_j$ can be corrected by collecting the outputs during the integration of $S_j$.

It is easy to see that this approach only requires communication between constituent systems at times $T_n$ and $T_{n+1}$ and that the integration of each subsystem can be done independently and in parallel \cite{Jackson1991}, using any numerical method with any step size control policy. 
The signals exchanged are functions in the interval $\brackets{ T_n, T_{n+1} }$.

The advantages of independent step size control policy become evident when one observes that many circuits have components that change at different rates. If the whole system were to be simulated, the simulation unit would have to use the smallest time step that ensures sufficient accuracy for the fastest changing component, which would be a huge waste of computational effort for the slow components.
This is the similarity to multi-rate numerical methods \cite{Gear1984a}.

To the best of our knowledge, dynamic iteration techniques and multi-rate numerical are the first to resemble co-simulation.
The coordination software that implements these techniques expect any number of subsystems but assumes that the subsystems are all specific in the same formalism: differential equations.

\subsection{Two Formalisms: Digital and Analog Co-simulation}

Co-simulation, in its modern definition, was applied to enable the virtual development of coupled software and hardware systems \cite{Gupta1992,Rowson1994,Zivojnovic1996,Chang1996}. 
In this application domain, co-simulation decreases the need to build prototype board circuits to validate the composition of the software and the hardware part. It enables software and hardware to be developed and validated concurrently.
To the best of our knowledge, this was one of the first uses of co-simulation in the modern sense.
The co-simulation frameworks developed in this application domain typically assumed two simulation units and two formalisms.

The hardware/software systems quickly became more complex and a new idea was introduced: use multiple models at different levels of abstraction of each subsystem. 
Simulations could be made arbitrarily faster in some intervals by solving the more abstract models, and arbitrarily accurate in other intervals, by solving the more detailed ones \cite{ElTahawy1993,Saleh2013,Martinez2001,Kleckner1984}.
In the particular case of analog-digital co-simulation, each level of abstraction was solved by a different tool: a continuous time tool and a discrete event tool.
The separation into continuous time and discrete event made the abstract synchronization problem and synchronization methods between simulation units in these two domains were developed \cite{Fey1997,Frey1998,Carothers1999a,Tseng1999}. We could call these some of the first master algorithms.

\subsection{Multi-abstraction/Multi-Formalism Co-simulation}

The heterogeneity aspect of co-simulation comes into play at this time: multiple formalisms can be used to describe the same subsystem at multiple levels of abstraction: state machines can describe a rough approximation of the modes, while differential equations can describe the detailed dynamics of the electronic circuit.
Depending on the purpose of the co-simulation, a subsystem and its neighbors can be solved in detail, whereas subsystems that are ``farther away'' can be simulated with higher levels of abstraction \cite{Hines1997,Hines1997a}.
For the domain of Hw/sw co-simulation, RTL and TLM classify the multiple abstraction levels of models \cite{Radetzki2008,Beltrame2007} and switching between these multiple levels of abstraction have been studied in \cite{Karner2010b}.

As the number and heterogeneity of simulation tools to be coupled increases, the need to provide a common interface to couple any number of tools is recognized in \citet{Zwolinski1995,Kuhl1999,Petrellis1998,Hickey2006} and later in \citet{Blochwitz2011}.

In parallel with the previous advancements, co-simulation has also been in use for heterogeneous physical systems, such as automotive \cite{LeMarrec1998,Gu2001,Kubler2000a,Schneider2014}, railway \cite{Dietz2002,Arnold2003} and HVAC \cite{Gu2004,Trcka2007}, to name just a few. 
The common motivation is the fact that co-simulation enables specialized simulation units to cooperatively simulated the system, with huge savings in time and cost, when compared to a monolithic modeling approach.

\subsection{Black-box Co-simulation}

Later, distributed and concurrent development processes, enabled by co-simulation, are studied and IP protection is identified as a desired characteristic \cite{VanderAuweraer2013,AlvarezCabrera2011} to enable suppliers and integrators to exchange co-simulation units without having to disclose sensitive information and avoiding \emph{vendor lock-in} contracts.

\subsection{Real-time Co-simulation}

Furthermore, co-simulation is used at every stage of the development process, from early system validation, to X-in-the-Loop co-simulation, bringing hard real-time constraints to the set of challenges \cite{Drenth2014}.

\subsection{Many simulation units: Large Scale Co-simulation}

More recently, with the acknowledgment that there is a need to be able to simulate even larger systems of systems, scale and distribution become inherent challenges in co-simulation \cite{Galtier2015,Awais2013a,Awais2013,Awais2013b}.
\todocasper{Can this be couped to 6.4? Go over articles and check}

	\section{State of the Art in Co-simulation Frameworks}
\label{cha:state_or_art}

This section provides the detailed classification of each reference.

\begin{Reference}{Co-Simulation of Distributed Engine Control System and Network Model using FMI \& SCNSL}{Pedersen2015}
This work describes a co-simulation master in the context of the maritime industry.

\catref{sr:causality:causal}
\catref{sr:rel_time:analytic}
\catref{sr:rollback:none}
\catref{sr:availability:local}
\catref{fr:results_visualization:post_mortem}
\catref{fr:alg_loop:explicit}
\catref{fr:sim_step_size:fixed}
\catref{fr:sim_rate:single}
\catref{fr:domain:ct}
\catref{fr:coupling_model:io_assignments}
\catref{fr:standard:fmi}
\catref{fr:communication_model:jacobi}
\catref{fr:num_sim:three_more}

\end{Reference}

\begin{Reference}{Power system and communication network co-simulation for smart grid applications}{Lin2011}
This work describes a co-simulation between power system and network simulator.

\catref{sr:info:predict_step_sizes}
\catref{sr:causality:causal}
\catref{sr:rel_time:analytic}
\catref{sr:rollback:none}
\catref{sr:availability:local}
\catref{fr:results_visualization:post_mortem}
\catref{fr:alg_loop:explicit}
\catref{fr:sim_rate:single}
\catref{fr:coupling_model:io_assignments}
\catref{fr:num_sim:two}
\catref{fr:domain:de}
\catref{fr:sim_step_size:variable}
\catref{fr:communication_model:gauss_seidel}

\end{Reference}

\begin{Reference}{Towards a Comprehensive Framework for Co- Simulation of Dynamic Models With an Emphasis on Time Stepping}{Hoepfer2011}
This work describes a co-simulation approach that finds an appropriate co-simulation step size.

\catref{nfr:performance}
\catref{nfr:accuracy}
\catref{nfr:ip_protection}
\catref{sr:availability:local}
\catref{sr:info:derivatives:out}
\catref{sr:info:derivatives:state}
\catref{sr:info:statevars}
\catref{sr:causality:causal}
\catref{sr:rollback:none}
\catref{sr:rel_time:analytic}
\catref{fr:num_sim:three_more}
\catref{fr:domain:ct}
\catref{fr:sim_rate:single}
\catref{fr:sim_step_size:variable}
\catref{fr:communication_model:jacobi}
\catref{fr:communication_model:gauss_seidel}
\catref{fr:alg_loop:explicit}
\catref{fr:results_visualization:post_mortem}

\end{Reference}

\begin{Reference}{Methods for real-time simulation of Cyber-Physical Systems: application to automotive domain}{Faure2011}
This work addresses co-simulation with real-time simulators.

\catref{nfr:performance}
\catref{nfr:parallelism}
\catref{sr:availability:local}
\catref{sr:rel_time:fixed_real_scaled_time_simulation}
\catref{sr:info:wcet}
\catref{sr:rollback:none}
\catref{fr:results_visualization:post_mortem}
\catref{fr:num_sim:three_more}
\catref{fr:coupling_model:io_assignments}
\catref{fr:domain:ct}
\catref{fr:communication_model:jacobi}
\catref{fr:alg_loop:explicit}
\catref{fr:sim_rate:single}
\catref{fr:sim_step_size:fixed}

\end{Reference}

\begin{Reference}{Simulation of multibody systems with the use of coupling techniques: a case study}{Tomulik2011}

This work discusses a co-simulation method for couplings with algebraic constraints.
One of the results is that this kind of coupling should be done with many derivatives of the coupling variables.

\catref{nfr:accuracy}
\catref{sr:rollback:single}
\catref{sr:availability:local}
\catref{sr:rel_time:analytic}
\catref{sr:causality:causal}
\catref{sr:info:derivatives:out}
\catref{sr:info:jacobian:out}
\catref{fr:results_visualization:post_mortem}
\catref{fr:communication_model:jacobi}
\catref{fr:alg_loop:implicit}
\catref{fr:coupling_model:algebraic_constraints}
\catref{fr:domain:ct}
\catref{fr:num_sim:three_more}
\catref{fr:sim_rate:single}
\catref{fr:sim_step_size:fixed}

\end{Reference}

\begin{Reference}{Combining Advantages of Specialized Simulation Tools and Modelica Models using Functional Mock-up Interface (FMI)}{Sun2011}

This work describes the application of co-simulation to the power production domain.

\catref{nfr:ip_protection}
\catref{nfr:performance}
\catref{sr:causality:causal}
\catref{sr:rel_time:analytic}
\catref{sr:rollback:none}
\catref{sr:availability:local}
\catref{fr:results_visualization:post_mortem}
\catref{fr:sim_rate:single}
\catref{fr:coupling_model:io_assignments}
\catref{fr:standard:fmi}
\catref{fr:domain:ct}
\catref{fr:communication_model:gauss_seidel}
\catref{fr:num_sim:two}
\catref{fr:alg_loop:explicit}
\catref{fr:sim_step_size:variable}

\end{Reference}

\begin{Reference}{Master for Co-Simulation Using FMI}{Bastian2011a}
This work describes a co-simulation approach.

\catref{nfr:ip_protection}
\catref{nfr:platform_independence}
\catref{nfr:parallelism}
\catref{sr:causality:causal}
\catref{sr:rollback:none}
\catref{sr:info:stateserial}
\catref{sr:info:jacobian:out}
\catref{sr:availability:local}
\catref{fr:results_visualization:post_mortem}
\catref{fr:coupling_model:io_assignments}
\catref{fr:sim_step_size:fixed}
\catref{fr:sim_rate:single}
\catref{fr:alg_loop:implicit}
\catref{fr:domain:ct}
\catref{fr:num_sim:three_more}
\catref{fr:standard:fmi}
\catref{fr:communication_model:jacobi}
\catref{fr:communication_model:gauss_seidel}

\end{Reference}

\begin{Reference}{Parallel Co-Simulation for Mechatronic Systems}{Friedrich2011}
This work describes a co-simulation framework based on the Jacobi iteration scheme.

\catref{nfr:parallelism}
\catref{nfr:performance}
\catref{nfr:distribution}
\catref{nfr:ip_protection}
\catref{sr:availability:remote}
\catref{sr:causality:causal}
\catref{sr:rel_time:analytic}
\catref{sr:rollback:none}
\catref{fr:results_visualization:post_mortem}
\catref{fr:alg_loop:explicit}
\catref{fr:domain:ct}
\catref{fr:coupling_model:io_assignments}
\catref{fr:coupling_model:algebraic_constraints}
\catref{fr:num_sim:three_more}
\catref{fr:communication_model:jacobi}
\catref{fr:sim_rate:single}
\catref{fr:sim_step_size:fixed}

\end{Reference}

\begin{Reference}{On the effect of multirate co-simulation techniques in the efficiency and accuracy of multibody system dynamics}{Gonzalez2011}

This work deals with multi-rate co-simulation.
Essentially, one of the simulators (the fast one) drives the simulation, while the slow one provides extrapolated inputs, to avoid excessive computation.

\catref{nfr:accuracy}
\catref{nfr:performance}
\catref{sr:causality:causal}
\catref{sr:rollback:none}
\catref{sr:availability:local}
\catref{sr:rel_time:analytic}
\catref{fr:results_visualization:post_mortem}
\catref{fr:alg_loop:explicit}
\catref{fr:communication_model:gauss_seidel}
\catref{fr:coupling_model:io_assignments}
\catref{fr:num_sim:two}
\catref{fr:domain:ct}
\catref{fr:sim_rate:multi}
\catref{fr:sim_step_size:fixed}

\end{Reference}

\begin{Reference}{Designing power system simulators for the smart grid: Combining controls, communications, and electro-mechanical dynamics}{Nutaro2011}
This work describes a tool that is formed by the coupling of a DEVS simulator with some other modules that wrap CT as DEVS simulators.

\catref{nfr:distribution}
\catref{nfr:accuracy}
\catref{sr:causality:causal}
\catref{sr:rollback:single}
\catref{sr:rel_time:analytic}
\catref{sr:availability:local}
\catref{fr:alg_loop:explicit}
\catref{fr:domain:de}
\catref{fr:communication_model:gauss_seidel}
\catref{fr:num_sim:two}
\catref{fr:sim_rate:single}
\catref{fr:sim_step_size:variable}
\catref{fr:coupling_model:io_assignments}
\catref{fr:results_visualization:post_mortem}

\end{Reference}

\begin{Reference}{Asynchronous method for the coupled simulation of mechatronic systems}{Busch2012}

This work describes co-simulation approaches between two simulation tools.
The main contribution is a semi-implicit method that applies a correction based on the jacobian of the sub-system's coupling variables.

\catref{nfr:accuracy}
\catref{nfr:distribution}
\catref{sr:causality:causal}
\catref{sr:rollback:single}
\catref{sr:availability:remote}
\catref{sr:info:jacobian:out}
\catref{sr:rel_time:analytic}
\catref{fr:sim_rate:single}
\catref{fr:sim_step_size:fixed}
\catref{fr:results_visualization:post_mortem}
\catref{fr:communication_model:gauss_seidel}
\catref{fr:alg_loop:semi_implicit}
\catref{fr:alg_loop:explicit}
\catref{fr:alg_loop:implicit}
\catref{fr:coupling_model:io_assignments}
\catref{fr:domain:ct}
\catref{fr:num_sim:two}

\end{Reference}

\begin{Reference}{Generating functional mockup units from software specifications}{Pohlmann2012}
This work describes an application of co-simulation to robotics.
\end{Reference}

\begin{Reference}{Convergence Study of Explicit Co-Simulation Approaches with Respect to Subsystem Solver Settings}{Schmoll2012}
\claudio{Missing summary}
\todocasper{missing summary}
This paper describes global error analysis for co-simulation, that takes into account sub-system solvers (instead of analytical solvers, as more commonly done).

\catref{nfr:accuracy}
\catref{sr:rollback:none}
\catref{sr:info:full_model}
\catref{sr:causality:causal}
\catref{sr:rel_time:analytic}
\catref{sr:availability:local}
\catref{fr:alg_loop:explicit}
\catref{fr:domain:ct}
\catref{fr:num_sim:two}
\catref{fr:coupling_model:io_assignments}
\catref{fr:communication_model:jacobi}
\catref{fr:sim_rate:single}
\catref{fr:sim_step_size:fixed}
\catref{fr:results_visualization:post_mortem}

\end{Reference}

\begin{Reference}{Hybrid systems modelling and simulation in DESTECS: a co-simulation approach}{Ni2012}

This work present a coupling of the tools Crescendo and 20-sim.

\catref{sr:causality:causal}
\catref{sr:rel_time:analytic}
\catref{sr:rollback:none}
\catref{sr:availability:local}
\catref{fr:communication_model:gauss_seidel}
\catref{fr:results_visualization:post_mortem}
\catref{fr:coupling_model:io_assignments}
\catref{fr:alg_loop:explicit}
\catref{fr:domain:ct}
\catref{fr:sim_rate:single}
\catref{fr:sim_step_size:fixed}
\catref{fr:num_sim:two}

\end{Reference}

\begin{Reference}{Matlab/SystemC for the New Co-Simulation Environment by JPEG Algorithm}{Hassairi2012}
This work introduces guidelines for the implementation of co-simulation between Matlab and SystemC.
The case study is the JPEG Algorithm.

\catref{sr:info:full_model}
\catref{sr:causality:causal}
\catref{sr:rel_time:analytic}
\catref{sr:availability:local}
\catref{fr:domain:ct}
\catref{fr:coupling_model:io_assignments}
\catref{fr:sim_step_size:fixed}
\catref{fr:sim_rate:single}
\catref{fr:alg_loop:explicit}
\catref{fr:communication_model:gauss_seidel}
\catref{fr:results_visualization:live}
\catref{fr:num_sim:two}

\end{Reference}

\begin{Reference}{Stabilized overlapping modular time integration of coupled differential-algebraic equations}{Schierz2012}

This work discusses co-simulation techniques for simulators coupled via algebraic constraints.

\catref{nfr:accuracy}
\catref{sr:availability:local}
\catref{sr:rel_time:analytic}
\catref{sr:causality:causal}
\catref{sr:rollback:none}
\catref{sr:info:full_model}
\catref{sr:info:jacobian:out}
\catref{fr:results_visualization:post_mortem}
\catref{fr:sim_rate:single}
\catref{fr:num_sim:three_more}
\catref{fr:alg_loop:explicit}
\catref{fr:sim_step_size:fixed}
\catref{fr:domain:ct}
\catref{fr:coupling_model:algebraic_constraints}
\catref{fr:communication_model:gauss_seidel}
\catref{fr:communication_model:jacobi}

\end{Reference}

\begin{Reference}{A Modular Technique for Automotive System Simulation}{Gunther2012}

This work describes the MDPCosim framework.

\catref{nfr:performance} The decomposition of the system for co-simulation is done for performance reasons.
\catref{nfr:parallelism}
\catref{nfr:accuracy}
\catref{sr:availability:local} IPC communication is used.
\catref{sr:causality:causal}
\catref{sr:info:derivatives:out}
\catref{sr:rollback:none}
\catref{sr:info:predict_step_sizes}
\catref{fr:results_visualization:post_mortem}
\catref{fr:alg_loop:explicit}
\catref{fr:coupling_model:io_assignments}
\catref{fr:domain:ct}
\catref{fr:communication_model:jacobi}
\catref{fr:sim_step_size:variable} The step size control approach is based on looking at the derivatives.
\catref{fr:sim_rate:single}
\catref{fr:num_sim:three_more}

\end{Reference}

\begin{Reference}{A SystemC/Matlab co-simulation tool for networked control systems}{Quaglia2012}
Work describing another tool coupling.

\catref{nfr:distribution}
\catref{sr:causality:causal}
\catref{sr:rel_time:analytic}
\catref{sr:rollback:none}
\catref{sr:availability:remote}
\catref{fr:results_visualization:post_mortem}
\catref{fr:alg_loop:explicit}
\catref{fr:coupling_model:io_assignments}
\catref{fr:domain:ct}
\catref{fr:num_sim:two}
\catref{fr:sim_rate:single}
\catref{fr:sim_step_size:fixed}
\catref{fr:communication_model:gauss_seidel}

\end{Reference}

\begin{Reference}{A comprehensive co-simulation platform for cyber-physical systems}{Al-Hammouri2012}
The work describes the integration of two tools: Modelica, and NS-2.

\catref{sr:causality:causal}
\catref{sr:rel_time:analytic}
\catref{sr:availability:local} Communication is done over named pipes.
\catref{sr:rollback:none}
\catref{fr:results_visualization:post_mortem}
\catref{fr:num_sim:two}
\catref{fr:coupling_model:io_assignments}
\catref{fr:alg_loop:explicit}
\catref{fr:domain:ct}
\catref{fr:communication_model:gauss_seidel}
\catref{fr:domain:de}
\catref{fr:sim_rate:single}
\catref{fr:sim_step_size:variable}

\end{Reference}

\begin{Reference}{NCSWT: An integrated modeling and simulation tool for networked control systems}{Eyisi2012}

This work describes the coupling of two tools: Matlab and NS-2.
The coupling is done through HLA standard.
The preliminary version of the tool is described in \cite{Riley2011}.

\catref{nfr:platform_independence}
\catref{nfr:performance}
\catref{nfr:distribution}
\catref{sr:info:predict_step_sizes}
\catref{sr:causality:causal}
\catref{sr:rel_time:analytic}
\catref{sr:rollback:none}
\catref{sr:availability:remote}
\catref{fr:alg_loop:explicit}
\catref{fr:num_sim:two}
\catref{fr:coupling_model:io_assignments}
\catref{fr:domain:de}
\catref{fr:standard:hla}
\catref{fr:communication_model:gauss_seidel}
\catref{fr:sim_rate:single}
\catref{fr:sim_step_size:variable}
\catref{fr:results_visualization:live}

\end{Reference}

\begin{Reference}{Networked Control System Wind Tunnel (NCSWT): An Evaluation Tool for Networked Multi-agent Systems}{Riley2011}

This work describes the coupling of two tools: Matlab and NS-2.
The coupling is done through HLA standard.

\catref{nfr:platform_independence}
\catref{nfr:performance}
\catref{nfr:distribution}
\catref{sr:info:predict_step_sizes}
\catref{sr:causality:causal}
\catref{sr:rel_time:analytic}
\catref{sr:rollback:none}
\catref{sr:availability:remote}
\catref{fr:alg_loop:explicit}
\catref{fr:num_sim:two}
\catref{fr:coupling_model:io_assignments}
\catref{fr:domain:de}
\catref{fr:standard:hla}
\catref{fr:communication_model:gauss_seidel}
\catref{fr:sim_rate:single}
\catref{fr:sim_step_size:variable}
\catref{fr:results_visualization:live}

\end{Reference}

\begin{Reference}{A Framework for Co-simulation of AI Tools with Power Systems Analysis Software}{Roche2012}
This work describes a co-simulation between two tools in the power grid domain with matlab running the co-simulation.

\catref{nfr:distribution}
\catref{nfr:open_source}
\catref{sr:causality:causal}
\catref{sr:rel_time:analytic}
\catref{sr:rollback:none}
\catref{sr:availability:remote}
\catref{fr:coupling_model:io_assignments}
\catref{fr:domain:ct}
\catref{fr:communication_model:gauss_seidel}
\catref{fr:alg_loop:explicit}
\catref{fr:results_visualization:post_mortem}
\catref{fr:num_sim:two}
\catref{fr:sim_rate:single}
\catref{fr:sim_step_size:fixed}

\end{Reference}

\begin{Reference}{Collaborative Modelling and Co-simulation in the Development of Dependable Embedded Systems}{Fitzgerald2010}
This work describes the coupling between two tools: Overture and 20-sim.

\catref{nfr:accuracy}
\catref{nfr:distribution}
\catref{nfr:platform_independence}
\catref{sr:availability:remote}
\catref{sr:causality:causal}
\catref{sr:rel_time:analytic}
\catref{fr:sim_step_size:variable}
\catref{fr:sim_rate:single}
\catref{fr:domain:de}
\catref{fr:domain:ct}
\catref{fr:num_sim:two}
\catref{fr:coupling_model:io_assignments}
\catref{fr:alg_loop:explicit}
\catref{fr:results_visualization:live}
\catref{fr:communication_model:gauss_seidel}

\end{Reference}

\begin{Reference}{A formal approach to collaborative modelling and co-simulation for embedded systems}{Fitzgerald2013}
This work describes the coupling between two tools: Overture and 20-sim; already described in \cite{Fitzgerald2010}.

\catref{nfr:accuracy}
\catref{nfr:distribution}
\catref{nfr:platform_independence}
\catref{sr:availability:remote}
\catref{sr:causality:causal}
\catref{sr:rel_time:analytic}
\catref{fr:sim_step_size:variable}
\catref{fr:sim_rate:single}
\catref{fr:domain:de}
\catref{fr:domain:ct}
\catref{fr:num_sim:two}
\catref{fr:coupling_model:io_assignments}
\catref{fr:alg_loop:explicit}
\catref{fr:results_visualization:live}
\catref{fr:communication_model:gauss_seidel}

\end{Reference}

\begin{Reference}{RoboNetSim: An integrated framework for multi-robot and network simulation}{Kudelski2013}
This work describes the integration of three simulators (ARGoS, NS-2 and NS-3) that can be used in co-simulation scenarios with two simulators.

\catref{nfr:distribution}
\catref{nfr:platform_independence}
\catref{sr:rollback:none}
\catref{sr:causality:causal}
\catref{sr:rel_time:analytic}
\catref{sr:availability:remote}
\catref{fr:results_visualization:post_mortem}
\catref{fr:alg_loop:explicit}
\catref{fr:communication_model:jacobi}
\catref{fr:domain:de}
\catref{fr:num_sim:two}
\catref{fr:sim_rate:single}
\catref{fr:sim_step_size:fixed}
\catref{fr:coupling_model:io_assignments}

\end{Reference}

\begin{Reference}{Determinate Composition of FMUs for Co-simulation}{Broman2013}
This work describes a master algorithm that ensures a determinate execution.

\catref{nfr:ip_protection}
\catref{sr:availability:local}
\catref{sr:rollback:none}
\catref{sr:rel_time:analytic}
\catref{sr:causality:causal}
\catref{sr:info:causality:feedthrough}
\catref{sr:info:predict_step_sizes}
\catref{sr:info:stateserial}
\catref{fr:coupling_model:io_assignments}
\catref{fr:standard:fmi}
\catref{fr:sim_step_size:variable}
\catref{fr:domain:ct}
\catref{fr:communication_model:jacobi}
\catref{fr:num_sim:three_more}
\catref{fr:sim_rate:single}
\catref{fr:alg_loop:explicit}
\catref{fr:results_visualization:post_mortem}

\end{Reference}

\begin{Reference}{Guidelines for the Application of a Coupling Method for Non-iterative Co-simulation}{Benedikt2013}

This work describes a co-simulation approach where energy information about the signals is used, and those errors are compensated in a corrector step.

\catref{nfr:accuracy}
\catref{nfr:performance}
\catref{sr:rollback:none}
\catref{sr:rel_time:analytic}
\catref{sr:causality:causal}
\catref{sr:info:record_outputs}
\catref{sr:availability:local}
\catref{fr:results_visualization:post_mortem}
\catref{fr:alg_loop:explicit}
\catref{fr:num_sim:two}
\catref{fr:coupling_model:io_assignments}
\catref{fr:communication_model:gauss_seidel}
\catref{fr:domain:ct}
\catref{fr:sim_rate:single}
\catref{fr:sim_step_size:fixed}

\end{Reference}

\begin{Reference}{Macro-step-size selection and monitoring of the coupoling error for weak coupled subsystems in the frequency-domain}{Benedikt2013b}

The work describes a method for finding appropriate communication step sizes in co-simulations between LTI systems.
Essentially, it provides rules of thumb to chose a communication step size based on the maximum instantaneous frequency of components.

\catref{nfr:accuracy}
\catref{sr:availability:local}
\catref{sr:rollback:none}
\catref{sr:rel_time:analytic}
\catref{sr:causality:causal}
\catref{sr:info:frequency_outputs}
\catref{fr:results_visualization:post_mortem}
\catref{fr:sim_rate:single}
\catref{fr:coupling_model:io_assignments}
\catref{fr:alg_loop:explicit}
\catref{fr:communication_model:gauss_seidel}
\catref{fr:domain:ct}
\catref{fr:sim_step_size:fixed}
\catref{fr:num_sim:two}

\end{Reference}

\begin{Reference}{Communication simulations for power system applications}{Fuller2013}
This work describes a co-simulation between two co-simulation tools (ns-3 and GridLAB-D) for smart grid development.

\catref{nfr:scalability}
\catref{nfr:faulttolerance}
\catref{nfr:ip_protection}
\catref{nfr:distribution} The tools keeps track of messages in transit.
\catref{sr:causality:causal}
\catref{sr:rel_time:analytic}
\catref{sr:availability:remote}
\catref{sr:rollback:none}
\catref{fr:coupling_model:io_assignments}
\catref{fr:alg_loop:explicit}
\catref{fr:domain:de}
\catref{fr:num_sim:two}
\catref{fr:sim_rate:single}
\catref{fr:sim_step_size:variable}
\catref{fr:results_visualization:post_mortem}
\catref{fr:communication_model:gauss_seidel}

\end{Reference}

\begin{Reference}{A model-driven co-simulation environment for heterogeneous systems}{Bombino2013}

This work describes the coupling between two simulation tools.

\catref{nfr:distribution}
\catref{sr:rollback:none}
\catref{sr:causality:causal}
\catref{sr:rollback:single}
\catref{sr:rel_time:dy_real_scaled_time_simulation}
\catref{sr:availability:remote}
\catref{fr:alg_loop:explicit}
\catref{fr:coupling_model:io_assignments}
\catref{fr:communication_model:gauss_seidel}
\catref{fr:domain:ct}
\catref{fr:num_sim:two}
\catref{fr:results_visualization:interactive_live}
\catref{fr:sim_rate:single}
\catref{fr:sim_step_size:fixed}

\end{Reference}

\begin{Reference}{HybridSim: A Modeling and Co-simulation Toolchain for Cyber-physical Systems}{Wang2013}
The approach described in this reference allows to arrange and process co-simulation units, Modelica models and TinyOS applications.
SysML is used to configure the co-simulation master.
The coordination of simulators is done through the FMI standard.

\catref{nfr:ip_protection}
\catref{nfr:config_reusability}
\catref{sr:causality:causal}
\catref{sr:rel_time:analytic}
\catref{sr:rollback:none}
\catref{sr:availability:local}
\catref{fr:coupling_model:io_assignments}
\catref{fr:alg_loop:explicit}
\catref{fr:communication_model:gauss_seidel}
\catref{fr:num_sim:two}
\catref{fr:domain:ct}
\catref{fr:standard:fmi}
\catref{fr:sim_rate:single}
\catref{fr:sim_step_size:variable}
\catref{fr:results_visualization:post_mortem}

\end{Reference}

\begin{Reference}{An Investigation on Loose Coupling Co-Simulation with the BCVTB}{Hafner2013}

This work discusses the consistency and stability of the Jacobi and Gauss-Seidel co-simulation methods.
Later, it presents a case study in HVAC systems.

\catref{nfr:accuracy}
\catref{sr:causality:causal}
\catref{sr:rel_time:analytic}
\catref{sr:rollback:none}
\catref{sr:availability:local}
\catref{fr:coupling_model:io_assignments}
\catref{fr:results_visualization:post_mortem}
\catref{fr:communication_model:jacobi}
\catref{fr:communication_model:gauss_seidel}
\catref{fr:alg_loop:explicit}
\catref{fr:domain:ct}
\catref{fr:num_sim:three_more}
\catref{fr:sim_rate:single}
\catref{fr:sim_step_size:fixed}

\end{Reference}

\begin{Reference}{Co-simulation research and application for Active Distribution Network based on Ptolemy II and Simulink}{Zhao2014}

This work describes the co-simulation between Ptolemy II and Simulink.

\catref{nfr:distribution}
\catref{sr:causality:causal}
\catref{sr:rel_time:analytic}
\catref{sr:rollback:none}
\catref{sr:availability:remote}
\catref{fr:coupling_model:io_assignments}
\catref{fr:domain:ct}
\catref{fr:communication_model:gauss_seidel}
\catref{fr:num_sim:two}
\catref{fr:sim_rate:single}
\catref{fr:sim_step_size:fixed}
\catref{fr:alg_loop:explicit}
\catref{fr:results_visualization:post_mortem}

\end{Reference}

\begin{Reference}{VPNET: A co-simulation framework for analyzing communication channel effects on power systems}{Li2011c}

This work describes the coupling of two simulation tools (VTB and OPNET) to achieve co-simulation.

\catref{sr:availability:local}
\catref{sr:rollback:none}
\catref{sr:causality:causal}
\catref{sr:rel_time:analytic}
\catref{fr:results_visualization:post_mortem}
\catref{fr:alg_loop:explicit}
\catref{fr:coupling_model:io_assignments}
\catref{fr:communication_model:gauss_seidel}
\catref{fr:domain:ct} The coordination is a sample discrete time system.
\catref{fr:num_sim:two}
\catref{fr:sim_rate:single}
\catref{fr:sim_step_size:fixed}

\end{Reference}

\begin{Reference}{Distributed hybrid simulation using the HLA and the Functional Mock-up Interface}{Awais2013b}

The main difference between this work and \cite{Awais2013a} is that this proposes a variable step size wrapper around CT components.
The approach taken to do this is quantization.

\catref{nfr:distribution}
\catref{nfr:parallelism}
\catref{sr:rollback:none}
\catref{sr:causality:causal}
\catref{sr:rel_time:analytic}
\catref{sr:availability:remote}
\catref{sr:availability:local}
\catref{fr:alg_loop:explicit}
\catref{fr:communication_model:gauss_seidel}
\catref{fr:communication_model:jacobi}
\catref{fr:standard:fmi}
\catref{fr:standard:hla}
\catref{fr:num_sim:three_more}
\catref{fr:sim_rate:multi}
\catref{fr:sim_step_size:variable}
\catref{fr:coupling_model:io_assignments}
\catref{fr:domain:de}
\catref{fr:results_visualization:post_mortem}

\end{Reference}

\begin{Reference}{Using the HLA for Distributed Continuous Simulations}{Awais2013a}
This work addresses the need to adapt CT simulators as DE simulators, in order to be used in a hybrid co-simulation scenario that is fundamentally DE oriented.

\catref{nfr:distribution}
\catref{nfr:parallelism}
\catref{sr:rollback:none}
\catref{sr:causality:causal}
\catref{sr:rel_time:analytic}
\catref{sr:availability:remote}
\catref{sr:availability:local}
\catref{fr:alg_loop:explicit}
\catref{fr:communication_model:gauss_seidel}
\catref{fr:communication_model:jacobi}
\catref{fr:standard:fmi}
\catref{fr:standard:hla}
\catref{fr:num_sim:three_more}
\catref{fr:sim_rate:multi}
\catref{fr:sim_step_size:fixed}
\catref{fr:coupling_model:io_assignments}
\catref{fr:domain:de}
\catref{fr:results_visualization:post_mortem}

\end{Reference}

\begin{Reference}{FERAL - Framework for simulator coupling on requirements and architecture level}{Kuhr2013}

They describe a framework that borrows many concepts from Ptolemy, but that is fundamentally event based co-simulation.
It allows for the specialization of basic directors for the semantic adaptation of simulation units.

\catref{nfr:ip_protection}
\catref{nfr:extensibility}
\catref{sr:info:signal}
\catref{sr:causality:causal}
\catref{sr:rollback:none}
\catref{sr:availability:local}
\catref{sr:rel_time:analytic}
\catref{fr:sim_rate:multi}
\catref{fr:communication_model:gauss_seidel}
\catref{fr:standard:fmi}
\catref{fr:domain:de}
\catref{fr:domain:ct}
\catref{fr:num_sim:three_more}
\catref{fr:sim_step_size:variable}

\end{Reference}

\begin{Reference}{Implementing stabilized co-simulation of strongly coupled systems using the Functional Mock-up Interface 2.0.}{Viel2014}
This work describes the implementation of the method described in \cite{Arnold2010} in the context of the FMI standard.

\catref{nfr:accuracy}
\catref{nfr:ip_protection}
\catref{sr:info:jacobian:out}
\catref{sr:info:input_extrapolation}
\catref{sr:info:record_outputs}
\catref{sr:info:stateserial}
\catref{sr:causality:causal}
\catref{sr:rel_time:analytic}
\catref{sr:availability:local}
\catref{sr:rollback:none}
\catref{fr:domain:ct}
\catref{fr:num_sim:three_more}
\catref{fr:sim_rate:single}
\catref{fr:sim_step_size:fixed}
\catref{fr:alg_loop:implicit}
\catref{fr:results_visualization:post_mortem}
\catref{fr:coupling_model:algebraic_constraints}
\catref{fr:communication_model:gauss_seidel}
\catref{fr:standard:fmi}
\end{Reference}

\begin{Reference}{Interface Jacobian-based Co-Simulation}{Sicklinger2014}
Describes a co-simulation method that makes use of the Jacobian information for fixed point computations.

\catref{nfr:performance}
\catref{nfr:accuracy}
\catref{sr:availability:local}
\catref{sr:rel_time:analytic}
\catref{sr:rollback:single}
\catref{sr:info:jacobian:out}
\catref{sr:causality:causal}
\catref{fr:results_visualization:post_mortem}
\catref{fr:communication_model:gauss_seidel}
\catref{fr:communication_model:jacobi}
\catref{fr:coupling_model:algebraic_constraints}
\catref{fr:domain:ct}
\catref{fr:alg_loop:implicit}
\catref{fr:num_sim:three_more}
\catref{fr:sim_rate:single}
\catref{fr:sim_step_size:fixed}

\end{Reference}

\begin{Reference}{A co-simulation framework for design of time-triggered automotive cyber physical systems}{Zhang2014}
The work describes a co-simulation that integrates SystemC and CarSim.

\catref{sr:availability:local}
\catref{sr:rollback:none}
\catref{sr:causality:causal}
\catref{sr:info:full_model}
\catref{sr:rel_time:analytic}
\catref{fr:coupling_model:io_assignments}
\catref{fr:domain:de}
\catref{fr:domain:ct}
\catref{fr:alg_loop:explicit}
\catref{fr:results_visualization:post_mortem}
\catref{fr:sim_step_size:fixed}
\catref{fr:sim_rate:single}
\catref{fr:num_sim:two}

\end{Reference}

\begin{Reference}{A microgrid co-simulation framework}{Kounev2015}

Describes the coupling of two simulators written in  MATLAB and OMNeT++.

\catref{nfr:performance}
\catref{sr:availability:local}
\catref{sr:rel_time:analytic}
\catref{sr:rollback:none} The DEV's orchestration is conservative.
\catref{sr:causality:causal}
\catref{sr:info:predict_step_sizes}
\catref{fr:coupling_model:io_assignments}
\catref{fr:results_visualization:post_mortem}
\catref{fr:communication_model:gauss_seidel}
\catref{fr:domain:de}
\catref{fr:num_sim:two}
\catref{fr:sim_rate:single}
\catref{fr:sim_step_size:variable}
\catref{fr:alg_loop:explicit}

\end{Reference}

\begin{Reference}{Co-Simulation of Hybrid Systems with SpaceEx and Uppaal}{Bogomolov2015}

The orchestration algorithm is the one described in \cite{Broman2013}.
The work exploits the standard by allowing zero step transitions.

\catref{sr:info:causality:feedthrough}
\catref{sr:rollback:none}
\catref{sr:info:stateserial}
\catref{sr:causality:causal}
\catref{sr:rel_time:analytic}
\catref{sr:availability:local}
\catref{fr:coupling_model:io_assignments}
\catref{fr:standard:fmi}
\catref{fr:num_sim:two}
\catref{fr:communication_model:jacobi}
\catref{fr:alg_loop:explicit}
\catref{fr:sim_rate:single}
\catref{fr:sim_step_size:variable} This is due to the rejection of steps, not due to accuracy.
\catref{fr:domain:ct}
\catref{fr:domain:de} They abuse the FMI standard to be able to support state transitions.
\catref{fr:results_visualization:post_mortem}

\end{Reference}

\begin{Reference}{Real-time co-simulation platform using OPAL-RT and OPNET for analyzing smart grid performance}{Bian2015}

Not many details are provided about the co-simulation orchestration.
However, due to the fact that it is real-time, we can infer certain features.

\catref{nfr:performance}
\catref{sr:rollback:none}
\catref{sr:causality:causal}
\catref{sr:rel_time:fixed_real_scaled_time_simulation}
\catref{sr:availability:remote} Communication is done through UDP.
\catref{fr:sim_step_size:fixed}
\catref{fr:sim_rate:single} Two simulators do not give more than this.
\catref{fr:domain:ct}
\catref{fr:coupling_model:io_assignments}
\catref{fr:num_sim:two}
\catref{fr:alg_loop:explicit}
\catref{fr:results_visualization:post_mortem}

\end{Reference}

\begin{Reference}{Coupling the multizone airflow and contaminant transport software CONTAM with EnergyPlus using co-simulation}{Dols2016}

The work described the coupling of the CONTAM and EnergyPlus tools to achieve HVAC simulation.
The coupling is done through FMI.
The coupling is done through the compiled binaries.
The case study highlights the problems with an explicit method for co-simulation, even if the Gauss-seidel. 
Instabilities occur.

\catref{nfr:ip_protection}
\catref{sr:rollback:none}
\catref{sr:rel_time:analytic}
\catref{sr:causality:causal}
\catref{sr:availability:remote}
\catref{fr:results_visualization:post_mortem}
\catref{fr:domain:ct}
\catref{fr:num_sim:two}
\catref{fr:standard:fmi}
\catref{fr:alg_loop:explicit}
\catref{fr:coupling_model:io_assignments}
\catref{fr:communication_model:gauss_seidel}
\catref{fr:sim_step_size:fixed} It uses a 5-minute synchronization step.
\catref{fr:sim_rate:single}

\end{Reference}

\begin{Reference}{Multicore simulation of powertrains using weakly synchronized model partitioning}{BenKhaled2012}

According to \cite{BenKhaled2014}, this work explores variable step solvers.

\catref{nfr:parallelism}
\catref{nfr:performance}
\catref{sr:info:causality:feedthrough}
\catref{sr:info:full_model}
\catref{sr:rel_time:fixed_real_scaled_time_simulation}
\catref{sr:rollback:none}
\catref{sr:availability:local}
\catref{fr:standard:fmi}
\catref{fr:coupling_model:io_assignments}
\catref{fr:num_sim:three_more}
\catref{fr:domain:ct}
\catref{fr:sim_rate:single}
\catref{fr:sim_step_size:fixed}
\catref{fr:alg_loop:explicit}
\catref{fr:results_visualization:post_mortem}
\catref{fr:communication_model:jacobi}
\catref{fr:communication_model:gauss_seidel}

\end{Reference}

\begin{Reference}{Fast multi-core co-simulation of Cyber-Physical Systems: Application to internal combustion engines}{BenKhaled2014}

This paper focus on the parallelization of co-simulation.
The approach is to start with a single model and partition it into multiple models, which are then executed in separate FMUs in parallel.
The partitioning is important for accuracy reasons (e.g., break the algebraic loops at less sensitive variables).

\catref{nfr:parallelism}
\catref{nfr:accuracy}
\catref{nfr:performance}
\catref{nfr:scalability}
\catref{sr:info:full_model}
\catref{sr:info:wcet}
\catref{sr:info:causality:feedthrough}
\catref{sr:causality:causal}
\catref{sr:rel_time:analytic}
\catref{sr:rollback:none}
\catref{sr:availability:local}
\catref{fr:communication_model:gauss_seidel}
\catref{fr:sim_step_size:fixed}
\catref{fr:sim_rate:single}
\catref{fr:standard:fmi}
\catref{fr:domain:ct}
\catref{fr:num_sim:three_more}
\catref{fr:alg_loop:explicit} It breaks the loops by establishing an order and delaying one of the variables in the loop.

\end{Reference}

\begin{Reference}{Acceleration of FMU Co-Simulation On Multi-core Architectures}{Saidi2016}

The paper addresses the problem of performance in FMI co-simulation.
The solution proposed is to go parallel.
The parallelization approach is the same as the one presented in \cite{BenKhaled2012}.
Since FMI does not enforce thread safety across multiple instances of the same FMU, the work presented ensures that these do not execute concurrently by using mutexes or changing the scheduling policy.

\catref{nfr:parallelism}
\catref{nfr:performance}
\catref{nfr:ip_protection}
\catref{nfr:scalability}
\catref{sr:info:wcet}
\catref{sr:info:causality:feedthrough}
\catref{sr:causality:causal}
\catref{sr:rel_time:analytic}
\catref{sr:rollback:none}
\catref{sr:availability:local}
\catref{fr:communication_model:jacobi}
\catref{fr:standard:fmi}
\catref{fr:domain:ct}
\catref{fr:num_sim:three_more}
\catref{fr:alg_loop:explicit}

\end{Reference}

\begin{Reference}{ADAS Virtual Prototyping using Modelica and Unity Co-simulation via OpenMETA}{Yamaura2016}

The co-simulation framework includes 4 tools.
The communication between the tools is realized using OpenMeta.
The work uses Unity for the modelling and simulation of the environment, allowing for live interaction.
Communication is over UDP but there is no report on extra caution due to network delays and failures.

\catref{nfr:parallelism}
\catref{sr:info:full_model}
\catref{sr:causality:causal}
\catref{sr:rel_time:analytic}
\catref{sr:rollback:none}
\catref{sr:availability:remote}
\catref{fr:num_sim:three_more}
\catref{fr:domain:ct}
\catref{fr:results_visualization:live}
\catref{fr:results_visualization:interactive_live}
\catref{fr:coupling_model:io_assignments}
\catref{fr:sim_rate:single}
\catref{fr:alg_loop:explicit}

\end{Reference}

\begin{Reference}{Combining DEVS with multi-agent concepts to design and simulate multi-models of complex systems (WIP)}{Camus2015}
This work is the preliminary description of \cite{Camus2016}.

\catref{nfr:parallelism}
\catref{nfr:distribution}
\catref{nfr:ip_protection}
\catref{nfr:accuracy}
\catref{sr:rollback:none}
\catref{sr:info:stateserial}
\catref{sr:causality:causal}
\catref{sr:rel_time:analytic}
\catref{sr:availability:local}
\catref{fr:domain:de}
\catref{fr:alg_loop:explicit}
\catref{fr:num_sim:three_more}
\catref{fr:sim_rate:multi}
\catref{fr:sim_step_size:variable}
\catref{fr:standard:fmi}
\catref{fr:coupling_model:io_assignments}
\catref{fr:communication_model:gauss_seidel}
\catref{fr:results_visualization:post_mortem}

\end{Reference}

\begin{Reference}{Hybrid Co-simulation of FMUs using DEV \& DESS in MECSYCO}{Camus2016}
It proposes to use a FMU wrapper around DEV and DESS models, meaning that the co-simulation proceeds using a DE approach.
It handles black box FMUs and the algorithm used to drive the co-simulation is the conservative parallel DEVS simulator.
It requires that the FMU is able to perform rollback (through the use of state set and get).

\catref{nfr:parallelism}
\catref{nfr:distribution}
\catref{nfr:ip_protection}
\catref{nfr:accuracy}
\catref{sr:rollback:none}
\catref{sr:info:stateserial}
\catref{sr:causality:causal}
\catref{sr:rel_time:analytic}
\catref{sr:availability:local}
\catref{fr:domain:de}
\catref{fr:alg_loop:explicit}
\catref{fr:num_sim:three_more}
\catref{fr:sim_rate:multi}
\catref{fr:sim_step_size:variable}
\catref{fr:standard:fmi}
\catref{fr:coupling_model:io_assignments}
\catref{fr:communication_model:gauss_seidel}
\catref{fr:results_visualization:post_mortem}

\end{Reference}

\begin{Reference}{FMI for Co-Simulation of Embedded Control Software}{Pedersen2016}
The paper describes the adaptation of an embedded system to comply with FMI and thus interface with other FMUs.
To validate the implementation, they run a co-simulation.

\catref{nfr:distribution}
\catref{nfr:parallelism}
\catref{sr:rel_time:fixed_real_scaled_time_simulation}
\catref{fr:domain:ct}
\catref{fr:num_sim:two}
\catref{fr:sim_rate:single}
\catref{fr:sim_step_size:fixed}
\catref{fr:communication_model:gauss_seidel}
\catref{fr:standard:fmi}
\catref{fr:coupling_model:io_assignments}
\catref{fr:alg_loop:explicit}
\catref{fr:results_visualization:live}

\end{Reference}

\begin{Reference}{A Co-Simulation Framework for Power System Analysis}{Oh2016}

The paper proposes a co-simulation framework that takes into account network delays and compensates for that.
It proposes to use cubic spline extrapolation to compensate for the delay but recognizes that if there are faults in the line (resulting in voltage drops), the derivatives used in the extrapolation assume gigantic proportions, thus wreaking havoc in the simulation.
To address that, the framework employes an algorithm to detect discontinuities.
The detection is simple: they check the derivative of the signal to see whether it exceeds a pre-determined empirically threshold.
Basically, it looks for and Dirac delta.
Figure 7 shows the effect of not handling a discontinuity.

\catref{nfr:distribution}
\catref{nfr:accuracy}
\catref{nfr:parallelism}
\catref{fr:num_sim:two}
\catref{fr:sim_rate:single}
\catref{fr:sim_step_size:fixed}
\catref{fr:communication_model:gauss_seidel} Due to the parallel interface protocol that they use.
\catref{fr:coupling_model:io_assignments}
\catref{fr:domain:ct}
\catref{fr:alg_loop:explicit}
\catref{sr:availability:remote}
\catref{sr:causality:causal}
\catref{sr:rel_time:analytic}
\catref{sr:rollback:none}

\end{Reference}

\begin{Reference}{Continuous-Mass-Model-Based Mechanical and Electrical Co-Simulation of SSR and Its Application to a Practical Shaft Failure Event}{Xie2016}

Between two simulators.
As it is explained in the paper, prior to co-simulation, the most common approach would be to run two simulations: one complete for one sub-system, and then another for the second sub-system, using the first as inputs. This is an open loop approach, whose results can be misleading due to ignoring the feedback loops.
Each simulator advances in parallel and their communication is made with a barrier.

\catref{nfr:parallelism}
\catref{fr:communication_model:jacobi}
\catref{fr:num_sim:two}
\catref{fr:domain:ct}
\catref{fr:sim_step_size:fixed}
\catref{fr:results_visualization:post_mortem}
\catref{fr:sim_rate:single}
\catref{sr:availability:local}
\catref{sr:causality:causal}
\catref{sr:rel_time:analytic}
\catref{sr:rollback:none}
\catref{fr:alg_loop:explicit}
\catref{fr:coupling_model:io_assignments}

\end{Reference}

\begin{Reference}{Impact of EV penetration on Volt–VAR Optimization of distribution networks using real-time co-simulation monitoring platform}{Manbachi2016}

It describes an application of co-simulation in the distribution of energy in smart grids, supported by a real-time co-simulation framework.
The simulators involved are the RTDS, which simulates the distribution network model, and the VVO Engine, coded in MATLAB.

\catref{sr:rel_time:fixed_real_scaled_time_simulation}
\catref{fr:num_sim:two}
\catref{sr:causality:causal}
\catref{sr:rollback:none}
\catref{sr:availability:local}
\catref{fr:coupling_model:io_assignments}
\catref{fr:sim_rate:single}
\catref{fr:sim_step_size:fixed}
\catref{fr:alg_loop:explicit}
\catref{fr:communication_model:jacobi}

\end{Reference}

\begin{Reference}{Co-simulation with communication step size control}{Schierz2012a}
Describes a master algorithm.
Does not allow for interpolation of inputs.
Needs rollback.
It touches upon accuracy, as it suggests an adaptive step size control mechanism.
It does not address algebraic loops.
It assumes that there is no feedthrough information.

\catref{nfr:performance}
\catref{nfr:accuracy}
\catref{sr:info:derivatives:out}
\catref{sr:info:stateserial}
\catref{sr:causality:causal}
\catref{sr:rel_time:analytic}
\catref{sr:rollback:single}
\catref{sr:availability:local}
\catref{fr:standard:fmi}
\catref{fr:coupling_model:io_assignments}
\catref{fr:num_sim:three_more}
\catref{fr:domain:ct}
\catref{fr:sim_rate:single}
\catref{fr:sim_step_size:variable}
\catref{fr:alg_loop:explicit}
\catref{fr:results_visualization:post_mortem}
\catref{fr:communication_model:jacobi}
\end{Reference}

\begin{Reference}{Co-simulation based platform for wireless protocols design explorations}{Fourmigue2009}
Application of co-simulation to wireless network development.
One of the simulators is the actual Linux operating system, and the other is represents a wireless network protocol simulator.
\catref{sr:causality:causal}
\catref{sr:rel_time:analytic}
\catref{sr:availability:local}
\catref{fr:coupling_model:io_assignments}
\catref{fr:num_sim:two}
\catref{fr:domain:de}
\catref{fr:communication_model:jacobi}
\end{Reference}

\begin{Reference}{Calculation of Wing Flutter by a Coupled Fluid-Structure Method}{Liu2001}

A fully implicit method, dealing with parallelism.

\catref{nfr:parallelism}
\catref{nfr:performance}
\catref{sr:causality:causal}
\catref{sr:rel_time:analytic}
\catref{sr:rollback:single}
\catref{sr:availability:remote}
\catref{fr:coupling_model:io_assignments}
\catref{fr:num_sim:two}
\catref{fr:domain:ct}
\catref{fr:alg_loop:implicit}
\catref{fr:results_visualization:post_mortem}

\end{Reference}

\begin{Reference}{Coupled simulation of flow-structure interaction in turbomachinery}{Carstens2003}

Relates to the application of a co-simulation algorithm to the simulation of the deformation in the blades of a transonic compressor rotor under airflow.
One of the simulators calculates deformation of the blades, while the other calculates the flow dynamics around the blades.

The communication of orchestration algorithm in use is shifted by half a step.

\catref{nfr:performance}
They highlight the need for it, because the computation of a rotor is just too expensive.
\catref{sr:info:derivatives:out}
\catref{sr:causality:causal}
\catref{sr:rel_time:analytic}
\catref{sr:rollback:none}
\catref{sr:availability:remote}
It seems that they perform the computation in separate computers.
\catref{fr:coupling_model:io_assignments}
\catref{fr:num_sim:two}
\catref{fr:domain:ct}
\catref{fr:sim_rate:single}
\catref{fr:sim_step_size:fixed}
\catref{fr:alg_loop:explicit}
\catref{fr:results_visualization:post_mortem}
\catref{fr:communication_model:gauss_seidel}
Although it is a gauss seidel shifted in time.
\end{Reference}

\begin{Reference}{Model-based coupling approach for non-iterative real-time co-simulation}{Stettinger2014}

Proposes to address the challenges in real-time co-simulation by using a model based coupling approach.
The master has to keep track of two values for each packet of data: receiving time delay $t_r$ -- the time it takes for a packet to reach the master from the simulator --, and sending time delay $t_s$ -- the time it takes for a packet to leave the master and reach the simulator.  When a sample is delayed, the master acts as a replacement for it.
Basically, it is a dead reckoning model.

\catref{nfr:performance}
\catref{nfr:parallelism}
\catref{nfr:accuracy}
\catref{sr:causality:causal}
\catref{fr:domain:ct}
\catref{fr:num_sim:two}
\catref{sr:availability:local}
\catref{fr:coupling_model:io_assignments}
\catref{sr:rollback:none}
\catref{sr:rel_time:fixed_real_scaled_time_simulation}
\catref{fr:sim_rate:single}
\catref{fr:sim_step_size:fixed}
\catref{fr:alg_loop:explicit}
\catref{fr:results_visualization:post_mortem}
\catref{fr:communication_model:jacobi}
\catref{fr:communication_model:gauss_seidel}

\end{Reference}

\begin{Reference}{Automated configuration for non-iterative co-simulation}{Benedikt2016}
Describes how a co-simulation master can configure some parameters throughout the co-simulation execution.
This is the idea behind adaptive master algorithms.

\catref{nfr:ip_protection}
\catref{nfr:accuracy}
\catref{sr:info:causality:feedthrough}
\catref{sr:causality:causal}
\catref{sr:availability:local}
\catref{sr:rollback:none}
\catref{sr:rel_time:analytic}
\catref{fr:domain:ct}
\catref{fr:num_sim:three_more}
\catref{fr:coupling_model:io_assignments}
\catref{fr:sim_rate:single}
\catref{fr:sim_step_size:variable}
\catref{fr:alg_loop:explicit}
\catref{fr:results_visualization:post_mortem}

\end{Reference}

\begin{Reference}{An explicit approach for controlling the macro-step size of co-simulation methods}{Busch2011}
Presents an approach to estimate the local truncation error caused by the extrapolations of inputs in a co-simulation.
The sub-systems are assumed to make no error.
It does not require rollback or re-initialization.

Categories:
\catref{nfr:accuracy}
Because they study the global error and control the local error.
\catref{nfr:ip_protection}
\catref{nfr:performance}
They control the step size, which increases performance. And they study how to get an optimal step size.
\catref{sr:info:causality:feedthrough}
\catref{sr:rollback:none}
\catref{sr:causality:causal}
\catref{fr:domain:ct}
\catref{fr:num_sim:three_more}
\catref{sr:rel_time:analytic}
\catref{fr:sim_rate:multi}
\catref{fr:sim_step_size:variable}
\catref{fr:alg_loop:explicit}
\catref{sr:availability:local}
\catref{fr:results_visualization:post_mortem}
\catref{fr:coupling_model:io_assignments}
\catref{fr:communication_model:jacobi}
In theory, they seem to support any communication model. In the paper they studied assuming the Jacobi.

\end{Reference}

\begin{Reference}{DEVS coupling of spatial and ordinary differential equations: VLE framework}{Quesnel2005}
Proposes a way to wrap a continuous time ODE simulator as a DEVS model. 
It requires that the state variables, and derivatives are available.

Categories:
\catref{nfr:hierarchy}
\catref{nfr:open_source}
\catref{sr:info:derivatives:out}
\catref{sr:info:statevars}
\catref{sr:info:predict_step_sizes}
\catref{sr:causality:causal}
\catref{fr:domain:de}
\catref{fr:num_sim:three_more}
\catref{sr:rel_time:analytic}
\catref{fr:sim_rate:multi}
Any discrete event framework is by definition multi-rate.
\catref{fr:sim_step_size:variable}
Any discrete event framework is by definition in this category.
\catref{fr:alg_loop:explicit}
\catref{sr:availability:local}
\catref{fr:results_visualization:post_mortem}
\catref{fr:coupling_model:io_assignments}
\catref{fr:communication_model:gauss_seidel}
	A discrete event framework is in this category as there is no extrapolation of inputs. Also, Gauss seidel does not violate the causality of inputs and outputs, because it sorts according to these dependencies. Events are processed to retain their causality.

\end{Reference}

\begin{Reference}{Error analysis for co-simulation with force-displacement coupling}{Arnold2014a}

Describes an FMI based master called SNiMoWrapper.

Categories:
\catref{nfr:accuracy}
Because they study the global error and control the local error.
\catref{nfr:ip_protection}
\catref{sr:info:causality:feedthrough}
\catref{sr:rollback:none}
\catref{sr:causality:causal}
\catref{sr:rel_time:analytic}
\catref{sr:availability:local}
\catref{fr:domain:ct}
\catref{fr:num_sim:three_more}
\catref{fr:sim_rate:multi}
\catref{fr:sim_step_size:fixed}
\catref{fr:alg_loop:explicit}
\catref{fr:results_visualization:post_mortem}
\catref{fr:coupling_model:io_assignments}
\catref{fr:communication_model:jacobi}

\catref{fr:standard:fmi}
\end{Reference}

\begin{Reference}{Error Analysis and Error Estimates for Co-simulation in FMI for Model Exchange and Co-Simulation v2.0}{Arnold2014}

Studies the error control method known as Richard's extrapolation.

Categories:
\catref{nfr:accuracy}
Because they study the global error and control the local error.
\catref{nfr:ip_protection}
\catref{sr:info:causality:feedthrough}
\catref{sr:info:statevars}
\catref{sr:rollback:single}
\catref{sr:causality:causal}
\catref{fr:domain:ct}
\catref{fr:num_sim:three_more}
\catref{sr:rel_time:analytic}
\catref{fr:sim_rate:multi}
\catref{fr:sim_step_size:variable}
\catref{fr:alg_loop:explicit}
\catref{sr:availability:local}
\catref{fr:results_visualization:post_mortem}
\catref{fr:coupling_model:io_assignments}
\catref{fr:communication_model:jacobi} 
\catref{fr:communication_model:gauss_seidel}

\catref{fr:standard:fmi}
\end{Reference}

\begin{Reference}{Preconditioned Dynamic Iteration for Coupled Differential-Algebraic Systems}{Arnold2001}

Studies the convergence of the Gauss-Seidel dynamic iteration method and proposes a way to ensure it.
The way to do it though, requires information from the model.

Categories:
\catref{nfr:accuracy}
Because they study the global error.
\catref{sr:info:jacobian:out}
\catref{sr:info:record_outputs}
\catref{sr:info:full_model}
\catref{sr:rollback:single}
\catref{sr:causality:causal}
\catref{sr:rel_time:analytic}
\catref{sr:availability:local}
\catref{fr:domain:ct}
\catref{fr:num_sim:three_more}
\catref{fr:sim_rate:multi}
\catref{fr:sim_step_size:fixed}
\catref{fr:alg_loop:implicit}
\catref{fr:results_visualization:post_mortem}
\catref{fr:coupling_model:algebraic_constraints}
\catref{fr:communication_model:gauss_seidel}

\end{Reference}

\begin{Reference}{Semi-implicit co-simulation approach for solver coupling}{Schweizer2014}

Proposes a predictor corrector master that evaluates the macro step twice and uses a perturbation on the inputs to get an estimate of the required partial derivatives.
This approach is then generalized to multiple kinds of joints in the mechanical domain.
A double pendulum, double mass-spring-damper and a slider crank mechanism are used as numerical examples.

Categories:
\catref{sr:rollback:single}
\catref{sr:causality:causal}
\catref{sr:rel_time:analytic}
\catref{sr:availability:local}
\catref{fr:domain:ct}
\catref{fr:num_sim:two}
\catref{fr:sim_rate:multi}
\catref{fr:sim_step_size:fixed}
\catref{fr:alg_loop:semi_implicit}
\catref{fr:results_visualization:post_mortem}
\catref{fr:coupling_model:algebraic_constraints}
\catref{fr:communication_model:jacobi}

\end{Reference}

\begin{Reference}{Stabilized implicit co-simulation methods: solver coupling based on constitutive laws}{Schweizer2015d}

It presents an implicit and semi-explicit methods for the co-simulation of scenarios coupled via applied forces.
The difference between this paper and the previous ones by the same author seems to be in the fact that the coupling constraints are integrated and differentiated, to enrich the information being used to ensure that the original coupling constraints are met.

Categories:
\catref{sr:rollback:single}
\catref{sr:causality:causal}
\catref{sr:rel_time:analytic}
\catref{sr:availability:local}
\catref{fr:domain:ct}
\catref{fr:num_sim:three_more}
\catref{fr:sim_rate:multi}
\catref{fr:sim_step_size:fixed}
\catref{fr:alg_loop:semi_implicit}
\catref{fr:results_visualization:post_mortem}
\catref{fr:coupling_model:algebraic_constraints}
\catref{fr:communication_model:jacobi}

\end{Reference}

\begin{Reference}{Energy Conservation and Power Bonds in Co-Simulations: Non-Iterative Adaptive Step Size Control and Error Estimation}{Sadjina2016}

Proposes a master for co-simulation that requires the identification of power bonds between sub-systems.
It assumes that the scenario is energy conserving and thus calculate the energy residual as an error to be minimized.
The step size is adapted via a PI-Controller. 
When the step size is reduced, it is only on the next co-simulation step, so the method is explicit.

\catref{nfr:ip_protection}
\catref{nfr:accuracy} Due to step size control.
\catref{sr:rollback:none}
\catref{sr:causality:causal}
\catref{sr:rel_time:analytic}
\catref{sr:availability:local}
\catref{fr:domain:ct}
\catref{fr:num_sim:three_more}
\catref{fr:sim_rate:multi}
\catref{fr:sim_step_size:variable}
\catref{fr:alg_loop:explicit}
\catref{fr:results_visualization:post_mortem}
\catref{fr:coupling_model:io_assignments}
\catref{fr:communication_model:jacobi}

\end{Reference}

\begin{Reference}{Continuous approximation techniques for co-simulation methods: Analysis of numerical stability and local error}{Busch2016}
Analyses the stability and local error of multiple co-simulation approaches with multiple extrapolation approaches for the inputs.
It considers Gauss-Seidel and Jacobi.
It also talks about a method called the extrapolated interpolation method, which ensure no discontinuities at the inputs of the subsystems.

\catref{sr:rollback:none}
The method is explicit.
\catref{sr:causality:causal}
\catref{fr:domain:ct}
\catref{fr:num_sim:three_more}
\catref{sr:rel_time:analytic}
\catref{fr:sim_rate:single}
\catref{fr:sim_step_size:fixed}
\catref{fr:alg_loop:explicit}
\catref{sr:availability:local}
\catref{fr:results_visualization:post_mortem}
\catref{fr:coupling_model:io_assignments}
\catref{fr:communication_model:jacobi}
\catref{fr:communication_model:gauss_seidel}

\end{Reference}

\begin{Reference}{Stability of Sequential Modular Time Integration Methods for Coupled Multibody System Models}{Arnold2010}

Studies stability of a gauss Seidel co-simulation method proposed in previous work: \cite{Arnold2001}.
Based on that analysis, it proposes an implicit stabilization technique that uses Gauss-Seidel iteration.
The resulting method is implicit but the equations that are being solved are linear.

Categories:
\catref{nfr:accuracy}
Because they study the global error.
\catref{sr:info:jacobian:out}
\catref{sr:info:record_outputs}
\catref{sr:info:full_model}
\catref{sr:rollback:single}
\catref{sr:causality:causal}
\catref{sr:rel_time:analytic}
\catref{sr:availability:local}
\catref{fr:domain:ct}
\catref{fr:num_sim:three_more}
\catref{fr:sim_rate:single}
\catref{fr:sim_step_size:fixed}
\catref{fr:alg_loop:implicit}
\catref{fr:results_visualization:post_mortem}
\catref{fr:coupling_model:algebraic_constraints}
\catref{fr:communication_model:gauss_seidel}

\end{Reference}

\begin{Reference}{Co-simulation of algebraically coupled dynamic subsystems}{Gu2001}
Describes a technique to deal with algebraically coupled sub-systems using a control theoretic approach.
The highlights of this method are: it supports scenarios of arbitrary index; the boundary condition coordinator is seen as a co-simulation unit (this is an elegant approach) and the method is explicit.
The beauty of making the BCC as a co-simulation unit, is that it can, just like any other sub-system be run at a different rate and in the paper they show that by running it at a higher rate, the stability of the co-simulation increases.

\catref{sr:rollback:none}
\catref{sr:causality:causal}
\catref{fr:domain:ct}
\catref{fr:num_sim:two}
\catref{sr:rel_time:analytic}
\catref{fr:sim_rate:multi}
\catref{fr:sim_step_size:fixed}
\catref{fr:alg_loop:explicit}
\catref{sr:availability:local}
\catref{fr:results_visualization:post_mortem}
\catref{fr:coupling_model:algebraic_constraints}
\catref{fr:communication_model:jacobi}

\end{Reference}

\begin{Reference}{Co-Simulation of Algebraically Coupled Dynamic Subsystems Without Disclosure of Proprietary Subsystem Models}{Gu2004}
Describes a technique to solve causal conflicts using a Boundary Condition Coordinator (BCC).
Causal conflicts arise naturally from the coupling of different sub-systems and they are a relevant challenge that needs to be overcome in order to perform correct co-simulation.
While in \cite{Gu2001}, the BCC requires the knowledge of the state variables of the simulations, in \cite{Gu2004}, some modifications are made to ensure that this information is not required.

\catref{nfr:ip_protection}
\catref{nfr:distribution}
\catref{sr:rollback:none}
\catref{sr:causality:causal}
\catref{fr:domain:ct}
\catref{fr:num_sim:three_more}
\catref{sr:rel_time:analytic}
\catref{fr:sim_rate:multi}
\catref{fr:sim_step_size:fixed}
\catref{fr:alg_loop:explicit}
\catref{sr:availability:local}
\catref{fr:results_visualization:post_mortem}
\catref{fr:coupling_model:algebraic_constraints}
\catref{fr:communication_model:jacobi}

\end{Reference}

\begin{Reference}{Co-simulation method for solver coupling with algebraic constraints incorporating relaxation techniques}{Schweizer2016}

A master algorithm capable of dealing with algebraic constraints is described.
It requires the derivatives of the coupled variables to be available.
It executes each communication step twice, being a semi-implicit method.
It uses a predict step and a corrector step. 
The final corrected coupling variables are obtained by polynomial extrapolation and relaxation (to avoid instabilities).

Categories:
\catref{sr:rollback:single}
\catref{sr:causality:causal}
\catref{fr:domain:ct}
\catref{fr:num_sim:three_more}
\catref{sr:rel_time:analytic}
\catref{fr:sim_rate:multi}
\catref{sr:info:jacobian:out}
\catref{fr:sim_step_size:fixed}
\catref{fr:alg_loop:semi_implicit}
\catref{sr:availability:local}
\catref{fr:results_visualization:post_mortem}
\catref{fr:coupling_model:algebraic_constraints}
\catref{fr:communication_model:jacobi}

\end{Reference}

\begin{Reference}{Predictor/corrector co-simulation approaches for solver coupling with algebraic constraints}{Schweizer2015}

Proposes a predictor corrector master that evaluates the macro step twice and uses a perturbation on the inputs to get an estimate of the required partial derivatives.

Categories:
\catref{sr:rollback:single}
\catref{sr:causality:causal}
\catref{fr:domain:ct}
\catref{fr:num_sim:three_more}
\catref{sr:rel_time:analytic}
\catref{fr:sim_rate:multi}
\catref{fr:sim_step_size:fixed}
\catref{fr:alg_loop:semi_implicit}
\catref{sr:info:jacobian:out}
\catref{sr:availability:local}
\catref{fr:results_visualization:post_mortem}
\catref{fr:coupling_model:algebraic_constraints}
\catref{fr:communication_model:jacobi}

\end{Reference}

\begin{Reference}{Stabilized index-2 co-simulation approach for solver coupling with algebraic constraints}{Schweizer2015a}

A master algorithm capable of dealing with algebraic constraints is described.
It requires the derivatives of the coupled variables to be available.
It executes each communication step twice, being a semi-implicit method.
It uses a predict step and a corrector step. The predictor step allows the method to estimate the sensitivity of the state variables with respect to the applied forces/torques.

Categories:
\catref{sr:rollback:single}
\catref{sr:causality:causal}
\catref{fr:domain:ct}
\catref{fr:num_sim:three_more}
\catref{sr:rel_time:analytic}
\catref{fr:sim_rate:multi}
\catref{sr:info:jacobian:out}
\catref{fr:sim_step_size:fixed}
\catref{fr:alg_loop:semi_implicit}
\catref{sr:availability:local}
\catref{fr:results_visualization:post_mortem}
\catref{fr:coupling_model:algebraic_constraints}
\catref{fr:communication_model:jacobi}

\end{Reference}

\begin{Reference}{Methods and Tools for Co-Simulation of Dynamic Systems with the Functional Mock-up Interface}{Andersson2016}

This is a Phd Thesis.
A linear extrapolation based master is proposed that is convergent and does not require fixed point iterations. 
Then, a modification to multi-step methods is proposed to increase their performance when executing in a co-simulation environment. 
This modification avoids the need to restart when dealing with discontinuities.

Categories:
\catref{nfr:ip_protection}
\catref{nfr:platform_independence}
\catref{nfr:open_source}
\catref{sr:rollback:none}
\catref{sr:causality:causal}
\catref{fr:domain:ct}
\catref{fr:num_sim:three_more}
\catref{sr:rel_time:analytic}
\catref{fr:sim_step_size:fixed}
\catref{sr:availability:local}
\catref{fr:results_visualization:post_mortem}
\catref{fr:coupling_model:io_assignments}
\catref{fr:communication_model:jacobi}

\catref{fr:standard:fmi}
\end{Reference}

\begin{Reference}{Model-Based Configuration of Automotive Co-Simulation Scenarios}{Krammer2015}
The language is further developed in \cite{Krammer2015} with the addition of three novel diagrams to represent different aspects of the co-simulation configuration:
\begin{compactitem}
  \item Architectural -- coupling of executable units;
  \item Tools -- assignment of tools to models;
  \item Connections -- connections (it was not clear what does this diagram do);
\end{compactitem}
In addition, they define a couple of well formedness properties that can be checked more easily with the model-based approach.
They give a brief summary of the tool ICOS.

Categories:
\catref{nfr:config_reusability}
\catref{nfr:parallelism}
\catref{nfr:hierarchy}
\catref{nfr:extensibility}
\catref{fr:domain:ct}
\catref{fr:num_sim:three_more}
\catref{sr:rel_time:analytic}
\catref{sr:availability:local}
\catref{fr:coupling_model:io_assignments}
\catref{fr:results_visualization:post_mortem}
\catref{sr:info:full_model}

\end{Reference}

\begin{Reference}{FMI-Based Distributed Multi-Simulation with DACCOSIM}{Galtier2015}
DACCOSIM is able to perform Distributed simulations and multi-core simulations. The term ``computation node'' is used for a
collection of FMU wrappers (which include an FMU) and a local master / global master. The FMU wrappers, and thereby not the masters, are responsible for passing outputs
to connected inputs. This is to avoid bottlenecks. A component node contains a master and some FMUs, which are wrapped in so-called ``FMU-wrappers''.
The masters take responsibility of coordinated step sizes in case an FMU needs to roll back. 

\catref{nfr:performance}
Because of the possibility of splitting the simulation over a cluster / multi-core and the focus on performance in the article.
Additionally because of their use of variable step size
\catref{nfr:config_reusability}
It is possible to create multiple co-simulation configuration files in the simulation configuration GUI. These can be stored and therefore reused.
\catref{nfr:ip_protection}
Some level of IP protection because of FMI..
\catref{nfr:parallelism}
Because of the possibility of splitting the simulation over a cluster
\catref{nfr:distribution}
Because a co-simulation can be executed on a cluster of computers

\catref{sr:availability:remote}
\catref{sr:availability:local}

\catref{nfr:hierarchy}
DACCOSIM is weak hierarchical because it has the notion of local and global masters.

\catref{nfr:scalability}
The framework is considered to be scalable because of the multi-core and distributed architecture.

\catref{nfr:platform_independence}
There are two versions of the DACCOSIM library. A cross-platform version relying on JAVA and a Windows version using C++ and QTronic SDK.

\catref{nfr:accuracy}
The article provides an example where the result of the co-simulation using DACCOSIM is compared to the simulation using Dymola and the results are
very close to each other. Accuracy is ensured by each FMU examining its outputs and estimating how far they are from the exact value.

\catref{nfr:open_source}
The framework is distributed under an open source license from January 2016.

\catref{sr:info:stateserial}
The framework can perform a single rollback using the state variable serialization.

\catref{sr:causality:causal}
Because the framework is based on FMI for co-simulation it is considered to be causal.

\catref{fr:domain:ct}
The framework supports multiple formalisms because it is based on FMI for co-simulation.

\catref{fr:num_sim:three_more}
The frameworks is capable of supporting many FMUs and thereby many simulation units.
DACCOSIM offers its own algorithm depending on global/local masters.

\catref{fr:coupling_model:io_assignments}

\catref{sr:rel_time:analytic}
There is no mentioning of any other time models than this in the article.

\catref{fr:sim_rate:single}
The simulation rate is the same for all FMUs.

\catref{fr:sim_step_size:variable}
The framework uses Coordinated Variable Step

\catref{fr:alg_loop:explicit}
The framework uses Euler's method and Richardson's method. Whether this is default, parameterizable or fully customizable is unknown based on this
article.

\catref{fr:communication_model:jacobi}
See Co-initialization bullet 2 in the article.

\catref{fr:standard:fmi}
It is based on the FMI standard.

\catref{fr:results_visualization:post_mortem}

\end{Reference}

\begin{Reference}{Parallel synchronization of continuous time discrete event simulators}{Fey1997}
Presents two synchronization approaches, detailed in three different synchronization protocols, to coordinate simulation scenarios that include one discrete event simulator and one continuous time simulator.
The discrete event simulator can implement any parallel simulation approach that we know, such as Time-Warp.
This means that, even internally, the DE simulator can be forced to rollback due to straggler messages.
The focus is on parallel approaches.

Categories:
\catref{nfr:performance}
\catref{nfr:ip_protection}
\catref{nfr:parallelism}
\catref{nfr:accuracy}
\catref{fr:domain:de}
\catref{fr:domain:ct}
\catref{fr:num_sim:two}
\catref{fr:sim_rate:multi}
\catref{fr:sim_step_size:fixed}
\catref{fr:results_visualization:post_mortem}

\catref{sr:rel_time:analytic}
\catref{sr:availability:local}
\catref{sr:rollback:single}
\catref{fr:coupling_model:io_assignments}
\catref{sr:info:full_model}
\end{Reference}

\begin{Reference}{Generation of an Optimised Master Algorithm for FMI Co-simulation}{Acker2015}
Essentially, this paper shows how a compiled approach increases the performance of the co-simulation.
It also shows that, because there are so many decisions to be made when designing the master, a compiled approach allows for a more elegant, and specifically tailored master, to be generated.

\catref{nfr:performance}
\catref{nfr:ip_protection}
\catref{nfr:config_reusability}
\catref{nfr:open_source}
\catref{sr:rel_time:analytic}
\catref{sr:availability:local}
\catref{sr:info:causality:feedthrough}
\catref{sr:info:statevars}
\catref{sr:rollback:none}
\catref{sr:causality:causal}
\catref{sr:info:preferred_step_sizes}
\catref{fr:domain:ct}
\catref{fr:num_sim:three_more}
\catref{fr:sim_rate:multi}
\catref{fr:sim_step_size:fixed}
\catref{fr:results_visualization:post_mortem}
\catref{fr:standard:fmi}
\catref{fr:communication_model:gauss_seidel}
\catref{fr:alg_loop:implicit}
\catref{fr:coupling_model:io_assignments}
\end{Reference}

\begin{Reference}{Functional Digital Mock-up and the Functional Mock-up Interface - Two Complementary Approaches for a Comprehensive Investigation of Heterogeneous Systems}{Enge-Rosenblatt2011}
The paper describes and compares two approaches to performing co-simulation of heterogeneous systems, namely the Functional Digital Mock-up (FDMU) and
the Functional Mock-up Interface (FMI). Besides describing these approaches it also introduces the ``FDMU framework'',  a framework that implements the
Functional Digital Mock-up approach. Furthermore, proposals are presented for combining FDMU and FMI approaches. 

The FDMU approach is a tool-independent and web service-based framework build on the Web Service standards. It is capable of coupling
different simulation tools and provide visualization based on CAD models.

FDMU consists of three main concepts: functional building blocks (FBB), wrappers, FDMU master, and FDMU Console.
The functional building block can wrap geometric information (CAD Models), behavioral models, and a simulator tool. It is the responsibility of the
wrappers to establish a connection between the different simulation tools and the FDMU Master Simulator. Finally, the FDMU master ensures correct
communication between the simulators. The FDMU Console is the user's front-end.

\catref{nfr:performance} Communication overhead of a web service-based approach.

\catref{nfr:ip_protection} Because of the the web service-based approach IP protection should be possible.

\catref{nfr:parallelism} Because of the web service-based approach it is parallel by nature.  It uses thread-safe queues and deadlock-free
transmission of data.

\catref{nfr:distribution} Because of the web service-based approach it is easy distributable.

\catref{nfr:scalability} The distributed systems paradigm ensures scalability.

\catref{nfr:platform_independence} web service-based approach.

\catref{nfr:extensibility} A new wrapper can be implemented.

\catref{sr:causality:causal} Every input of an FBB must have an appropriate output belonging to another FBB.

\catref{fr:domain:ct}

\catref{fr:coupling_model:io_assignments}

\catref{fr:num_sim:three_more} 

\catref{fr:sim_rate:multi}

\catref{sr:availability:remote}

\catref{fr:results_visualization:live} The framework provides an interactive 3D visualization based on CAD.

\catref{fr:standard:fdmu}

\end{Reference}

\begin{Reference}{Heterogeneous co-simulation platform for the efficient analysis of FlexRay-based automotive distributed embedded systems}{Karner2010a}
Motivation: FlexRay is a wired network for automotive high-speed control applications and no solutions exist that simulates all parts of the network.

What: a co-simulation platform called TEODACS FlexRayExprt.Sim. The simulation approach used in the platform covers mechanics and all parts of the
network from physical layer to application layer, which is not done by other solutions. The framework CISC SyAD is used to perform the
microelectronics co-simulation, CarMaker/AVL InMotion for the mechanics, and they are bridged by TEODACS FlexRayEprt.Sim. The platform uses a very
interesting approach to faster co-simulations, namely the use of model switching, where a less detailed model replaces a more detailed model in parts
of the simulation.

The paper provides an overview of existing approaches such as transaction based modeling, HDLs such as SystemC and Verilog, and cable harness and
topology modeling along with why these contain shortcomings to this domain. Furthermore, the paper provides some details of the implementation of the
models used in the co-simulation and showcases how the platform can analyse a system with specific examples.

\catref{nfr:accuracy}
Because of model switching.
\catref{nfr:performance} 
Because of model switching.
\catref{fr:dynamic_structure} 
Because the structure of the co-simulation is changed via model switching.
\catref{fr:domain:ct} 
Because SyAD supprots multiple formalisms and CarMaker / AVL InMotion.
\catref{fr:domain:de}
\catref{fr:coupling_model:io_assignments}
\catref{fr:num_sim:three_more} 
Multiple FlexRay nodes can be added.
\catref{sr:rel_time:analytic} 
From model switching and similar it is clear that analytic simulation is used.
\catref{fr:results_visualization:post_mortem}

\end{Reference}

\begin{Reference}{MOKA: An Object-Oriented Framework for FMI Co-Simulation}{Aslan2015}
The paper describes MOKA, which is a framework for performing co-simulations and creating FMUs using FMI 2.0 for co-simulation.
The framework turns the creation of FMUs into an object-oriented process by using C++. 
An FMU is created by inheriting one of the classes and implementing virtual functions thereby avoiding writing boilerplate code. The implementation of FMUs is realised by the concepts of FMUBlock, which is to be inherited, FMUPort, and FMUStateVariables. FMUBlock is to be extended by a concrete FMU slave and implements common computation phase functions for slaves. It contains FMUPort for data exchange and FMUStateVariables for state tracking during the simulation. 
The FMUPort classes provides the data exchange interface of a slave. It abstracts the value references by automatically assigning a value reference to the variable. 
The BaseStateVariable class also functions as base that is to be extended. It provides virtual functions for state variable services. The StateVariable inherits from BaseStateVariable and represents state variables for the slave. The framework also provides a template for the FMU Master so the master code changes minimally for different scenarios.

The article exemplifies an application of the MOKA framework where two FMUs are used: The bouncing ball and integer counter example from the QTronic SDK, where the bouncing ball has been re-developed with MOKA.

In future work it is stated that development of a DSL in a current study, so that different scenarios can be executed without altering the master code.

\catref{nfr:ip_protection}
\catref{nfr:config_reusability}
\catref{sr:causality:causal}
\catref{sr:rel_time:analytic}
\catref{sr:rollback:none}
\catref{sr:availability:local}
\catref{fr:coupling_model:io_assignments}
\catref{fr:num_sim:three_more}
\catref{fr:standard:fmi}
\catref{fr:domain:ct}
\catref{fr:alg_loop:explicit}
\catref{fr:sim_rate:single}
\catref{fr:communication_model:gauss_seidel}
\catref{fr:results_visualization:post_mortem}
\end{Reference}

\begin{Reference}{Co-simulation of building energy and control systems with the Building Controls Virtual Test Bed}{Wetter2010}
Describes a co-simulation framework called Building Controls Virtual Test Bed (BCVTB) that can be used for real-time simulation and co-simulation. 
It is a modular extensible open-source platform to interface different simulation programs with each other. The intention is to give users the option to use the best tools suited to model various aspects of building energy and control systems, or use programs where they have expertise.
The middleware to couple any number of simulation programs also provides libraries such that it can be extended.
Furthermore, the paper describes how they gathered capabilities the framework should support.
The framework is based on Ptolemy II, which is extended by some java packages. The simulator package adds functionality that allows an actor to perform system calls to start any executable on Windows, OSX or Linux. It simply starts a simulation program, sends input tokens to the simulation program, receives new values and sends them to its output port.
Algorithms are also provided on how simulators are coupled. These are also exemplified with specific simulators. It is also described how to connect client programs. The article describes how the interfaces are created for simulink, matlab, modelica, and system cals.  Furthermore, a specific example is presented.

\catref{nfr:config_reusability}
\catref{nfr:distribution}
\catref{nfr:platform_independence} 
\catref{nfr:extensibility}
\catref{nfr:open_source}
\catref{sr:causality:causal}
\catref{sr:rel_time:analytic}
\catref{sr:rel_time:fixed_real_scaled_time_simulation}
\catref{fr:domain:ct} In the paper, their explanation is focused on the CT domain.
\catref{fr:num_sim:three_more}
\catref{fr:sim_rate:single}
\catref{fr:sim_step_size:fixed}
\catref{fr:results_visualization:post_mortem}
\catref{fr:coupling_model:io_assignments}
\catref{fr:communication_model:jacobi}
\end{Reference}

\begin{Reference}{Model-based integration platform for FMI co-simulation and heterogeneous simulations of cyber-physical systems}{Neema2014}
The article concerns integrating FMI as an HLA federate and extending the Command
and Control Wind Tunnel (C2WT) metamodel to include FMI-specifics. This enables the C2WT tool to use FMUs as part of a simulation. The C2WT tool is
describes as a multi-model integration platform that allows users to model and synthesize complex, heterogeneous, command and control simulations. The
tool therefore has support for multiple simulation engines and an introduction to the tool is given in the paper.
Furthermore, a case study on Vehicle Thermal Management using FMUs are presented and The focompared to a simulation in a different environment.
The work is sponsored by the US DoD.

\catref{nfr:platform_independence}
\catref{nfr:distribution}
\catref{nfr:config_reusability}
\catref{fr:domain:de}
\catref{fr:num_sim:three_more}
\catref{fr:sim_rate:multi}
\catref{fr:sim_step_size:fixed}
\catref{fr:sim_step_size:variable}
\catref{sr:rel_time:fixed_real_scaled_time_simulation}
\catref{sr:rel_time:analytic}
\catref{fr:results_visualization:live}
\catref{fr:standard:hla}
\catref{fr:standard:fmi}
\catref{fr:communication_model:jacobi}
\end{Reference}

\begin{Reference}{Integrated Tool Chain for Model-Based Design of Cyber-Physical Systems}{Larsen16c}
This article presents an overview of the INTO-CPS project and thereby a Co-Simulation tool. The projects concerns production of a well-founded tool chain for model-based design of CPSs, and therefore consists of a semantic foundation and several baseline tools such as Modelio, Overture, 20-sim, OpenModelica and RT-Tester. Furthermore, an application called the INTO-CPS application is the entry point for configuring co-simulations and uses the co-simulation orchestration engine (COE) to perform the actual simulations. This COE is based on the FMI standard. The entire tool chain and the semantic foundation are presented in this paper. This is related to \cite{Larsen16c}.


\catref{nfr:config_reusability} 
\catref{nfr:ip_protection} 
\catref{sr:info:nominal_values:output} 
\catref{sr:info:nominal_values:state} 
\catref{sr:info:derivatives:out}
\catref{sr:info:derivatives:state}
\catref{sr:info:jacobian:out}
\catref{sr:info:jacobian:state}
\catref{sr:info:preferred_step_sizes}
\catref{sr:info:causality:feedthrough}
\catref{sr:causality:causal} 
\catref{sr:availability:local}
\catref{sr:info:statevars}
\catref{sr:info:record_outputs}
\catref{sr:rel_time:analytic}
\catref{sr:info:stateserial} 
\catref{fr:standard:fmi}
\catref{sr:info:signal}
\catref{fr:num_sim:three_more}
\catref{fr:domain:de}
\catref{fr:domain:ct}
\catref{fr:sim_rate:single}
\catref{fr:results_visualization:post_mortem}
\catref{fr:results_visualization:live}
\end{Reference}

\begin{Reference}{Design of the INTO-CPS Platform}{INTOCPSD41d}
This is an EU deliverable related to the INTO-CPS project. It contains the technical documentation of the INTO-CPS platform at the end of 2015 (the first year of the project). Part of this project is the Co-simulation Orchestration Engine (COE). This is related to \cite{Larsen16c}.
\catref{sr:info:predict_step_sizes} Supports the FMI suggested extension fmi2getMaxStepSize
\catref{nfr:performance} The COE supports variable step size, which can increase performance.
\catref{sr:rollback:single} Supports rollback to last successful state.
\catref{nfr:performance} The COE supports variable step size, which can increase performance.
\catref{nfr:accuracy} Contains various constraints such as zero-crossing, bounded difference and sampling rate. Furthermore, support for allowed min and max values.
\end{Reference}

	\section{Co-Simulation Scenario Categorization}

This section describes each category and lists the references that belong to that category, classified in the previous section.

\subsection{Non-Functional Requirements}

\subsubsection{Fault Tolerance}
\label{nfr:faulttolerance}

A co-simulation platform is fault tolerant if, for example, when one simulation unit fails, other can take its place.
This is particularly important for long running simulations.
To be fault tolerant, certain features need to available: periodically store the state of simulation units; record all inputs to each simulation unit.
If a simulation unit fails and the state is periodically stored, then the simulation can be paused while the state is restored in a new instance of the simulation unit. 
The history of input values passed to each simulation unit can be used to bring the simulation unit to the current state.

References in this category:

\refref{Fuller2013}


\subsubsection{Configuration Reusability}
\label{nfr:config_reusability}

This category refers to the fact that frameworks can provide a way to configure co-simulation scenarios that can be reused.
This means that the configuration is considered external to the execution of the co-simulation.
External in this context means that the configuration can reused without altering the binaries for the co-simulation application.

If a tool/frame does not provide a way to reuse configurations for co-simulation, 
then it is a time-consuming, error-prone and non-trivial process to set up co-simulations \cite{Krammer2015}.

References in this category:

\refref{Wang2013}

\refref{Krammer2015}

\refref{Galtier2015}

\refref{Acker2015}

\refref{Aslan2015}

\refref{Wetter2010}

\refref{Neema2014}


\subsubsection{Performance}
\label{nfr:performance}

Performance is a relative measure: a co-simulation platform is performant when it is able to simulate a great deal in a short amount of time while needing little resources.
This can be achieved by using variable step integration methods and signal extrapolation techniques.
Parallelism also plays a role but mostly on the time aspect of performance.

References in this category:

\refref{Hoepfer2011}

\refref{Faure2011}

\refref{Sun2011}

\refref{Friedrich2011}

\refref{Gonzalez2011}

\refref{Gunther2012}

\refref{Eyisi2012}

\refref{Riley2011}

\refref{Benedikt2013}

\refref{Sicklinger2014}

\refref{Kounev2015}

\refref{Bian2015}

\refref{BenKhaled2012}

\refref{BenKhaled2014}

\refref{Saidi2016}

\refref{Schierz2012a}

\refref{Liu2001}

\refref{Carstens2003}

\refref{Stettinger2014}

\refref{Busch2011}

\refref{Galtier2015}

\refref{Fey1997}

\refref{Acker2015}

\refref{Enge-Rosenblatt2011}

\refref{Karner2010a}


\subsubsection{IP Protection}
\label{nfr:ip_protection}

IP Protection deals with not requiring the models participating in the co-simulation to provide detailed structure, variables, etc\ldots
There are multiple levels of protection ranging from fully protected to not protected at all.
A good IP protection enables component suppliers to provide the system integrators with detailed simulations of their components avoiding expensive lock-in contracts.

There are multiple techniques can be be employed to ensure some degree of protection.
For instance, making the models (and corresponding simulation units) available as a web service is a possible solution.
Another example is any framework that implements the FMI Standard \cite{Blochwitz2011,Blochwitz2012}, which allows models and simulation units to be exported as a single functional unit, in binary format, that can be imported into a co-simulation.

References in this category:

\refref{Hoepfer2011}

\refref{Sun2011}

\refref{Bastian2011a}

\refref{Friedrich2011}

\refref{Broman2013}

\refref{Fuller2013}

\refref{Wang2013}

\refref{Kuhr2013}

\refref{Viel2014}

\refref{Dols2016}

\refref{Saidi2016}

\refref{Camus2015}

\refref{Camus2016}

\refref{Benedikt2016}

\refref{Busch2011}

\refref{Arnold2014a}

\refref{Arnold2014}

\refref{Sadjina2016}

\refref{Gu2004}

\refref{Andersson2016}

\refref{Galtier2015}

\refref{Fey1997}

\refref{Acker2015}

\refref{Enge-Rosenblatt2011}

\refref{Aslan2015}


\subsubsection{Parallelism}
\label{nfr:parallelism}

A co-simulation framework is parallel when it makes use of multiple processes/threads to perform the co-simulation.
This is typically in the same computer or same local network.

Techniques such as signal extrapolation help improve the speed-up gained from parallelism.
Furthermore waveform relaxation techniques and the Jacobi iterations promote parallelism \cite{McCalla1987}.

References in this category:

\refref{Faure2011}

\refref{Bastian2011a}

\refref{Friedrich2011}

\refref{Gunther2012}

\refref{Awais2013b}

\refref{Awais2013a}

\refref{BenKhaled2012}

\refref{BenKhaled2014}

\refref{Saidi2016}

\refref{Yamaura2016}

\refref{Camus2015}

\refref{Camus2016}

\refref{Pedersen2016}

\refref{Oh2016}

\refref{Xie2016}

\refref{Liu2001}

\refref{Stettinger2014}

\refref{Krammer2015}

\refref{Galtier2015}

\refref{Fey1997}

\refref{Enge-Rosenblatt2011}


\subsubsection{Distribution}
\label{nfr:distribution}

A co-simulation framework is parallel and distributed when it allows each simulation unit to be remote, across a wide area network.

This is very important since suppliers can, instead of transferring the simulation units in executable form across computers, can make them available over the web.
This offers much more control over how the simulation units are used.

The same techniques used in parallelism can be used to promote distribution, but fault tolerance is also important.

References in this category:

\refref{Friedrich2011}

\refref{Nutaro2011}

\refref{Busch2012}

\refref{Quaglia2012}

\refref{Eyisi2012}

\refref{Riley2011}

\refref{Roche2012}

\refref{Fitzgerald2010}

\refref{Fitzgerald2013}

\refref{Kudelski2013}

\refref{Fuller2013}

\refref{Bombino2013}

\refref{Zhao2014}

\refref{Awais2013b}

\refref{Awais2013a}

\refref{Camus2015}

\refref{Camus2016}

\refref{Pedersen2016}

\refref{Oh2016}

\refref{Gu2004}

\refref{Galtier2015}

\refref{Enge-Rosenblatt2011}

\refref{Wetter2010}

\refref{Neema2014}


\subsubsection{Hierarchy}
\label{nfr:hierarchy}

A hierarchical co-simulation framework is able to abstract a co-simulation scenario as a black box simulation unit.
This is very intuitive and promotes abstraction of complex systems.

References in this category:

\refref{Quesnel2005}

\refref{Krammer2015}

\refref{Galtier2015}


\subsubsection{Scalability}
\label{nfr:scalability}

A co-simulation framework is scalable when it supports a large number of simulation units.
It is intimately related to performance and paralelism.

References in this category:

\refref{Fuller2013}

\refref{BenKhaled2014}

\refref{Saidi2016}

\refref{Galtier2015}

\refref{Enge-Rosenblatt2011}


\subsubsection{Platform Independence}
\label{nfr:platform_independence}

A co-simulation framework is platform independent when it works on multiple computing platforms.
For this to be achieved, a platform independent language, such as Java, can be used to coordinate the simulation.

References in this category:

\refref{Bastian2011a}

\refref{Eyisi2012}

\refref{Riley2011}

\refref{Fitzgerald2010}

\refref{Fitzgerald2013}

\refref{Kudelski2013}

\refref{Andersson2016}

\refref{Galtier2015}

\refref{Enge-Rosenblatt2011}

\refref{Wetter2010}

\refref{Neema2014}


\subsubsection{Extensibility}
\label{nfr:extensibility}

A co-simulation framework is extensible when it can be easily extended to support new kinds of simulation units, with new kinds of capabilities.
A higher level, domain specific, language can be used to specify the behaviour in a platform agnostic way.
Code is then generated from this description.
The hypothesis is that the high level description can be more easily extended to describe new behaviour and that the code generation process can be adapted accordingly.

References in this category:

\refref{Kuhr2013}

\refref{Krammer2015}

\refref{Enge-Rosenblatt2011}

\refref{Wetter2010}


\subsubsection{Accuracy}
\label{nfr:accuracy}

A co-simulation is accurate when the error between the trace produced and the correct trace is minimal.
This can be achieved by error control mechanisms.

References in this category:

\refref{Hoepfer2011}

\refref{Tomulik2011}

\refref{Gonzalez2011}

\refref{Nutaro2011}

\refref{Busch2012}

\refref{Schmoll2012}

\refref{Schierz2012}

\refref{Gunther2012}

\refref{Fitzgerald2010}

\refref{Fitzgerald2013}

\refref{Benedikt2013}

\refref{Benedikt2013b}

\refref{Hafner2013}

\refref{Viel2014}

\refref{Sicklinger2014}

\refref{BenKhaled2014}

\refref{Camus2015}

\refref{Camus2016}

\refref{Oh2016}

\refref{Schierz2012a}

\refref{Stettinger2014}

\refref{Benedikt2016}

\refref{Busch2011}

\refref{Arnold2014a}

\refref{Arnold2014}

\refref{Arnold2001}

\refref{Sadjina2016}

\refref{Arnold2010}

\refref{Galtier2015}

\refref{Fey1997}

\refref{Karner2010a}


\subsubsection{Open source}
\label{nfr:open_source}

We consider open source the frameworks that make available the source code under certain licenses that are not paid for in any way.

References in this category:

\refref{Roche2012}

\refref{Quesnel2005}

\refref{Andersson2016}

\refref{Galtier2015}

\refref{Acker2015}

\refref{Wetter2010}


\subsection{Simulator Requirements}

This sub section covers the taxonomy that focuses on individual simulators' capabilities.

\subsubsection{Information Exposed}

\paragraph{Frequency of State}
\label{sr:info:frequency_state}

The instantaneous frequency of the state of the sub-system can be used to adjust the communication step size.


\paragraph{Frequency of Outputs}
\label{sr:info:frequency_outputs}

The instantaneous frequency of the output of the sub-system can be used to adjust the communication step size, as is done in \cite{Benedikt2013b}.

References in this category:

\refref{Benedikt2013b}


\paragraph{Detailed Model}
\label{sr:info:full_model}

Simulators that make the equations of the dynamic system available fall into this category.

References in this category:

\refref{Schmoll2012}

\refref{Hassairi2012}

\refref{Schierz2012}

\refref{Zhang2014}

\refref{BenKhaled2012}

\refref{BenKhaled2014}

\refref{Yamaura2016}

\refref{Arnold2001}

\refref{Arnold2010}

\refref{Krammer2015}

\refref{Fey1997}


\paragraph{Nominal Values of Outputs}
\label{sr:info:nominal_values:output}

This information indicates the order of magnitude of output signals.


\paragraph{Nominal Values of State}
\label{sr:info:nominal_values:state}

This information indicates the order of magnitude of state signals.


\paragraph{I/O Signal Kind}
\label{sr:info:signal}
The kind of output signal helps the master algorithm understand what assumptions are in a signal.

References in this category:

\refref{Kuhr2013}


\paragraph{Time Derivative}

\subparagraph{Output}
\label{sr:info:derivatives:out}

References in this category:

\refref{Hoepfer2011}

\refref{Tomulik2011}

\refref{Gunther2012}

\refref{Schierz2012a}

\refref{Carstens2003}

\refref{Quesnel2005}


\subparagraph{State}
\label{sr:info:derivatives:state}

References in this category:

\refref{Hoepfer2011}


\paragraph{Jacobian}

\subparagraph{Output}
\label{sr:info:jacobian:out}

References in this category:

\refref{Tomulik2011}

\refref{Bastian2011a}

\refref{Busch2012}

\refref{Schierz2012}

\refref{Viel2014}

\refref{Sicklinger2014}

\refref{Arnold2001}

\refref{Arnold2010}

\refref{Schweizer2016}

\refref{Schweizer2015}

\refref{Schweizer2015a}


\subparagraph{State}
\label{sr:info:jacobian:state}


\paragraph{Discontinuity Indicator}

A discontinuity indicator is a signal that indicates the presence of a discontinuity in the output of the simulation unit.

\paragraph{Deadreckoning model}

A deadreckoning model is a function that can be used by other simulation units to extrapolate the behavior of this simulation unit.

\paragraph{Preferred Step Size}
\label{sr:info:preferred_step_sizes}

References in this category:

\refref{Acker2015}


\paragraph{Next Step Size}
\label{sr:info:predict_step_sizes}

The next step size indicates the next communication time that is appropriate for the current simulator.

References in this category:

\refref{Lin2011}

\refref{Gunther2012}

\refref{Eyisi2012}

\refref{Riley2011}

\refref{Broman2013}

\refref{Kounev2015}

\refref{Quesnel2005}


\paragraph{Order of Accuracy}

The order of accuracy can be used to determine the appropriate input extrapolation functions to be used in a co-simulation scenario.

\paragraph{I/O Causality}

\emph{Feedthrough}
\label{sr:info:causality:feedthrough}

References in this category:

\refref{Broman2013}

\refref{Bogomolov2015}

\refref{BenKhaled2012}

\refref{BenKhaled2014}

\refref{Saidi2016}

\refref{Benedikt2016}

\refref{Busch2011}

\refref{Arnold2014a}

\refref{Arnold2014}

\refref{Acker2015}


\emph{Propagation Delay}

The propagation delay indicates how many micro-steps have to be performed before a change in the input affects an output.
In Simulink, this is the number of delay blocks in chain from an input to an output.

\paragraph{Input Extrapolation}
\label{sr:info:input_extrapolation}

This information denotes the kind of input extrapolation being performed by the simulation unit.

References in this category:

\refref{Viel2014}


\paragraph{State Variables}
\label{sr:info:statevars}

References in this category:

\refref{Hoepfer2011}

\refref{Quesnel2005}

\refref{Arnold2014}

\refref{Acker2015}


\paragraph{Serialized State}
\label{sr:info:stateserial}

References in this category:

\refref{Bastian2011a}

\refref{Broman2013}

\refref{Viel2014}

\refref{Bogomolov2015}

\refref{Camus2015}

\refref{Camus2016}

\refref{Schierz2012a}

\refref{Galtier2015}


\paragraph{Micro-Step Outputs}
\label{sr:info:record_outputs}

This information denotes the output of the simulation unit, evaluated at each of the micro-steps.

References in this category:

\refref{Benedikt2013}

\refref{Viel2014}

\refref{Arnold2001}

\refref{Arnold2010}


\paragraph{WCET}
\label{sr:info:wcet}

This denotes the worst case excution time.

References in this category:

\refref{Faure2011}

\refref{BenKhaled2014}

\refref{Saidi2016}


\subsubsection{Causality}

\paragraph{Causal}
\label{sr:causality:causal}

References in this category:

\refref{Pedersen2015}

\refref{Lin2011}

\refref{Hoepfer2011}

\refref{Tomulik2011}

\refref{Sun2011}

\refref{Bastian2011a}

\refref{Friedrich2011}

\refref{Gonzalez2011}

\refref{Nutaro2011}

\refref{Busch2012}

\refref{Schmoll2012}

\refref{Ni2012}

\refref{Hassairi2012}

\refref{Schierz2012}

\refref{Gunther2012}

\refref{Quaglia2012}

\refref{Al-Hammouri2012}

\refref{Eyisi2012}

\refref{Riley2011}

\refref{Roche2012}

\refref{Fitzgerald2010}

\refref{Fitzgerald2013}

\refref{Kudelski2013}

\refref{Broman2013}

\refref{Benedikt2013}

\refref{Benedikt2013b}

\refref{Fuller2013}

\refref{Bombino2013}

\refref{Wang2013}

\refref{Hafner2013}

\refref{Zhao2014}

\refref{Li2011c}

\refref{Awais2013b}

\refref{Awais2013a}

\refref{Kuhr2013}

\refref{Viel2014}

\refref{Sicklinger2014}

\refref{Zhang2014}

\refref{Kounev2015}

\refref{Bogomolov2015}

\refref{Bian2015}

\refref{Dols2016}

\refref{BenKhaled2014}

\refref{Saidi2016}

\refref{Yamaura2016}

\refref{Camus2015}

\refref{Camus2016}

\refref{Oh2016}

\refref{Xie2016}

\refref{Manbachi2016}

\refref{Schierz2012a}

\refref{Fourmigue2009}

\refref{Liu2001}

\refref{Carstens2003}

\refref{Stettinger2014}

\refref{Benedikt2016}

\refref{Busch2011}

\refref{Quesnel2005}

\refref{Arnold2014a}

\refref{Arnold2014}

\refref{Arnold2001}

\refref{Schweizer2014}

\refref{Schweizer2015d}

\refref{Sadjina2016}

\refref{Busch2016}

\refref{Arnold2010}

\refref{Gu2001}

\refref{Gu2004}

\refref{Schweizer2016}

\refref{Schweizer2015}

\refref{Schweizer2015a}

\refref{Andersson2016}

\refref{Galtier2015}

\refref{Acker2015}

\refref{Enge-Rosenblatt2011}

\refref{Aslan2015}

\refref{Wetter2010}


\paragraph{A-Causal}
\label{sr:causality:acausal}


\subsubsection{Time Constraints}

\paragraph{Analytic Simulation}
\label{sr:rel_time:analytic}

References in this category:

\refref{Pedersen2015}

\refref{Lin2011}

\refref{Hoepfer2011}

\refref{Tomulik2011}

\refref{Sun2011}

\refref{Friedrich2011}

\refref{Gonzalez2011}

\refref{Nutaro2011}

\refref{Busch2012}

\refref{Schmoll2012}

\refref{Ni2012}

\refref{Hassairi2012}

\refref{Schierz2012}

\refref{Quaglia2012}

\refref{Al-Hammouri2012}

\refref{Eyisi2012}

\refref{Riley2011}

\refref{Roche2012}

\refref{Fitzgerald2010}

\refref{Fitzgerald2013}

\refref{Kudelski2013}

\refref{Broman2013}

\refref{Benedikt2013}

\refref{Benedikt2013b}

\refref{Fuller2013}

\refref{Wang2013}

\refref{Hafner2013}

\refref{Zhao2014}

\refref{Li2011c}

\refref{Awais2013b}

\refref{Awais2013a}

\refref{Kuhr2013}

\refref{Viel2014}

\refref{Sicklinger2014}

\refref{Zhang2014}

\refref{Kounev2015}

\refref{Bogomolov2015}

\refref{Dols2016}

\refref{BenKhaled2014}

\refref{Saidi2016}

\refref{Yamaura2016}

\refref{Camus2015}

\refref{Camus2016}

\refref{Oh2016}

\refref{Xie2016}

\refref{Schierz2012a}

\refref{Fourmigue2009}

\refref{Liu2001}

\refref{Carstens2003}

\refref{Benedikt2016}

\refref{Busch2011}

\refref{Quesnel2005}

\refref{Arnold2014a}

\refref{Arnold2014}

\refref{Arnold2001}

\refref{Schweizer2014}

\refref{Schweizer2015d}

\refref{Sadjina2016}

\refref{Busch2016}

\refref{Arnold2010}

\refref{Gu2001}

\refref{Gu2004}

\refref{Schweizer2016}

\refref{Schweizer2015}

\refref{Schweizer2015a}

\refref{Andersson2016}

\refref{Krammer2015}

\refref{Galtier2015}

\refref{Fey1997}

\refref{Acker2015}

\refref{Karner2010a}

\refref{Aslan2015}

\refref{Wetter2010}

\refref{Neema2014}


\paragraph{Scaled Real Time Simulation}
\hfill \\
\emph{Fixed}
\label{sr:rel_time:fixed_real_scaled_time_simulation}

A simulator is fixed scaled real time when it simulated time progresses according to a fixed linear relationship with the real time.

References in this category:

\refref{Faure2011}

\refref{Bian2015}

\refref{BenKhaled2012}

\refref{Pedersen2016}

\refref{Manbachi2016}

\refref{Stettinger2014}

\refref{Wetter2010}

\refref{Neema2014}


\emph{Dynamic}
\label{sr:rel_time:dy_real_scaled_time_simulation}

A simulator is dynamic scaled real time when the relation between the simulated time and the real time can be changed throughout the simulation.

References in this category:

\refref{Bombino2013}


\subsubsection{Rollback Support}

\paragraph{None}
\label{sr:rollback:none}

References in this category:

\refref{Pedersen2015}

\refref{Lin2011}

\refref{Hoepfer2011}

\refref{Faure2011}

\refref{Sun2011}

\refref{Bastian2011a}

\refref{Friedrich2011}

\refref{Gonzalez2011}

\refref{Schmoll2012}

\refref{Ni2012}

\refref{Schierz2012}

\refref{Gunther2012}

\refref{Quaglia2012}

\refref{Al-Hammouri2012}

\refref{Eyisi2012}

\refref{Riley2011}

\refref{Roche2012}

\refref{Kudelski2013}

\refref{Broman2013}

\refref{Benedikt2013}

\refref{Benedikt2013b}

\refref{Fuller2013}

\refref{Bombino2013}

\refref{Wang2013}

\refref{Hafner2013}

\refref{Zhao2014}

\refref{Li2011c}

\refref{Awais2013b}

\refref{Awais2013a}

\refref{Kuhr2013}

\refref{Viel2014}

\refref{Zhang2014}

\refref{Kounev2015}

\refref{Bogomolov2015}

\refref{Bian2015}

\refref{Dols2016}

\refref{BenKhaled2012}

\refref{BenKhaled2014}

\refref{Saidi2016}

\refref{Yamaura2016}

\refref{Camus2015}

\refref{Camus2016}

\refref{Oh2016}

\refref{Xie2016}

\refref{Manbachi2016}

\refref{Carstens2003}

\refref{Stettinger2014}

\refref{Benedikt2016}

\refref{Busch2011}

\refref{Arnold2014a}

\refref{Sadjina2016}

\refref{Busch2016}

\refref{Gu2001}

\refref{Gu2004}

\refref{Andersson2016}

\refref{Acker2015}

\refref{Aslan2015}


\paragraph{Single}
\label{sr:rollback:single}

Single rollback means that the simulation unit is capable of saving a certain state in the past (simulated time) and revert to that state.
Once reverted, the simulation unit cannot revert further in the past.

References in this category:

\refref{Tomulik2011}

\refref{Nutaro2011}

\refref{Busch2012}

\refref{Bombino2013}

\refref{Sicklinger2014}

\refref{Schierz2012a}

\refref{Liu2001}

\refref{Arnold2014}

\refref{Arnold2001}

\refref{Schweizer2014}

\refref{Schweizer2015d}

\refref{Arnold2010}

\refref{Schweizer2016}

\refref{Schweizer2015}

\refref{Schweizer2015a}

\refref{Fey1997}


\paragraph{Multiple}
\label{sr:rollback:multiple}

Multiple rollback means that the simulation unit is capable of saving a certain state in the past (simulated time) and revert to that state.
Once reverted, the simulation unit revert further into the past as many times as necessary.


\subsubsection{Availability}

\paragraph{Remote}
\label{sr:availability:remote}

References in this category:

\refref{Friedrich2011}

\refref{Busch2012}

\refref{Quaglia2012}

\refref{Eyisi2012}

\refref{Riley2011}

\refref{Roche2012}

\refref{Fitzgerald2010}

\refref{Fitzgerald2013}

\refref{Kudelski2013}

\refref{Fuller2013}

\refref{Bombino2013}

\refref{Zhao2014}

\refref{Awais2013b}

\refref{Awais2013a}

\refref{Bian2015}

\refref{Dols2016}

\refref{Yamaura2016}

\refref{Oh2016}

\refref{Liu2001}

\refref{Carstens2003}

\refref{Galtier2015}

\refref{Enge-Rosenblatt2011}


\paragraph{Local}
\label{sr:availability:local}

References in this category:

\refref{Pedersen2015}

\refref{Lin2011}

\refref{Hoepfer2011}

\refref{Faure2011}

\refref{Tomulik2011}

\refref{Sun2011}

\refref{Bastian2011a}

\refref{Gonzalez2011}

\refref{Nutaro2011}

\refref{Schmoll2012}

\refref{Ni2012}

\refref{Hassairi2012}

\refref{Schierz2012}

\refref{Gunther2012}

\refref{Al-Hammouri2012}

\refref{Broman2013}

\refref{Benedikt2013}

\refref{Benedikt2013b}

\refref{Wang2013}

\refref{Hafner2013}

\refref{Li2011c}

\refref{Awais2013b}

\refref{Awais2013a}

\refref{Kuhr2013}

\refref{Viel2014}

\refref{Sicklinger2014}

\refref{Zhang2014}

\refref{Kounev2015}

\refref{Bogomolov2015}

\refref{BenKhaled2012}

\refref{BenKhaled2014}

\refref{Saidi2016}

\refref{Camus2015}

\refref{Camus2016}

\refref{Xie2016}

\refref{Manbachi2016}

\refref{Schierz2012a}

\refref{Fourmigue2009}

\refref{Stettinger2014}

\refref{Benedikt2016}

\refref{Busch2011}

\refref{Quesnel2005}

\refref{Arnold2014a}

\refref{Arnold2014}

\refref{Arnold2001}

\refref{Schweizer2014}

\refref{Schweizer2015d}

\refref{Sadjina2016}

\refref{Busch2016}

\refref{Arnold2010}

\refref{Gu2001}

\refref{Gu2004}

\refref{Schweizer2016}

\refref{Schweizer2015}

\refref{Schweizer2015a}

\refref{Andersson2016}

\refref{Krammer2015}

\refref{Galtier2015}

\refref{Fey1997}

\refref{Acker2015}

\refref{Aslan2015}


\subsection{Framework Requirements}

\subsubsection{Standard}

\paragraph{High Level Architecture}
\label{fr:standard:hla}

References in this category:

\refref{Eyisi2012}

\refref{Riley2011}

\refref{Awais2013b}

\refref{Awais2013a}

\refref{Neema2014}


\paragraph{Functional Mock-up Interface}
\label{fr:standard:fmi}

References in this category:

\refref{Pedersen2015}

\refref{Sun2011}

\refref{Bastian2011a}

\refref{Broman2013}

\refref{Wang2013}

\refref{Awais2013b}

\refref{Awais2013a}

\refref{Kuhr2013}

\refref{Viel2014}

\refref{Bogomolov2015}

\refref{Dols2016}

\refref{BenKhaled2012}

\refref{BenKhaled2014}

\refref{Saidi2016}

\refref{Camus2015}

\refref{Camus2016}

\refref{Pedersen2016}

\refref{Schierz2012a}

\refref{Arnold2014a}

\refref{Arnold2014}

\refref{Andersson2016}

\refref{Galtier2015}

\refref{Acker2015}

\refref{Aslan2015}

\refref{Neema2014}


\paragraph{Functional Digital Mock-up}
\label{fr:standard:fdmu}

References in this category:

\refref{Enge-Rosenblatt2011}


\subsubsection{Coupling}

\paragraph{Input/Output Assignments}
\label{fr:coupling_model:io_assignments}

References in this category:

\refref{Pedersen2015}

\refref{Lin2011}

\refref{Faure2011}

\refref{Sun2011}

\refref{Bastian2011a}

\refref{Friedrich2011}

\refref{Gonzalez2011}

\refref{Nutaro2011}

\refref{Busch2012}

\refref{Schmoll2012}

\refref{Ni2012}

\refref{Hassairi2012}

\refref{Gunther2012}

\refref{Quaglia2012}

\refref{Al-Hammouri2012}

\refref{Eyisi2012}

\refref{Riley2011}

\refref{Roche2012}

\refref{Fitzgerald2010}

\refref{Fitzgerald2013}

\refref{Kudelski2013}

\refref{Broman2013}

\refref{Benedikt2013}

\refref{Benedikt2013b}

\refref{Fuller2013}

\refref{Bombino2013}

\refref{Wang2013}

\refref{Hafner2013}

\refref{Zhao2014}

\refref{Li2011c}

\refref{Awais2013b}

\refref{Awais2013a}

\refref{Zhang2014}

\refref{Kounev2015}

\refref{Bogomolov2015}

\refref{Bian2015}

\refref{Dols2016}

\refref{BenKhaled2012}

\refref{Yamaura2016}

\refref{Camus2015}

\refref{Camus2016}

\refref{Pedersen2016}

\refref{Oh2016}

\refref{Xie2016}

\refref{Manbachi2016}

\refref{Schierz2012a}

\refref{Fourmigue2009}

\refref{Liu2001}

\refref{Carstens2003}

\refref{Stettinger2014}

\refref{Benedikt2016}

\refref{Busch2011}

\refref{Quesnel2005}

\refref{Arnold2014a}

\refref{Arnold2014}

\refref{Sadjina2016}

\refref{Busch2016}

\refref{Andersson2016}

\refref{Krammer2015}

\refref{Galtier2015}

\refref{Fey1997}

\refref{Acker2015}

\refref{Enge-Rosenblatt2011}

\refref{Karner2010a}

\refref{Aslan2015}

\refref{Wetter2010}


\paragraph{Algebraic Constraints}
\label{fr:coupling_model:algebraic_constraints}

References in this category:

\refref{Tomulik2011}

\refref{Friedrich2011}

\refref{Schierz2012}

\refref{Viel2014}

\refref{Sicklinger2014}

\refref{Arnold2001}

\refref{Schweizer2014}

\refref{Schweizer2015d}

\refref{Arnold2010}

\refref{Gu2001}

\refref{Gu2004}

\refref{Schweizer2016}

\refref{Schweizer2015}

\refref{Schweizer2015a}


\subsubsection{Number of Simulation Units}

\paragraph{Two}
\label{fr:num_sim:two}

References in this category:

\refref{Lin2011}

\refref{Sun2011}

\refref{Gonzalez2011}

\refref{Nutaro2011}

\refref{Busch2012}

\refref{Schmoll2012}

\refref{Ni2012}

\refref{Hassairi2012}

\refref{Quaglia2012}

\refref{Al-Hammouri2012}

\refref{Eyisi2012}

\refref{Riley2011}

\refref{Roche2012}

\refref{Fitzgerald2010}

\refref{Fitzgerald2013}

\refref{Kudelski2013}

\refref{Benedikt2013}

\refref{Benedikt2013b}

\refref{Fuller2013}

\refref{Bombino2013}

\refref{Wang2013}

\refref{Zhao2014}

\refref{Li2011c}

\refref{Zhang2014}

\refref{Kounev2015}

\refref{Bogomolov2015}

\refref{Bian2015}

\refref{Dols2016}

\refref{Pedersen2016}

\refref{Oh2016}

\refref{Xie2016}

\refref{Manbachi2016}

\refref{Fourmigue2009}

\refref{Liu2001}

\refref{Carstens2003}

\refref{Stettinger2014}

\refref{Schweizer2014}

\refref{Gu2001}

\refref{Fey1997}


\paragraph{Three or More}
\label{fr:num_sim:three_more}

References in this category:

\refref{Pedersen2015}

\refref{Hoepfer2011}

\refref{Faure2011}

\refref{Tomulik2011}

\refref{Bastian2011a}

\refref{Friedrich2011}

\refref{Schierz2012}

\refref{Gunther2012}

\refref{Broman2013}

\refref{Hafner2013}

\refref{Awais2013b}

\refref{Awais2013a}

\refref{Kuhr2013}

\refref{Viel2014}

\refref{Sicklinger2014}

\refref{BenKhaled2012}

\refref{BenKhaled2014}

\refref{Saidi2016}

\refref{Yamaura2016}

\refref{Camus2015}

\refref{Camus2016}

\refref{Schierz2012a}

\refref{Benedikt2016}

\refref{Busch2011}

\refref{Quesnel2005}

\refref{Arnold2014a}

\refref{Arnold2014}

\refref{Arnold2001}

\refref{Schweizer2015d}

\refref{Sadjina2016}

\refref{Busch2016}

\refref{Arnold2010}

\refref{Gu2004}

\refref{Schweizer2016}

\refref{Schweizer2015}

\refref{Schweizer2015a}

\refref{Andersson2016}

\refref{Krammer2015}

\refref{Galtier2015}

\refref{Acker2015}

\refref{Enge-Rosenblatt2011}

\refref{Karner2010a}

\refref{Aslan2015}

\refref{Wetter2010}

\refref{Neema2014}


\subsubsection{Domain}

\paragraph{CT}
\label{fr:domain:ct}

References in this category:

\refref{Pedersen2015}

\refref{Hoepfer2011}

\refref{Faure2011}

\refref{Tomulik2011}

\refref{Sun2011}

\refref{Bastian2011a}

\refref{Friedrich2011}

\refref{Gonzalez2011}

\refref{Busch2012}

\refref{Schmoll2012}

\refref{Ni2012}

\refref{Hassairi2012}

\refref{Schierz2012}

\refref{Gunther2012}

\refref{Quaglia2012}

\refref{Al-Hammouri2012}

\refref{Roche2012}

\refref{Fitzgerald2010}

\refref{Fitzgerald2013}

\refref{Broman2013}

\refref{Benedikt2013}

\refref{Benedikt2013b}

\refref{Bombino2013}

\refref{Wang2013}

\refref{Hafner2013}

\refref{Zhao2014}

\refref{Li2011c}

\refref{Kuhr2013}

\refref{Viel2014}

\refref{Sicklinger2014}

\refref{Zhang2014}

\refref{Bogomolov2015}

\refref{Bian2015}

\refref{Dols2016}

\refref{BenKhaled2012}

\refref{BenKhaled2014}

\refref{Saidi2016}

\refref{Yamaura2016}

\refref{Pedersen2016}

\refref{Oh2016}

\refref{Xie2016}

\refref{Schierz2012a}

\refref{Liu2001}

\refref{Carstens2003}

\refref{Stettinger2014}

\refref{Benedikt2016}

\refref{Busch2011}

\refref{Arnold2014a}

\refref{Arnold2014}

\refref{Arnold2001}

\refref{Schweizer2014}

\refref{Schweizer2015d}

\refref{Sadjina2016}

\refref{Busch2016}

\refref{Arnold2010}

\refref{Gu2001}

\refref{Gu2004}

\refref{Schweizer2016}

\refref{Schweizer2015}

\refref{Schweizer2015a}

\refref{Andersson2016}

\refref{Krammer2015}

\refref{Galtier2015}

\refref{Fey1997}

\refref{Acker2015}

\refref{Enge-Rosenblatt2011}

\refref{Karner2010a}

\refref{Aslan2015}

\refref{Wetter2010}


\paragraph{DE}
\label{fr:domain:de}

References in this category:

\refref{Lin2011}

\refref{Nutaro2011}

\refref{Al-Hammouri2012}

\refref{Eyisi2012}

\refref{Riley2011}

\refref{Fitzgerald2010}

\refref{Fitzgerald2013}

\refref{Kudelski2013}

\refref{Fuller2013}

\refref{Awais2013b}

\refref{Awais2013a}

\refref{Kuhr2013}

\refref{Zhang2014}

\refref{Kounev2015}

\refref{Bogomolov2015}

\refref{Camus2015}

\refref{Camus2016}

\refref{Fourmigue2009}

\refref{Quesnel2005}

\refref{Fey1997}

\refref{Karner2010a}

\refref{Neema2014}


\subsubsection{Dynamic structure}
\label{fr:dynamic_structure}

References in this category:

\refref{Karner2010a}


\subsubsection{Co-simulation Rate}

\paragraph{Single}
\label{fr:sim_rate:single}

References in this category:

\refref{Pedersen2015}

\refref{Lin2011}

\refref{Hoepfer2011}

\refref{Faure2011}

\refref{Tomulik2011}

\refref{Sun2011}

\refref{Bastian2011a}

\refref{Friedrich2011}

\refref{Nutaro2011}

\refref{Busch2012}

\refref{Schmoll2012}

\refref{Ni2012}

\refref{Hassairi2012}

\refref{Schierz2012}

\refref{Gunther2012}

\refref{Quaglia2012}

\refref{Al-Hammouri2012}

\refref{Eyisi2012}

\refref{Riley2011}

\refref{Roche2012}

\refref{Fitzgerald2010}

\refref{Fitzgerald2013}

\refref{Kudelski2013}

\refref{Broman2013}

\refref{Benedikt2013}

\refref{Benedikt2013b}

\refref{Fuller2013}

\refref{Bombino2013}

\refref{Wang2013}

\refref{Hafner2013}

\refref{Zhao2014}

\refref{Li2011c}

\refref{Viel2014}

\refref{Sicklinger2014}

\refref{Zhang2014}

\refref{Kounev2015}

\refref{Bogomolov2015}

\refref{Bian2015}

\refref{Dols2016}

\refref{BenKhaled2012}

\refref{BenKhaled2014}

\refref{Yamaura2016}

\refref{Pedersen2016}

\refref{Oh2016}

\refref{Xie2016}

\refref{Manbachi2016}

\refref{Schierz2012a}

\refref{Carstens2003}

\refref{Stettinger2014}

\refref{Benedikt2016}

\refref{Busch2016}

\refref{Arnold2010}

\refref{Galtier2015}

\refref{Aslan2015}

\refref{Wetter2010}


\paragraph{Multi}
\label{fr:sim_rate:multi}

Multi-rate co-simulation denotes that the framework distinguishes between slow and fast sub-systems and dimensions the communication step size accordingly, providing for interpolation/extrapolation of the slow systems.

References in this category:

\refref{Gonzalez2011}

\refref{Awais2013b}

\refref{Awais2013a}

\refref{Kuhr2013}

\refref{Camus2015}

\refref{Camus2016}

\refref{Busch2011}

\refref{Quesnel2005}

\refref{Arnold2014a}

\refref{Arnold2014}

\refref{Arnold2001}

\refref{Schweizer2014}

\refref{Schweizer2015d}

\refref{Sadjina2016}

\refref{Gu2001}

\refref{Gu2004}

\refref{Schweizer2016}

\refref{Schweizer2015}

\refref{Schweizer2015a}

\refref{Fey1997}

\refref{Acker2015}

\refref{Enge-Rosenblatt2011}

\refref{Neema2014}


\subsubsection{Communication Step Size}

\paragraph{Fixed}
\label{fr:sim_step_size:fixed}

References in this category:

\refref{Pedersen2015}

\refref{Faure2011}

\refref{Tomulik2011}

\refref{Bastian2011a}

\refref{Friedrich2011}

\refref{Gonzalez2011}

\refref{Busch2012}

\refref{Schmoll2012}

\refref{Ni2012}

\refref{Hassairi2012}

\refref{Schierz2012}

\refref{Quaglia2012}

\refref{Roche2012}

\refref{Kudelski2013}

\refref{Benedikt2013}

\refref{Benedikt2013b}

\refref{Bombino2013}

\refref{Hafner2013}

\refref{Zhao2014}

\refref{Li2011c}

\refref{Awais2013a}

\refref{Viel2014}

\refref{Sicklinger2014}

\refref{Zhang2014}

\refref{Bian2015}

\refref{Dols2016}

\refref{BenKhaled2012}

\refref{BenKhaled2014}

\refref{Pedersen2016}

\refref{Oh2016}

\refref{Xie2016}

\refref{Manbachi2016}

\refref{Carstens2003}

\refref{Stettinger2014}

\refref{Arnold2014a}

\refref{Arnold2001}

\refref{Schweizer2014}

\refref{Schweizer2015d}

\refref{Busch2016}

\refref{Arnold2010}

\refref{Gu2001}

\refref{Gu2004}

\refref{Schweizer2016}

\refref{Schweizer2015}

\refref{Schweizer2015a}

\refref{Andersson2016}

\refref{Fey1997}

\refref{Acker2015}

\refref{Wetter2010}

\refref{Neema2014}


\paragraph{Variable}
\label{fr:sim_step_size:variable}

References in this category:

\refref{Lin2011}

\refref{Hoepfer2011}

\refref{Sun2011}

\refref{Nutaro2011}

\refref{Gunther2012}

\refref{Al-Hammouri2012}

\refref{Eyisi2012}

\refref{Riley2011}

\refref{Fitzgerald2010}

\refref{Fitzgerald2013}

\refref{Broman2013}

\refref{Fuller2013}

\refref{Wang2013}

\refref{Awais2013b}

\refref{Kuhr2013}

\refref{Kounev2015}

\refref{Bogomolov2015}

\refref{Camus2015}

\refref{Camus2016}

\refref{Schierz2012a}

\refref{Benedikt2016}

\refref{Busch2011}

\refref{Quesnel2005}

\refref{Arnold2014}

\refref{Sadjina2016}

\refref{Galtier2015}

\refref{Neema2014}


\subsubsection{Strong Coupling Support}

\paragraph{None -- Explicit Method}
\label{fr:alg_loop:explicit}

References in this category:

\refref{Pedersen2015}

\refref{Lin2011}

\refref{Hoepfer2011}

\refref{Faure2011}

\refref{Sun2011}

\refref{Friedrich2011}

\refref{Gonzalez2011}

\refref{Nutaro2011}

\refref{Busch2012}

\refref{Schmoll2012}

\refref{Ni2012}

\refref{Hassairi2012}

\refref{Schierz2012}

\refref{Gunther2012}

\refref{Quaglia2012}

\refref{Al-Hammouri2012}

\refref{Eyisi2012}

\refref{Riley2011}

\refref{Roche2012}

\refref{Fitzgerald2010}

\refref{Fitzgerald2013}

\refref{Kudelski2013}

\refref{Broman2013}

\refref{Benedikt2013}

\refref{Benedikt2013b}

\refref{Fuller2013}

\refref{Bombino2013}

\refref{Wang2013}

\refref{Hafner2013}

\refref{Zhao2014}

\refref{Li2011c}

\refref{Awais2013b}

\refref{Awais2013a}

\refref{Zhang2014}

\refref{Kounev2015}

\refref{Bogomolov2015}

\refref{Bian2015}

\refref{Dols2016}

\refref{BenKhaled2012}

\refref{BenKhaled2014}

\refref{Saidi2016}

\refref{Yamaura2016}

\refref{Camus2015}

\refref{Camus2016}

\refref{Pedersen2016}

\refref{Oh2016}

\refref{Xie2016}

\refref{Manbachi2016}

\refref{Schierz2012a}

\refref{Carstens2003}

\refref{Stettinger2014}

\refref{Benedikt2016}

\refref{Busch2011}

\refref{Quesnel2005}

\refref{Arnold2014a}

\refref{Arnold2014}

\refref{Sadjina2016}

\refref{Busch2016}

\refref{Gu2001}

\refref{Gu2004}

\refref{Galtier2015}

\refref{Aslan2015}


\paragraph{Partial -- Semi-Implicit Method}
\label{fr:alg_loop:semi_implicit}

References in this category:

\refref{Busch2012}

\refref{Schweizer2014}

\refref{Schweizer2015d}

\refref{Schweizer2016}

\refref{Schweizer2015}

\refref{Schweizer2015a}


\paragraph{Full -- Implicit Method}
\label{fr:alg_loop:implicit}

References in this category:

\refref{Tomulik2011}

\refref{Bastian2011a}

\refref{Busch2012}

\refref{Viel2014}

\refref{Sicklinger2014}

\refref{Liu2001}

\refref{Arnold2001}

\refref{Arnold2010}

\refref{Acker2015}


\subsubsection{Results Visualization}

\paragraph{Postmortem}
\label{fr:results_visualization:post_mortem}
The results are available after the simulation.
References in this category:

\refref{Pedersen2015}

\refref{Lin2011}

\refref{Hoepfer2011}

\refref{Faure2011}

\refref{Tomulik2011}

\refref{Sun2011}

\refref{Bastian2011a}

\refref{Friedrich2011}

\refref{Gonzalez2011}

\refref{Nutaro2011}

\refref{Busch2012}

\refref{Schmoll2012}

\refref{Ni2012}

\refref{Schierz2012}

\refref{Gunther2012}

\refref{Quaglia2012}

\refref{Al-Hammouri2012}

\refref{Roche2012}

\refref{Kudelski2013}

\refref{Broman2013}

\refref{Benedikt2013}

\refref{Benedikt2013b}

\refref{Fuller2013}

\refref{Wang2013}

\refref{Hafner2013}

\refref{Zhao2014}

\refref{Li2011c}

\refref{Awais2013b}

\refref{Awais2013a}

\refref{Viel2014}

\refref{Sicklinger2014}

\refref{Zhang2014}

\refref{Kounev2015}

\refref{Bogomolov2015}

\refref{Bian2015}

\refref{Dols2016}

\refref{BenKhaled2012}

\refref{Camus2015}

\refref{Camus2016}

\refref{Xie2016}

\refref{Schierz2012a}

\refref{Liu2001}

\refref{Carstens2003}

\refref{Stettinger2014}

\refref{Benedikt2016}

\refref{Busch2011}

\refref{Quesnel2005}

\refref{Arnold2014a}

\refref{Arnold2014}

\refref{Arnold2001}

\refref{Schweizer2014}

\refref{Schweizer2015d}

\refref{Sadjina2016}

\refref{Busch2016}

\refref{Arnold2010}

\refref{Gu2001}

\refref{Gu2004}

\refref{Schweizer2016}

\refref{Schweizer2015}

\refref{Schweizer2015a}

\refref{Andersson2016}

\refref{Krammer2015}

\refref{Galtier2015}

\refref{Fey1997}

\refref{Acker2015}

\refref{Karner2010a}

\refref{Aslan2015}

\refref{Wetter2010}


\paragraph{Live}
\label{fr:results_visualization:live}

References in this category:

\refref{Hassairi2012}

\refref{Eyisi2012}

\refref{Riley2011}

\refref{Fitzgerald2010}

\refref{Fitzgerald2013}

\refref{Yamaura2016}

\refref{Pedersen2016}

\refref{Enge-Rosenblatt2011}

\refref{Neema2014}


\paragraph{Interactive}
\label{fr:results_visualization:interactive_live}

References in this category:

\refref{Bombino2013}

\refref{Yamaura2016}


\subsubsection{Communication Approach}

\paragraph{Jacobi}
\label{fr:communication_model:jacobi}

References in this category:

\refref{Pedersen2015}

\refref{Hoepfer2011}

\refref{Faure2011}

\refref{Tomulik2011}

\refref{Bastian2011a}

\refref{Friedrich2011}

\refref{Schmoll2012}

\refref{Schierz2012}

\refref{Gunther2012}

\refref{Kudelski2013}

\refref{Broman2013}

\refref{Hafner2013}

\refref{Awais2013b}

\refref{Awais2013a}

\refref{Sicklinger2014}

\refref{Bogomolov2015}

\refref{BenKhaled2012}

\refref{Saidi2016}

\refref{Xie2016}

\refref{Manbachi2016}

\refref{Schierz2012a}

\refref{Fourmigue2009}

\refref{Stettinger2014}

\refref{Busch2011}

\refref{Arnold2014a}

\refref{Arnold2014}

\refref{Schweizer2014}

\refref{Schweizer2015d}

\refref{Sadjina2016}

\refref{Busch2016}

\refref{Gu2001}

\refref{Gu2004}

\refref{Schweizer2016}

\refref{Schweizer2015}

\refref{Schweizer2015a}

\refref{Andersson2016}

\refref{Galtier2015}

\refref{Wetter2010}

\refref{Neema2014}


\paragraph{Gauss-Seidel}
\label{fr:communication_model:gauss_seidel}

References in this category:

\refref{Lin2011}

\refref{Hoepfer2011}

\refref{Sun2011}

\refref{Bastian2011a}

\refref{Gonzalez2011}

\refref{Nutaro2011}

\refref{Busch2012}

\refref{Ni2012}

\refref{Hassairi2012}

\refref{Schierz2012}

\refref{Quaglia2012}

\refref{Al-Hammouri2012}

\refref{Eyisi2012}

\refref{Riley2011}

\refref{Roche2012}

\refref{Fitzgerald2010}

\refref{Fitzgerald2013}

\refref{Benedikt2013}

\refref{Benedikt2013b}

\refref{Fuller2013}

\refref{Bombino2013}

\refref{Wang2013}

\refref{Hafner2013}

\refref{Zhao2014}

\refref{Li2011c}

\refref{Awais2013b}

\refref{Awais2013a}

\refref{Kuhr2013}

\refref{Viel2014}

\refref{Sicklinger2014}

\refref{Kounev2015}

\refref{Dols2016}

\refref{BenKhaled2012}

\refref{BenKhaled2014}

\refref{Camus2015}

\refref{Camus2016}

\refref{Pedersen2016}

\refref{Oh2016}

\refref{Carstens2003}

\refref{Stettinger2014}

\refref{Quesnel2005}

\refref{Arnold2014}

\refref{Arnold2001}

\refref{Busch2016}

\refref{Arnold2010}

\refref{Acker2015}

\refref{Aslan2015}


	\newpage
\section{List of Acronyms}

\begin{longtable}{ll}
CPS & Cyber-Physical System\\
CT    & Continuous Time\\
DE	&Discrete Event\\
DEVS & Discrete Event System Specification\\
DTSS & Discrete Time System Specification\\
FMI & Functional Mock-up Interface \\
FR & Framework Requirement\\
GVT & Global Virtual Time\\
IP & Intellectual Property\\
IVP & Initial Value Problem\\
NFR & Non-Functional Requirement \\
ODE & Ordinary Differential Equation\\
SR & Simulator Requirement\\
\end{longtable}

\fi

\end{document}